\pdfoutput=1
\documentclass[UKenglish,cleveref,autoref]{jsat-arxiv}

\usepackage{amsmath}
\usepackage{amssymb}
\usepackage{xspace}
\usepackage{tikz}
\usetikzlibrary{positioning,calc}
\usetikzlibrary{shapes.geometric,arrows,shapes}
\usetikzlibrary{decorations.pathreplacing}
\usetikzlibrary{backgrounds,fit}

\newenvironment{proofsketch}{\renewcommand{\proofname}{Proof Sketch}\proof}{\endproof}

\usetikzlibrary{arrows.meta,shapes.misc,backgrounds}
\definecolor{figInk}{HTML}{1F2937}
\definecolor{figGrey}{HTML}{6B7280}
\definecolor{figNeutralFill}{HTML}{F3F4F6}
\definecolor{figNeutralDraw}{HTML}{4B5563}
\definecolor{figSatFill}{HTML}{FDF0C2}
\definecolor{figSatDraw}{HTML}{D4A017}
\definecolor{figRecFill}{HTML}{DCEBFB}
\definecolor{figRecDraw}{HTML}{3A7BD5}
\definecolor{figConfFill}{HTML}{FCE1E1}
\definecolor{figConfDraw}{HTML}{D14343}
\definecolor{figBand}{HTML}{F7F8FA}
\tikzset{
	fig/.style={font=\small, line width=0.7pt, draw=figInk, text=figInk, >={Stealth[length=2.4mm,width=1.8mm]}},
	fbox/.style={rectangle, rounded corners=3pt, draw=figNeutralDraw, fill=figNeutralFill, inner xsep=5pt, inner ysep=3.5pt, align=center},
	fsat/.style={fbox, draw=figSatDraw, fill=figSatFill},
	frec/.style={fbox, draw=figRecDraw, fill=figRecFill},
	fconf/.style={fbox, draw=figConfDraw, fill=figConfFill},
	fwhite/.style={fbox, fill=white},
	fterm/.style={rounded rectangle, draw=figNeutralDraw, fill=white, inner xsep=7pt, inner ysep=3.5pt, align=center},
	fdec/.style={diamond, aspect=2.6, inner sep=1.5pt, draw=figNeutralDraw, fill=white, align=center},
	fedge/.style={->, draw=figInk},
	flabel/.style={font=\scriptsize, text=figGrey, fill=white, inner sep=1.2pt},
}

\allowdisplaybreaks

\DeclareMathOperator{\ante}{ante}
\DeclareMathOperator{\red}{red}
\DeclareMathOperator{\lv}{lv}
\DeclareMathOperator{\game}{game}
\DeclareMathOperator{\fru}{fru}
\DeclareMathOperator{\cw}{cw}
\DeclareMathOperator{\var}{var}
\DeclareMathOperator{\block}{Block}
\DeclareMathOperator{\Img}{Image}
\DeclareMathOperator{\sat}{sat}

\newcommand{\abs}{\fru}
\newcommand{\resop}[1]{\stackrel{#1}{\bowtie}}
\newcommand{\pf}{\mathfrak{R}}
\newcommand{\N}{\mathbb{N}}
\newcommand{\uc}[1]{\left[#1\right]}

\newcommand{\proofsystem}[1]{{\sf{#1}}\xspace}
\newcommand{\udl}{\proofsystem{UDL}}
\newcommand{\qcdcl}{\proofsystem{QCDCL}}
\newcommand{\qres}{\proofsystem{Q-Res}}
\newcommand{\levordqres}{\proofsystem{level-ordered Q-Res}}
\newcommand{\qresnp}{\text{\proofsystem{Q}}\ensuremath{\overset{\text{\proofsystem{NP}}}{\text{--}}}\text{\proofsystem{Res}}}
\newcommand{\farpr}{\proofsystem{\forall Red + Res}}
\newcommand{\qcon}{\proofsystem{Q-Con}}
\newcommand{\cdcl}{\proofsystem{CDCL}}
\newcommand{\ldqres}{\proofsystem{LD-Q-Res}}
\newcommand{\ldqcon}{\proofsystem{LD-Q-Con}}
\newcommand{\Resolution}{\proofsystem{Res}}
\newcommand{\caqe}{\ensuremath{\mathsf{CAQE}^{\text{\textsf{\scriptsize C\tiny ORE}}}}\xspace}
\newcommand{\Rex}[1]{\ensuremath{G^{\exists}_{#1}}}
\newcommand{\Rall}[1]{\ensuremath{G^{\forall}_{#1}}}

\title{Towards Understanding the Complexity of CAQE: A Proof-Theoretic Analysis of its Core Procedure}
\titlerunning{Towards Understanding the Complexity of CAQE}

\author{Benjamin {B\"{o}hm}}{Friedrich Schiller University Jena, Germany}{benjamin.boehm@uni-jena.de}{https://orcid.org/0000-0002-6098-5572}{}
\author{Olaf {Beyersdorff}}{Friedrich Schiller University Jena, Germany}{olaf.beyersdorff@uni-jena.de}{https://orcid.org/0000-0002-2870-1648}{Carl-Zeiss Foundation and DFG grant BE 4209/3-2}
\authorrunning{B.\,B\"{o}hm and O.\,Beyersdorff}
\Copyright{Benjamin B\"{o}hm and Olaf Beyersdorff}
\category{}
\relatedversion{}
\relatedversiondetails[cite=BohmB26-SAT]{Conference Version}{https://doi.org/10.4230/LIPIcs.SAT.2026.7}
\acknowledgements{We thank Leroy Chew for asking a question at SAT'25 on the proof-theoretic strength of CAQE, which led to this work. We used a large language model (Claude) as a writing assistant for this article; all results and proofs were checked by the authors.}

\nolinenumbers

\begin{document}

\maketitle

\begin{abstract}
CAQE (Clausal Abstraction for Quantifier Elimination) is currently the most successful algorithmic paradigm for solving Quantified Boolean Formulas (QBF) in practice, clearly dominating recent QBF competitions. While apparently being a very strong solver, not much is known about CAQE in theory. We propose a framework for formalising runs of the basic CAQE algorithm (where failed assumptions are not considered) as proofs in a proof system \caqe, which we use to analyse the algorithm proof-theoretically and which provides methods for future work to separate it from optimised versions of CAQE that are used in practice.
We show that one can perform strategy extraction on the \caqe proof system in such a way that strategy size serves as a lower bound for basic CAQE runs. Furthermore, we introduce a measure, which we call CAQE width, which not only acts as a lower bound on \caqe proofs, but -- with the number of $\exists\forall$ alternations as exponent -- as an upper bound as well. Using this analysis, we prove that Q-resolution with an NP oracle simulates \caqe on QCNFs of bounded quantifier alternation. The same holds for QCDCL (Quantified Conflict-Driven Clause Learning, formalised as a proof system) on $\Sigma_3^b$ QCNFs when purely propositional CDCL steps are not counted, after a simple preprocessing step. In this setting, both systems are indeed strictly stronger than \caqe, where the separation of QCDCL from \caqe uses cube learning.
\end{abstract}

\section{Introduction}

Quantified Boolean Formulas (QBF) extend the satisfiability problem (SAT) by quantification over Boolean variables. This  increases the complexity of the problem (from NP to PSPACE), the versatility of expressions (many problems are more succinctly and more naturally encoded into QBF), and the range of applications~\cite{ShuklaBPS19}.

Different from SAT, where conflict-driven clause learning (CDCL) is the ruling algorithmic paradigm, there are a number of competing  approaches to QBF solving. Quantified conflict-driven clause learning (QCDCL) lifts the CDCL paradigm~\cite{sathandbookcdcl} to QBF~\cite{ZM02,sathandbookqbf}. QCDCL can be further enhanced by dependency schemes~\cite{PeitlSS19b} and dependency learning~\cite{PeitlSS19} and is implemented in a number of QBF solvers (Qute~\cite{PeitlSS19-Qute}, DepQBF~\cite{LonsingE17}). A second approach comes through expansion of universal variables in the counter-example guided abstraction refinement (CEGAR) framework, implemented in RAReQS~\cite{JanotaKMC16,JM15}.  CAQE (Clausal Abstraction for Quantifier Elimination) is a third algorithmic paradigm, introduced in 2015~\cite{RabeT15,Tentrup19} and implemented in the eponymous solver. CAQE also utilises CEGAR, but in a fundamentally different way compared to expansion.

With these various and seemingly orthogonal approaches, it is stunning that since 2017 versions of CAQE have won each QBF evaluation in the prenex CNF track, sometimes by quite a big margin~\cite{PulinaS19,QBFEval}. Experimental evaluations of different solvers as in QBF evaluations are an impactful driver for algorithmic progress. A systematic and theoretically concise comparison of different algorithmic solving techniques -- in SAT, QBF and further fields --  comes through proof complexity~\cite{sathandbookpc,sathandbookqbf,BeyersdorffHS24}. Proof complexity of QBF has been hugely advanced over the past decade (cf.~\cite{M4CQBF} for a survey) and many theoretically clean and appealing QBF proof systems have been introduced together with proof methods to analyse them (e.g.~\cite{BBCP20,BeyersdorffBMP024,BeyersdorffBMP23,BeyersdorffCJ19,BeyersdorffBKMS25,BWJ14}).

When aiming to compare different solvers, we need proof systems tightly modelling solvers and such proof systems are typically quite different from rule-based calculi such as resolution~\cite{KBKF95,Gelder12,ELW13}, Frege or strong systems for proof logging~\cite{BBCP20,HeuleSB17,ChewH22}. A rigorous study of this kind has been performed on different versions of QCDCL and their relations to QBF resolution systems, unearthing surprising incomparability results -- e.g.\ between standard QCDCL and Q-resolution~\cite{BB23-LMCS} -- as well as identifying potential improvements for solving (e.g.\ out-of-order decisions~\cite{BohmB24-JAIR,BPBSAT22journal} or cube learning and pure literal elimination~\cite{BPB22IJCAIjournal}).

Such a tight theoretical model and comparison to other QBF solvers and proof systems is largely missing for the award-winning solver CAQE. An investigation into the proof-theoretic strength of CAQE was started in the paper~\cite{Tentrup17}, focussing on the comparison between CAQE and other CEGAR approaches such as RAReQS, and establishing their incomparability. However, the relation to QCDCL -- arguably the other main QBF solving approach -- has been open for the past decade.

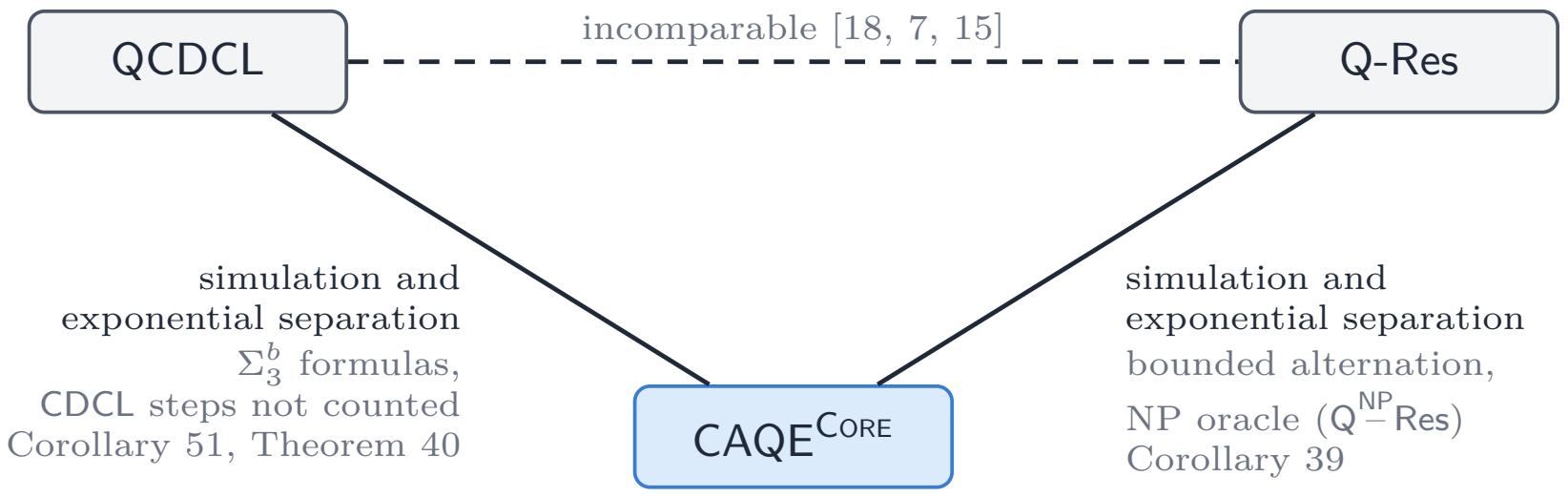
\begin{figure}
	\centering
	\begin{tikzpicture}[fig]
		\node[fbox, minimum width=2.2cm, minimum height=7mm] (QCDCL) at (0,0) {\qcdcl};
		\node[fbox, minimum width=2.2cm, minimum height=7mm] (QRes) at (8.4,0) {\qres};
		\node[frec, minimum width=2.2cm, minimum height=7mm] (CAQE) at (4.2,-2.6) {\caqe};
		\draw[line width=0.9pt] (QCDCL) -- coordinate[midway] (mL) (CAQE);
		\draw[line width=0.9pt] (QRes) -- coordinate[midway] (mR) (CAQE);
		\draw[line width=0.9pt, dash pattern=on 4pt off 3pt] (QCDCL) -- node[flabel, above=1pt, fill=none] {incomparable~\cite{BPB22IJCAIjournal,BB23-LMCS,BBSAT2021journal}} (QRes);
		\node[flabel, fill=none, align=right, anchor=north east, text=figInk] at ($(mL)+(-0.15,-0.05)$) {simulation and\\exponential separation\\[1pt] {\color{figGrey}$\Sigma_3^b$ formulas,}\\{\color{figGrey}\cdcl steps not counted}\\{\color{figGrey}\cref{Cor:QCDCLsimulatesCAQE}, \cref{thm:eq-sep}}};
		\node[flabel, fill=none, align=left, anchor=north west, text=figInk] at ($(mR)+(0.15,-0.05)$) {simulation and\\exponential separation\\[1pt] {\color{figGrey}bounded alternation,}\\{\color{figGrey}NP oracle ($\qresnp$)}\\{\color{figGrey}\cref{Cor:QresNPsimulatesCAQE}}};
	\end{tikzpicture}
	\caption{Hasse diagram of the simulation order of \qcdcl, \qres and \caqe (stronger systems on top). Solid lines denote a simulation together with an exponential separation. Both simulations hold in the oracle model of this paper: for \qres with an NP oracle ($\qresnp$) on QCNFs of bounded quantifier alternation, and for \qcdcl on $\Sigma_3^b$ QCNFs when purely propositional CDCL steps are not counted (and after a preprocessing step that establishes the XT-property); the separation of \qcdcl from \caqe uses cube learning. The dashed line indicates the known mutual exponential separations of \qcdcl and \qres in the standard (non-oracle) setting.}
	\label{Fig:SimulationOrderCAQE}
\end{figure}

\subparagraph*{Our results.}
We provide a theoretical proof-complexity analysis of the core procedure of CAQE  and compare it to QCDCL and Q-Resolution, resulting in the situation depicted in Figure~\ref{Fig:SimulationOrderCAQE}. We do not consider potential optimisations of CAQE such as failed assumptions -- we plan to address those in future work.
\begin{alphaenumerate}
\item As a first step, we tightly model the basic algorithmic paradigm of CAQE (without failed assumptions) as introduced in~\cite{RabeT15} as a proof system \caqe  such that runs of the basic CAQE algorithm can be translated into \caqe proofs and conversely, each tree-like proof gives rise to a run of the basic algorithm (\Cref{thm:caqe-alg-ps}). We note that a proof system $\farpr$ was introduced in~\cite{Tentrup17}  that likewise allows one to translate runs of the basic algorithm of CAQE  into $\farpr$, but the converse direction fails (\Cref{thm:farpr-caqe} and \Cref{Thm:SeparationCAQEfarpr}).

\item In our main result we compare the core procedure of CAQE to QCDCL and show the possibly surprising result that QCDCL simulates this basic version of CAQE on $\Sigma_3^b$ QCNFs (\Cref{Cor:QCDCLsimulatesCAQE}). This holds after a simple preprocessing step, which makes the formulas fulfil the \emph{XT-property}, and in a model where purely propositional CDCL steps are not counted (see below). Furthermore, we show that QCDCL with cube learning is exponentially stronger on selected formulas (\Cref{thm:eq-sep}).  Our theoretical comparison utilises our new  \caqe proof system and the \qcdcl proof systems developed in~\cite{BPB22IJCAIjournal,BB23-LMCS}.
\end{alphaenumerate}
As CAQE uses a massive number of SAT calls,  we believe that conceptually the fairest comparison is by discounting SAT computations.
This approach aligns with large parts of the QBF literature where propositional hardness is filtered out by allowing NP oracles in proofs~\cite{BHP20} (whereby lower bounds measure ``genuine'' QBF hardness and become more challenging).  For \caqe this is fairly simple: we just count a SAT call as one step, whereas for \qcdcl this is somewhat more intricate as QCDCL does not use SAT calls, but has all CDCL functionality ``built in''. We therefore define a \qcdcl model where purely propositional parts of the traces are not counted (imagine, e.g.\ that QCDCL stumbles upon a hard propositional formula not containing any universal variables).

Further, we show that our proof system \caqe is  simulated by Q-Resolution (\cref{Cor:QresNPsimulatesCAQE}), again in the oracle model and on bounded quantifier alternation. While $\farpr$ is known to be simulated by Q-Resolution~\cite{Tentrup17},  this result does not discount SAT calls, which means that it does not imply our stronger simulation. Q-Resolution is a weak system and in particular known to be incomparable to \qcdcl~\cite{BPB22IJCAIjournal,BB23-LMCS,BBSAT2021journal}.

Conceptually, the simulation order in Figure~\ref{Fig:SimulationOrderCAQE}  means that, from a proof-complexity perspective and in the oracle setting described above, the core procedure of CAQE without failed assumptions is weaker than Q-Resolution on QCNFs of bounded quantifier alternation and weaker than QCDCL on $\Sigma_3^b$ QCNFs.
Hence its practical outperformance appears to result from its clever and heavy SAT usage (cf.\ also the discussion in Section~\ref{sec:concl}) as well as from optimisations such as failed assumptions and the reuse of refinement clauses across different calls (cf.\ the discussion after \cref{thm:caqe-alg-ps}) -- which we did not consider in this work.

Technically, we achieve our results by a characterisation of \caqe proof size by a measure we call \emph{CAQE width}, which yields lower as well as upper bounds on \caqe proof size (\Cref{Prop:CAQEwidthUpperBoundFalse}, \Cref{Prop:LowerBoundCAQEwidth}). This characterisation gives rise to exponential lower bounds and the separations as well as to the extraction of short decision lists (\Cref{Prop:cwUDL}) and consequently the simulations.

\subparagraph*{Relation to the conference version.} A preliminary version of this article appeared in the proceedings of SAT 2026~\cite{BohmB26-SAT}. The present article contains complete proofs of all results, most of which were only sketched or omitted in the conference version. In the course of this, we have also corrected some errors and inaccuracies of the conference version.

\subparagraph*{Organisation.} After setting up notation in Section~\ref{sec:prelim} we explain the basic CAQE algorithm in Section~\ref{sec:caqe} and formalise it as a proof system \caqe in Section~\ref{Sec:Formalization}. In Section~\ref{Sec:LowerBoundsStrategySize} we show that \caqe is sound and allows extraction of winning strategies.
Section~\ref{Sec:CAQEwidth} introduces our main technical tool: CAQE width that we use to obtain lower and upper bounds and to prove the completeness of \caqe. In
Section~\ref{Sec:UDL} we show that decision lists for winning strategies can be extracted from \caqe proofs, yielding the simulation of \caqe by $\qresnp$.
Section~\ref{Sec:CAQEandQCDCL} compares \caqe and \qcdcl. We conclude with a discussion in Section~\ref{sec:concl}.

\section{Preliminaries}
\label{sec:prelim}

Variables $x$ and negated variables $\bar x$ are called \emph{literals}.
We denote the corresponding variable as $\var(x):=\var(\bar{x}):=x$.
A \emph{clause} is a disjunction of literals.
A \emph{unit clause} $(\ell)$ is a clause  consisting of only one literal. The \emph{empty clause} $(\bot)$ has zero literals.
A \emph{cube}  is a conjunction of literals, sometimes  viewed as a set of literals.
We define a \emph{unit cube} of a literal $\ell$, denoted by $\left[\ell \right]$, and the \emph{empty cube} $\uc{\top}$ with ``empty literal'' $\top$.
If $C$ is a clause or a cube, we define $\var(C):=\{ \var(\ell):\:\ell\in C \}$.

An \emph{assignment} $\sigma$ of a set of variables $V$ is a function $\sigma:\: V\rightarrow \{0,1\}$, which can also be interpreted as the set $\{ \ell:\: \var(\ell)\in V \wedge \sigma(\ell)=1 \}$. We denote the set of all assignments of $V$ as $\langle V\rangle$.
 For a cube $D$ we write $\neg D$ for the clause that consists of the negations of the literals of $D$. A clause $C$ is \emph{satisfied} by $\sigma$ if $C\cap \sigma\neq\emptyset$. A cube $D$ is \emph{satisfied} by $\sigma$ if $D\subseteq\sigma$, and it is \emph{falsified} by  $\sigma$ if $\neg D\cap \sigma\neq \emptyset$.   {A clause $C$ that is not satisfied by $\sigma$ can be \emph{restricted} by $\sigma$, defined as $C|_{\sigma}:=\bigvee_{\ell\in C, \bar{\ell}\not\in  \sigma}\ell$. Similarly we can restrict a non-falsified cube  $D$ as $D|_\sigma:=\bigwedge_{\ell\in D\backslash\sigma}\ell$.} We can restrict a clause $C$ to a set of variables $V$ by $C|_V:=\bigvee_{\ell\in C, \var(\ell)\in V}\ell$.  We can restrict a cube $D$ to a set of variables $V$ by $D|_V:=\bigwedge_{\ell\in D, \var(\ell)\in V}\ell$. Note the difference between the restriction $C|_\sigma$ by an assignment and the restriction $C|_V$ to a set of variables -- it will always be clear from the context which one is meant.

A \emph{CNF} ({conjunctive normal form}) is a conjunction of clauses and a \emph{DNF} (disjunctive normal form) is a disjunction of cubes.
A CNF (DNF) is \emph{satisfied} (\emph{falsified}) by $\sigma$ if all its clauses (cubes) are satisfied (falsified) by $\sigma$.

We use the notation $\sigma \vDash \psi$ for a propositional formula $\psi$ and assignment $\sigma$ if $\sigma$ satisfies $\psi$. We write $\varphi\vDash \psi$ for propositional formulas $\varphi$ and $\psi$ if each satisfying assignment of $\varphi$ satisfies $ \psi$.

A \emph{QBF} (quantified Boolean formula) $\Phi=\mathcal{Q}\cdot \varphi$ consists of a propositional formula $\varphi$,  called the \emph{matrix}, and a \emph{prefix} $\mathcal{Q}$. A \emph{prefix} $\mathcal{Q}={Q}_1V_1\dots{Q}_dV_d$ consists of non-empty and pairwise disjoint sets of variables $V_1,\dots,V_d$ (the \emph{quantifier blocks}) and quantifiers ${Q}_1,\ldots,{Q}_d\in \{\exists,\forall\}$ with ${Q}_i\neq {Q}_{i+1}$ for $i< d$. A QBF is \emph{propositional} if its prefix is empty or consists of a single existential block. In this case, the QBF is true if and only if its matrix is satisfiable, and we identify the QBF with its matrix.

A \emph{QCNF} is a QBF $\Phi=\mathcal{Q}\cdot \varphi$ with CNF $\varphi$.
We define $\var(\Phi):=\bigcup_{C\in \Phi}\var(C)$. We often identify a CNF with the set of its clauses and write $\varphi\subseteq\psi$ if every clause of $\varphi$ is a clause of $\psi$. Analogously, for QCNFs $\Phi=\mathcal{Q}\cdot\varphi$ and $\Psi=\mathcal{Q}'\cdot \psi$ with the same prefix we write $\Phi\subseteq \Psi$ if $\varphi\subseteq \psi$.
We restrict a CNF $\varphi$ by an assignment $\sigma$ as $\varphi|_{\sigma}:=\bigwedge_{C\in\varphi \text{ non-satisfied}}C|_\sigma$ and
 a QCNF  $\Phi=\mathcal{Q}\cdot \varphi$ as $\Phi|_\sigma:=\mathcal{Q}|_\sigma\cdot \varphi|_\sigma$,  where $\mathcal{Q}|_\sigma$ is obtained by deleting all variables from $\mathcal{Q}$ that  appear in $\sigma$. Note that variables that no longer occur in $\varphi|_\sigma$ remain in the prefix -- such variables are simply irrelevant for the truth value. Hence, restricting a QCNF by an assignment of a proper subset of its first block never deletes a quantifier block.

Throughout the paper we assume that the prefix of a QCNF ends with an existential block. This is no restriction, because the literals of a universal block at the end of the prefix can be removed from all clauses without changing the truth value (this is the universal reduction rule recalled below), after which the block can be dropped. Hence, we usually write the prefix of a QCNF as $\forall U_0\exists X_1\forall U_1\ldots \exists X_m \forall U_m \exists X_{m+1}$, where only the block $U_0$ is allowed to be empty. In this notation, $m$ is the number of universal blocks that are preceded by an existential block, that is, the number of alternations from $\exists$ to $\forall$.

Let $\Phi$ be a QCNF with the prefix $\forall U_0\exists X_1\forall U_1\ldots \exists X_m \forall U_m \exists X_{m+1}$. A \emph{(universal)  strategy} is a tuple of functions $s=(s_{U_0},\ldots,s_{U_m})$ with
$s_{U_j}:\: \langle X_1\cup \ldots\cup X_j \rangle \longrightarrow \langle  U_j \rangle  $
for $j=0,\ldots,m$.  If  for each $\gamma=(\gamma_1,\ldots, \gamma_{m+1})\in \langle X_1\cup \ldots\cup X_{m+1} \rangle$ the assignment
$	\gamma \cup \bigcup_{j=0}^{m} s_{U_j}\left(   \bigcup_{i=1}^j \gamma_i \right) $
falsifies the matrix of $\Phi$, then $s$ is called a \emph{universal winning strategy}. We define the size of a universal (winning) strategy as 	$|s|:=\sum_{j=0}^{m}|\Img(s_{U_j})|$.

Dually, an \emph{(existential)  strategy} is a tuple of functions $s=(s_{X_1},\ldots,s_{X_{m+1}})$ with
$	s_{X_j}:\: \langle U_0\cup \ldots\cup U_{j-1} \rangle \longrightarrow \langle  X_{j} \rangle  $
for $j=1,\ldots,m+1$. If for each $\gamma=(\gamma_0,\ldots, \gamma_{m})\in \langle U_0\cup \ldots\cup U_{m} \rangle$ the assignment
$\gamma \cup \bigcup_{j=1}^{m+1} s_{X_j}\left(   \bigcup_{i=0}^{j-1} \gamma_i \right) $
satisfies the matrix of $\Phi$, then $s$ is called an \emph{existential winning strategy}. The size of an existential (winning) strategy is defined as $|s|:=\sum_{j=1}^{m+1}|\Img(s_{X_j})|$.
We will  simply write $s(\gamma)$ instead of $\bigcup_{j=0}^{m} s_{U_j}\bigl(   \bigcup_{i=1}^j \gamma_i\bigr)$ or $\bigcup_{j=1}^{m+1} s_{X_j}\bigl(   \bigcup_{i=0}^{j-1} \gamma_i\bigr)$, respectively. It is well known that a QCNF is false if and only if it has a universal winning strategy, and true if and only if it has an existential winning strategy.

Often we need to follow the two-player game along a strategy. For this purpose we introduce the following notation.
\begin{definition}[$\game$ notation]\label{Def:game}
Let $s$ be a universal strategy for a QCNF with the prefix
$\forall U_0\exists X_1\forall U_1\ldots \exists X_m \forall U_m \exists X_{m+1}$,
where $U_0$ is allowed to be empty, and let $\gamma=(\gamma_1,\ldots,\gamma_j)\in \langle X_1\cup\ldots\cup X_j\rangle$ ($\gamma$ may also be an assignment of a superset of these variables). For $i\leq j$ we define
\begin{align*}
\game(s,\gamma,X_i)&:=\gamma_1\cup \ldots\cup \gamma_i \cup \bigcup_{z=0}^{i-1} s_{U_z}(\gamma_1\cup\ldots\cup \gamma_z) \qquad\text{and}\\
\game(s,\gamma,U_i)&:=\gamma_1\cup \ldots\cup \gamma_i \cup \bigcup_{z=0}^{i} s_{U_z}(\gamma_1\cup\ldots\cup \gamma_z).
\end{align*}
In other words, $\game(s,\gamma,X_i)\in \langle U_0\cup X_1\cup \ldots \cup U_{i-1}\cup X_i\rangle$ and $\game(s,\gamma,U_i)\in \langle U_0\cup X_1\cup \ldots \cup X_i\cup U_i\rangle$ are the assignments that are created during the two-player game up to the block $\exists X_i$ (resp. $\forall U_i$) if we follow the strategy $s$ and the assignment $\gamma$ for the existential variables. For convenience, we also set $\game(s,\gamma,X_0):=\emptyset$, so that $\Phi|_{\game(s,\gamma,X_0)}=\Phi$.
Analogously, for an existential strategy $s$ and $\gamma=(\gamma_0,\ldots,\gamma_j)\in\langle U_0\cup \ldots\cup U_j\rangle$ we define $\game(s,\gamma,U_i):=\gamma_0\cup\ldots\cup\gamma_i\cup \bigcup_{z=1}^{i}s_{X_z}(\gamma_0\cup\ldots\cup\gamma_{z-1})$ and $\game(s,\gamma,X_i):=\gamma_0\cup\ldots\cup\gamma_{i-1}\cup \bigcup_{z=1}^{i}s_{X_z}(\gamma_0\cup\ldots\cup\gamma_{z-1})$ for $i\leq j$ (resp. $i\leq j+1$).
\end{definition}

A \emph{proof system}  (cf.~\cite{CR79})  for a language $L$ is a poly-time computable function $\proofsystem{P}:\: \{0,1\}^*\rightarrow \{ 0,1 \}^*$ with $\Img(\proofsystem{P})=L$. A \emph{$\proofsystem{P}$-proof} $\pi$ of some $\Phi\in L$ is an element of $\{0,1\}^*$ such that $\proofsystem{P}(\pi)=\Phi$. The inclusion $\Img(\proofsystem{P})\subseteq L$ is  called \emph{soundness}, while $\Img(\proofsystem{P})\supseteq L$ is the \emph{completeness} of $\proofsystem{P}$. Here we consider the languages of unsatisfiable CNFs, false QCNFs and true QCNFs.
A proof   system $\proofsystem{P}_1$ \emph{simulates} a proof system $\proofsystem{P}_2$ if there exists a polynomial $p$ such that for each $\proofsystem{P}_2$-proof $\pi$ of a formula $\Phi$ there is a $\proofsystem{P}_1$-proof $\pi'$ of $\Phi$ with $|\pi'|\leq p(|\pi|+|\Phi|)$. Here $|\Phi|$ is the size of the formula. We include it because we measure the size of proofs by the number of their lines or steps, and a proof may be much smaller than the formula it proves. We say that $\proofsystem{P}_1$ \emph{p-simulates} $\proofsystem{P}_2$ (cf.~\cite{CR79}) if $\pi'$ can even be computed from $\pi$ by a poly-time computable function $f$. In this case we have $\proofsystem{P}_1(f(\pi))=\proofsystem{P}_2(\pi)$.
If $\proofsystem{P}_1$ and $\proofsystem{P}_2$ p-simulate each other, we write $\proofsystem{P}_1 \equiv_p \proofsystem{P}_2$. Two proof systems are \emph{incomparable} if neither simulates the other. We call $\proofsystem{P}_2$ \emph{exponentially weaker} than $\proofsystem{P}_1$ if $\proofsystem{P}_1$ simulates $\proofsystem{P}_2$ and there is a family of formulas that requires $\proofsystem{P}_2$-proofs of exponential size, while it has $\proofsystem{P}_1$-proofs of polynomial size. We also use these notions relative to a class of formulas, for instance for QCNFs of bounded quantifier alternation.

We  often add an \emph{NP oracle} to a proof system. While the details depend on the proof system, the idea is that $\proofsystem{P}$ proofs can evaluate SAT calls  for free~\cite{BHP20}. For such systems, the size of a proof counts a SAT call as a single step, and the propositional refutations that justify the calls are not counted (the proofs stay polynomially verifiable if these refutations are included). All simulations and separations in this paper that involve such a system refer to this measure.

Resolution (\Resolution) is a proof system for unsatisfiable CNFs. A \Resolution proof (or \emph{refutation}) $\pi$ of a CNF $\varphi$ is a sequence of clauses $\pi=(C_1,\ldots,C_k)$, where each $C_i$ is either an  axiom $C_i\in \varphi$ or derived by the
\emph{Resolution rule} $C_i=(C_a\vee C_b)\backslash\{ x,\bar{x} \}$ with $x\in C_a$, $\bar{x}\in C_b$ and $a,b<i$, and where $C_k=(\bot)$ is the empty clause. We define the size $|\pi|$ of $\pi$ as the number $k$ of its clauses.

\Resolution can be lifted to a proof system Q-Resolution (\qres)~\cite{KBKF95} for false QCNFs by restricting the Resolution rule to existentially quantified literals $x$, disallowing the derivation of tautological clauses, and adding the \emph{Reduction rule} $C_i=C_j\backslash\{u_1,\ldots,u_n\}$, where $u_1,\ldots,u_n$ are universal literals such that there are no existential literals in $C_j$ that are quantified right of each $u_h$. As for \Resolution, a \qres refutation ends with the empty clause.

We can add an NP-oracle to \qres and obtain $\qresnp$~\cite{BeyersdorffBMP23} by replacing the Resolution rule with the following \emph{Oracle rule}, which derives a non-tautological clause $C_i$ such that for some $\mathcal{G}\subseteq \{ C_1,\ldots,C_{i-1} \}$:

		$\bigwedge_{C\in \mathcal{G}}C^\exists \vDash C_i^\exists$ and
		$C^\forall$ is a subclause of $C_i^\forall$ for each $C\in \mathcal{G}$.
	Here $C^\exists$ (resp. $C^\forall$) denotes the existential (resp. universal) subclause of $C$.

\section{CAQE: the basic algorithmic paradigm}
\label{sec:caqe}

CAQE, which stands for \emph{Clausal Abstraction for Quantifier Elimination}, was introduced in 2015~\cite{RabeT15}. It is a recursive algorithm that  works on the first quantifier block of a QCNF and eliminates the quantifiers one by one. The idea is  simple, but elegant: Suppose we have a QCNF $\Phi=\exists V \mathcal{Q}\cdot \varphi$, such that $\exists V$ is the first quantifier block and $\mathcal{Q}$ are the remaining quantifiers. Our goal is to find out if $\Phi$ is true or false. Obviously, $\Phi$ is true if and only if there exists an assignment $\alpha\in \langle V\rangle$ such that $\Phi|_\alpha=\mathcal{Q}\cdot \varphi|_\alpha$ is true, whereas the truth value of $\Phi|_\alpha$ can be checked recursively. If we can find at least one such $\alpha$, we are done and we have proven the truth of $\Phi$.

However, if $\Phi$ is a false formula, then we need to check that for all assignments $\alpha\in\langle V\rangle$ the formula $\Phi|_\alpha$ is indeed false. If we  go through all of these assignments, we  end up with a runtime exponential in $|V|$. The trick is now to pick specific assignments $\alpha$ that suffice to show the falsity of~$\Phi$. More precisely, we want to pick the ``optimal'' or ``best'' assignments $\alpha$ which satisfy the ``highest number of'' clauses from $\Phi$. If these optimal assignments do not succeed in making $\Phi$ true, then neither will the assignments that are ``worse''. Let us demonstrate this on an example:

\begin{example}\label{Example:OptimalAssignment}
Let $\Phi$ be the QCNF with prefix $\exists x_1,x_2\forall u_1,u_2\exists y_1,y_2$ and the matrix

$\begin{array}{lll}
	x_1\vee u_1\vee y_1 & x_1\vee x_2\vee u_2\vee y_2 & \bar{y}_1\vee \bar{y}_2 \\
	\bar{x}_1\vee\bar{u}_1\vee y_1 &\bar{x}_1\vee\bar{u}_2\vee y_2 &  \\
	&\bar{x}_2\vee\bar{u}_2\vee y_2. &  \\
\end{array}$

\noindent This formula is false since the universal player wins by setting $u_1\equiv x_1$ and $u_2\equiv x_1\vee x_2$.
Now, there are four assignments for $\{x_1,x_2\}$ that would need to be tested: $\alpha_1=(x_1\mapsto 0,x_2\mapsto 0)$, $\alpha_2=(x_1\mapsto 0,x_2\mapsto 1)$, $\alpha_3=(x_1\mapsto 1,x_2\mapsto 0)$ and $\alpha_4=(x_1\mapsto 1,x_2\mapsto 1)$.

However, we can observe that $\alpha_3$  already satisfies  all the clauses that are satisfied by $\alpha_4$. Even worse, $\alpha_4$ satisfies fewer clauses than $\alpha_3$.  Hence, if we have shown that $\Phi|_{\alpha_3}$ is false, this also implies that $\Phi|_{\alpha_4}$ must be false as well.  Therefore, if we test $\alpha_3$ before $\alpha_4$, we can omit testing $\alpha_4$ altogether. In other words, checking $\alpha_1,\alpha_2,\alpha_3$ suffices to prove the falsity of $\Phi$.
\end{example}

An important aspect of CAQE is the direct inclusion of a SAT solver. Essentially,  the task of the SAT solver, usually implemented as CDCL, is to return an assignment $\alpha_j\in \langle V \rangle$ such that $\alpha_j$ is not ``worse'' than  or equal to one of the previously tested $\alpha_1,\ldots,\alpha_{j-1}$. Or more precisely, the SAT solver will return an $\alpha_j$ such that for each $i=1,\ldots,j-1$ the assignment $\alpha_j$ satisfies at least one clause that is not satisfied by $\alpha_i$. That also means that if the SAT solver cannot find such an $\alpha_j$ we can conclude that the formula is false.

The procedure for formulas $\Phi=\forall V \mathcal{Q}\cdot \varphi$ that start with a universal quantifier $\forall V$  is similar. The main difference is that the roles of universal and existential blocks are swapped and ``best'' or optimal assignments $\alpha\in \langle V\rangle$ are assignments that satisfy as few clauses as possible.

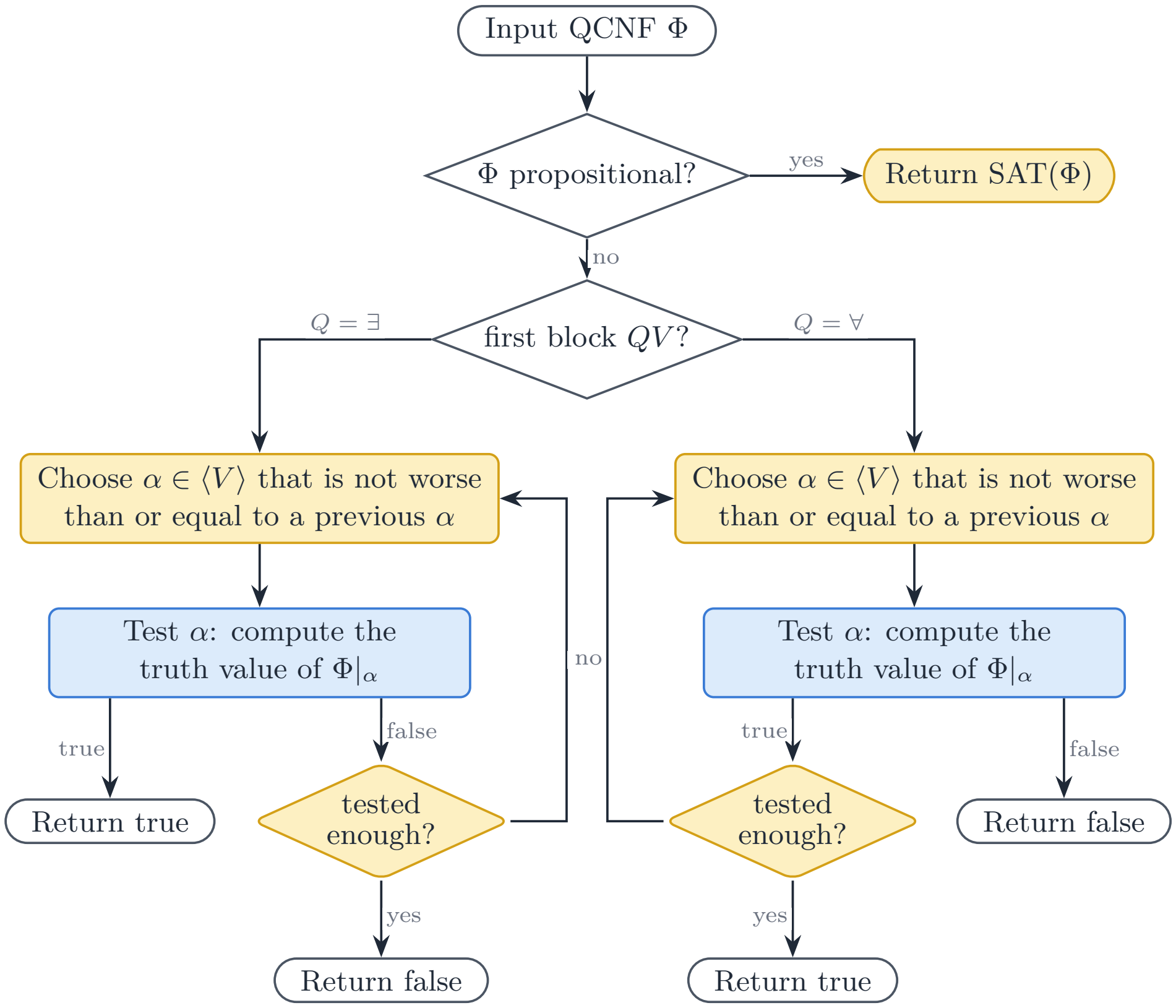
\begin{figure}[t]
	\centering
	\begin{tikzpicture}[fig]
		\node[fterm] (input) at (0,0) {Input QCNF $\Phi$};
		\node[fdec] (prop) at (0,-1.55) {$\Phi$ propositional?};
		\node[fsat, rounded rectangle] (satret) at (4.3,-1.55) {Return $\mathrm{SAT}(\Phi)$};
		\node[fdec] (block) at (0,-3.3) {first block $QV$?};
		\node[fsat, minimum width=4.5cm] (choL) at (-3.5,-5.0) {Choose $\alpha\in\langle V\rangle$ that is not worse\\than or equal to a previous $\alpha$};
		\node[frec, minimum width=4.5cm] (testL) at (-3.5,-6.65) {Test $\alpha$: compute the\\truth value of $\Phi|_\alpha$};
		\node[fterm] (trueL) at (-5.1,-8.45) {Return true};
		\node[fsat, diamond, aspect=2.2, inner sep=1pt] (enL) at (-2.2,-8.45) {tested\\[-2pt]enough?};
		\node[fterm] (falseL) at (-2.2,-10.15) {Return false};
		\node[fsat, minimum width=4.5cm] (choR) at (3.5,-5.0) {Choose $\alpha\in\langle V\rangle$ that is not worse\\than or equal to a previous $\alpha$};
		\node[frec, minimum width=4.5cm] (testR) at (3.5,-6.65) {Test $\alpha$: compute the\\truth value of $\Phi|_\alpha$};
		\node[fterm] (falseR) at (5.1,-8.45) {Return false};
		\node[fsat, diamond, aspect=2.2, inner sep=1pt] (enR) at (2.2,-8.45) {tested\\[-2pt]enough?};
		\node[fterm] (trueR) at (2.2,-10.15) {Return true};
		\draw[fedge] (input) -- (prop);
		\draw[fedge] (prop) -- node[flabel, above] {yes} (satret);
		\draw[fedge] (prop) -- node[flabel, right] {no} (block);
		\draw[fedge] (block.west) -| node[flabel, above, pos=0.25] {$Q=\exists$} (choL.north);
		\draw[fedge] (block.east) -| node[flabel, above, pos=0.25] {$Q=\forall$} (choR.north);
		\draw[fedge] (choL) -- (testL);
		\draw[fedge] (choR) -- (testR);
		\draw[fedge] (testL.south -| trueL) -- node[flabel, left] {true} (trueL);
		\draw[fedge] (testL.south -| enL) -- node[flabel, right] {false} (enL);
		\draw[fedge] (testR.south -| falseR) -- node[flabel, right] {false} (falseR);
		\draw[fedge] (testR.south -| enR) -- node[flabel, left] {true} (enR);
		\draw[fedge] (enL) -- node[flabel, right] {yes} (falseL);
		\draw[fedge] (enR) -- node[flabel, left] {yes} (trueR);
		\draw[fedge] (enL.east) -- ++(0.6,0) coordinate (loopL) -- node[flabel, right] {no} (loopL |- choL.east) -- (choL.east);
		\draw[fedge] (enR.west) -- ++(-0.6,0) coordinate (loopR) -- node[flabel, left] {no} (loopR |- choR.east) -- (choR.west);
	\end{tikzpicture}
	\caption{High-level flowchart of the basic CAQE algorithm. Yellow nodes are performed by the SAT solver. Blue nodes are performed recursively.}
	\label{Fig:CAQE}
\end{figure}

In the next section we will explain in further detail how CAQE compares the $\alpha$ in terms of optimality. A high-level overview of the functionality of CAQE is given in  \cref{Fig:CAQE}.

\section{Formalising basic CAQE as a proof system}\label{Sec:Formalization}

Intuitively, each clause $C_r$ of the given QCNF $\Phi$ is split
into two parts: the top part and the bottom part. The top part, which we denote as $C_{r,1}$, consists of variables from the first quantifier block  and contains all the information that we need to find all optimal assignments of that block. The bottom part, denoted as $C_{r,>1}$, contains the rest of the clause and is needed for the recursive call. Furthermore, $C_{r,1}$ and $C_{r,>1}$ are connected by an auxiliary variable $b_r$ (sometimes also called a \emph{bottom variable}) which we need to encode whether or not an assignment has satisfied $C_r$ on the first block. For each clause $C_r$ we have a separate variable $b_r$, and we define $B$ as the set of all auxiliary variables.

\begin{definition}\label{Def:Splitting}
	Let $\Phi:=QV_1\mathcal{Q}\cdot C_1\wedge \ldots \wedge C_k$ with clauses $C_1,\ldots,C_k$. We define auxiliary variables $B:=\{b_1,\ldots,b_k\}$ with $B\cap \var(\Phi)=\emptyset$. If $Q=\exists$, we define
	\begin{align*}
		C_{r,1} := \bigvee_{\substack{\ell \in C_r \\ \var(\ell) \in V_1}} \ell \vee b_r \qquad \mbox{and}\qquad
		C_{r,>1} := \bigvee_{\substack{\ell \in C_r \\ \var(\ell) \not\in V_1}} \ell \vee \bar{b}_r.
	\end{align*}
We set
\begin{align*}\Phi_{\exists V_1}:=\bigwedge_{r=1}^kC_{r,1} \qquad \mbox{and}\qquad \Phi_{>1}:=\mathcal{Q}\cdot\bigwedge_{r=1}^kC_{r,>1}.\end{align*}
Dually, if $Q=\forall$, we define
	$C_{r,1} := \bigwedge_{\substack{\ell \in C_r \\ \var(\ell) \in V_1}} \bigl(\bar{\ell} \vee \bar{b}_r\bigr)$ and
$	C_{r,>1} := \bigvee_{\substack{\ell \in C_r \\ \var(\ell) \not\in V_1}} \ell \vee \bar{b}_r$
and set $\Phi_{\forall V_1}:=\bigwedge_{r=1}^kC_{r,1}$ and $\Phi_{>1}:=\mathcal{Q}\cdot\bigwedge_{r=1}^kC_{r,>1}$.

For an assignment $\beta\in \langle B \rangle$, we define
\begin{align*}
	\Phi_{>1}|_{ \beta}:=
	\mathcal{Q} \cdot \bigwedge_{\substack{1\leq r\leq k}}C_{r,>1}|_{\beta}
		=
		\mathcal{Q} \cdot \bigwedge_{\substack{1\leq r\leq k\\\beta(b_r)=1}}C_{r,>1}|_\beta.
\end{align*}
Note that $\Phi_{\exists V_1}$ and $\Phi_{\forall V_1}$  are propositional formulas, while both $\Phi_{>1}|_{\beta}$ and $\Phi_{>1}$ are QCNFs.\footnote{In fact, we will  almost never use $\Phi_{>1}$ without a $\beta$. Hence it is unimportant whether or not $B$ is quantified.}
\end{definition}

In the whole paper, clause indices refer to the clauses $C_1,\ldots,C_k$ of the QCNF $\Phi$ we are currently considering. All formulas that we obtain from $\Phi$ by restrictions and by the splitting of \cref{Def:Splitting} consist of subclauses of the clauses $C_r$. Such a subclause keeps the index $r$ of the clause it comes from. In the same way, the auxiliary variables $B=\{b_1,\ldots,b_k\}$ are the same on all levels of the recursion. Hence, we can speak of ``the clause $C_r$'' of a restricted formula, and we can say that the clause $C_r$ is present in or satisfied in a restricted formula. Accordingly, we regard a restricted formula as its prefix together with the set of its indexed clauses: two formulas obtained from $\Phi$ by restrictions are equal if they have the same prefix and, for every index $r$, either both do not contain $C_r$ or both contain the same subclause of $C_r$ under the index $r$. For a restricted formula whose clauses have the index set $J\subseteq\{1,\ldots,k\}$, the auxiliary variables are the $b_r$ with $r\in J$, and a canonical assignment (\cref{Def:canonical} below) only assigns these variables.

The auxiliary $B$-variables that are assigned by $\beta$ indicate whether a clause is already satisfied in the first quantifier block, or if it might need to be satisfied using the remaining blocks.  The assignment $b_r\mapsto 1$ encodes the information ``$C_r$ might need to be satisfied in the bottom part'', while $b_r\mapsto 0$ means ``$C_r$ is already satisfied and does not need to be satisfied by the bottom part''.    More precisely, the propositional formula $\Phi_{ \exists V_1}$ (resp. $\Phi_{\forall V_1}$) is given to a SAT solver which returns a satisfying assignment $\alpha\cup \beta \in \langle V_1\cup B\rangle$. Such an assignment always exists (cf.\ \cref{Lemma:alphaBetaSatisfies} below). In the case $\exists V_1$, the variable $b_r$ must be set to $1$ if $\alpha$ does not satisfy $C_r$, otherwise it is allowed (but not forced) to be set to $0$.

Those clauses $C_r$ for which $b_r$ is set to $1$ are precisely the clauses that remain in $\Phi_{>1}$, hence $\Phi_{>1}|_\beta$ contains at least all the clauses that are not satisfied by the corresponding $\alpha$. That means if $\Phi_{>1}|_\beta$ is true, $\Phi$ turns out to be true as well.

In the case $\forall V_1$, the $b_r$ are interpreted a bit differently. Here $b_r$ needs to be set to $0$ if $\alpha$ satisfies $C_r$, and $b_r$ is allowed (but not forced) to be set to $1$ if $\alpha$ does not satisfy $C_r$. Therefore all clauses that remain in $\Phi_{>1}|_\beta$ are clauses that are not satisfied by $\alpha$. If $\Phi_{>1}|_\beta$ turns out to be false, then this also shows the falsity of $\Phi$ itself.

In both cases there is a special choice of $\beta$ that sets exactly the auxiliary variables of the clauses not satisfied by $\alpha$ to $1$. In the existential case, this is the minimal satisfying assignment $b_{\min}$ of $\Phi_{\exists V_1}$ in the terminology of~\cite{RabeT15}, and in the universal case it is the maximal satisfying assignment $b_{\max}$ of $\Phi_{\forall V_1}$.

\begin{definition}[canonical assignments]\label{Def:canonical}
	Let $\Phi$ be a QCNF with clauses $C_1,\ldots,C_k$ such that its prefix starts with the block $QV_1$, and let $\alpha\in \langle V_1\rangle$. The assignment $\beta\in \langle B \rangle$ is called \emph{canonical for $\alpha$} if for all $r=1,\ldots,k$ it holds that $\beta(b_r)=0$ if and only if $\alpha$ satisfies $C_r$.
\end{definition}

\begin{lemma}\label{Lemma:alphaBetaSatisfies}
	Let $\Phi$ be a QCNF with the matrix $C_1\wedge \ldots\wedge C_k$ that starts with the quantifier block $QV_1$ with $Q\in \{ \exists,\forall \}$, let $\alpha\in \langle V_1\rangle$ be arbitrary and let $\beta\in\langle B\rangle$ be canonical for $\alpha$. Then $\alpha\cup \beta$ satisfies $\Phi_{QV_1}$.
\end{lemma}
\begin{proof}
	Let us first consider the case $Q=\exists$. Then we have $\Phi_{\exists V_1}=\bigwedge_{r=1}^k C_{r,1}$ with
	$C_{r,1}=\bigvee_{\ell \in C_r,\, \var(\ell) \in V_1} \ell \vee b_r$
	for $r=1,\ldots,k$. If $\alpha$ satisfies $C_r$, then it does so by a literal from $V_1$, hence it also satisfies $C_{r,1}$. If $\alpha$ does not satisfy $C_r$, then we have $\beta(b_r)=1$, therefore $\beta$ satisfies $C_{r,1}$. Hence, $\alpha\cup \beta$ satisfies all $C_{r,1}$ for $r=1,\ldots,k$, and therefore also $\Phi_{QV_1}=\Phi_{ \exists V_1}$.

	In the case $Q=\forall$ we have $\Phi_{\forall V_1}=\bigwedge_{r=1}^k C_{r,1}$ with
	$C_{r,1} = \bigwedge_{\ell \in C_r,\, \var(\ell) \in V_1} (\bar{\ell} \vee \bar{b}_r)$
	for $r=1,\ldots,k$. If $\alpha$ satisfies $C_r$, then we have $\beta(b_r)=0$ and therefore $\beta$ satisfies $C_{r,1}$. If $\alpha$ does not satisfy $C_r$, then for all $\ell\in C_r$ with $\var(\ell)\in V_1$ we have $\alpha(\ell)=0$. Hence, $\alpha$ satisfies $C_{r,1}$. We conclude that $\alpha \cup \beta $ satisfies all $C_{r,1}$ for $r=1,\ldots,k$ and therefore also $\Phi_{QV_1}=\Phi_{\forall V_1}$.
\end{proof}

\begin{example}\label{Example:Splitting}
	Let $\Phi$ be the formula from  \cref{Example:OptimalAssignment}. Then $\Phi_{ \exists x_1,x_2}$ consists of the clauses
	\[	\begin{array}{llllll}
			x_1\vee b_1 & x_1\vee x_2\vee b_2 & (b_3)& \bar{x}_1\vee b_4 &\bar{x}_1\vee b_5 &\bar{x}_2\vee b_6
		\end{array}\]
and $\Phi_{>1}$ is the QCNF with the prefix $\forall u_1,u_2\exists y_1,y_2$ and the matrix
	 \[	\begin{array}{llllll}
	 		\bar{b}_1\vee u_1\vee y_1 & \bar{b}_2\vee u_2\vee y_2 & \bar{b}_3\vee\bar{y}_1\vee \bar{y}_2 & \bar{b}_4\vee\bar{u}_1\vee y_1 &\bar{b}_5\vee\bar{u}_2\vee y_2 & \bar{b}_6\vee\bar{u}_2\vee y_2.
	 	\end{array}\]
	 For $\alpha_3=(x_1\mapsto 1, x_2\mapsto 0)$, the canonical assignment is $\beta=\bar{b}_1\wedge\bar{b}_2\wedge b_3\wedge b_4\wedge b_5\wedge \bar{b}_6$. Then $\Phi_{>1}|_\beta$ is the QCNF $\forall u_1,u_2\exists y_1,y_2\cdot (\bar{y}_1\vee \bar{y}_2)\wedge(\bar{u}_1\vee y_1)\wedge(\bar{u}_2\vee y_2)=\Phi|_{\alpha_3}$.
\end{example}

\begin{remark}\label{Rem:TopVariables}
	In the introduction of CAQE in~\cite{RabeT15} the algorithm not only makes use of bottom variables $b_r$ but also of so-called \emph{top variables} $t_r$. Essentially, the $t_r$ are used for the clauses $C_{r,>1}$ instead of $\bar{b}_r$ with the side condition that each $t_r$ is equivalent to $\bar{b}_r$. This distinction between bottom and top variables is beneficial for the implementation of the recursive calls as the  auxiliary variables of successive recursion levels need to be tracked. However, we will omit these top variables since for our proof-theoretical formalisation and analysis of CAQE it suffices to use $\bar{b}_r$ instead of $t_r$.
\end{remark}

So far we have only explained how to obtain a solution $\alpha\cup \beta$ for the first iteration when no other $\alpha,\beta$ have been tested yet. Remember that we only want to test $\alpha$ that are not worse than previous ones. Because this concept of optimality does not primarily depend on $\alpha$ syntactically, we will instead use the $\beta$ for that as $\beta$  describes the effect of $\alpha$ on the clauses. In the $\exists V_1$ case, we want to make sure that for each previous $\alpha$ the  new $\alpha$ satisfies at least one clause that the previous $\alpha$ did not. In other words, the new $\beta$ has to flip at least one $1$ from each previous $\beta$ to $0$. If $\beta_1,\ldots,\beta_{j-1}$ are the previously tested assignments and $\beta_j$ is the new one, then $\beta_j$ has to satisfy the formula $\bigwedge_{a=1}^{j-1}\left( \bigvee_{b\in \beta_a^{-1}(1)}\bar{b}  \right)$ which we will simply add to $\Phi_{ \exists V_1}$ by conjunction.

Analogously, in the case $\forall V_1$ we require that the new $\beta$ flips at least one $0$ from each previous $\beta$ to~$1$. This can be enforced by  $\bigwedge_{a=1}^{j-1}\left( \bigvee_{b\in \beta_a^{-1}(0)}{b}  \right)$ which we add to $\Phi_{ \forall V_1}$ by conjunction.\footnote{Note that we consider a potentially weaker system of CAQE in this paper. In practical CAQE, the algorithm benefits from a feature of modern SAT solvers called \emph{failed assumptions}. To put it simply, such a SAT solver is called under a set of assumptions (here, literals over the auxiliary variables that select the clauses to be passed down) and, in case of unsatisfiability, returns a subset of these assumptions that already suffices for unsatisfiability -- in other words, a subset of the selected clauses. CAQE is then able to utilise this information when generating the constraints $\bigvee_{b\in \beta_a^{-1}(1)}\bar{b}$ $\left( \text{ resp. } \bigvee_{b\in \beta_a^{-1}(0)}{b}\right)$  by omitting the $B$-variables that were not necessary for the refutation. This obviously shortens those clauses in some cases and can lead to a significant improvement in runtime. We plan to add this optimisation to our framework in future work and separate it from the core procedure, but for now we will assume that the SAT solver always returns the full set of clauses as failed assumptions.}
We call these additional clauses \emph{refinement clauses} and introduce the notation
\begin{align*}
	\Rex{\beta}:=\bigvee_{b\in \beta^{-1}(1)}\bar{b} \qquad \text{and}\qquad \Rall{\beta}:=\bigvee_{b\in \beta^{-1}(0)}{b}
\end{align*}
for $\beta\in \langle B\rangle$. It is easy to see that the refinement clauses can be reformulated as follows.

\begin{lemma}\label{Lemma:RefinementReformulation}
	Let $\beta_1,\ldots,\beta_g\in \langle B\rangle$. Then
	\begin{enumerate}
		\item $\beta\in \langle B \rangle $ satisfies $\bigwedge_{a=1}^g\Rex{\beta_a} $ if and only if for each $a=1,\ldots,g$ there exists some $b\in B$ such that $\beta(b)<\beta_a(b)$, which means $\beta_a^{-1}(1)\not\subseteq \beta^{-1}(1)$,
		\item $\beta\in \langle B \rangle $ satisfies $\bigwedge_{a=1}^g\Rall{\beta_a} $ if and only if for each $a=1,\ldots,g$ there exists some $b\in B$ such that $\beta(b)>\beta_a(b)$, which means $\beta_a^{-1}(0)\not\subseteq \beta^{-1}(0)$.
	\end{enumerate}
\end{lemma}
\begin{proof}
	The statements become trivial by reformulating them: the assignment
	$\beta\in \langle B \rangle $ satisfies $\bigwedge_{a=1}^g\bigvee_{b\in \beta_a^{-1}(1)}\bar{b} $ if and only if for each $a=1,\ldots,g$ there exists some $b\in B$ such that $\beta(b)=0$ and  $\beta_a(b)=1$. Analogously, $\beta\in \langle B \rangle $ satisfies $\bigwedge_{a=1}^g\bigvee_{b\in \beta_a^{-1}(0)}{b} $ if and only if for each $a=1,\ldots,g$ there exists some $b\in B$ such that $\beta(b)=1$ and $\beta_a(b)=0$.
\end{proof}

Our goal to design a proof system that characterises the basic CAQE algorithm amounts to defining a structure that stores all the relevant information for each recursive call of the algorithm. We call these structures \emph{slices}. A slice contains the following information: The current formula for that call, all $\alpha$ that were tested, all $\beta$ that correspond to $\alpha$, and references to further recursive calls for $\Phi_{>1}|_{\beta}$ (i.e. references to further slices) and  a \Resolution refutation $\rho$ in case the propositional formula given to the SAT solver turns out to be unsatisfiable at some point. There are two exceptions: If $\Phi$ itself is an unsatisfiable propositional formula, a slice simply consists of a \Resolution refutation\footnote{One could use any refutationally complete propositional proof system.}, while for a satisfiable propositional formula a slice contains a satisfying assignment only. This is due to the fact that CAQE simply calls the SAT solver once only a propositional formula remains (i.e.\ when we reach the last quantifier block).

Each recursive call of CAQE either proves that its formula is false or that it is true. Therefore, we distinguish between \emph{refutation slices} and \emph{verification slices}. We define the two notions by mutual recursion, in the same way as the algorithm recurses. Let us first consider a slice for a formula that starts with an existential block. It is a refutation slice if all recursive calls returned ``false'' and the SAT solver finally reported that no further assignment needs to be tested. It is a verification slice if the last recursive call returned ``true''. For universal blocks the roles of true and false are swapped. We summarise this formally as follows.

\begin{definition}[slices]\label{Def:slices}
Let $\Phi$ be a QCNF with  prefix
$	Q_1V_1Q_2V_2\ldots Q_dV_d$,
where $V_i$ are disjoint sets of variables, $Q_i\in \{\exists,\forall\}$ and $Q_i\neq Q_{i+1}$ for all $i$, and with clauses $C_1,\ldots,C_k$. A \emph{slice} $\mathcal{S}$ for $\Phi$ together with its size $|\mathcal{S}|$ is one of the following. Each slice is either a \emph{refutation slice} or a \emph{verification slice}.
\begin{enumerate}
	\item $\mathcal{S}=(\Phi,\pi)$, where $\Phi$ is propositional and $\pi$ is a \Resolution refutation of $\Phi$; $|\mathcal{S}|:=1$. $\mathcal{S}$ is a refutation slice.
	\item $\mathcal{S}=(\Phi,\sigma)$, where $\Phi$ is propositional and $\sigma\in\langle\var(\Phi)\rangle$ satisfies $\Phi$; $|\mathcal{S}|:=1$. $\mathcal{S}$ is a verification slice.
	\item $\Phi$ is not propositional and $\mathcal{S}=(\Phi,(\alpha_1,\ldots,\alpha_g),(\beta_1,\ldots,\beta_g),(\mathcal{S}_1,\ldots,\mathcal{S}_g),\rho)$ with $g\geq 1$, $\alpha_j\in\langle V_1\rangle$, $\beta_j\in\langle B\rangle$, each $\mathcal{S}_j$ a slice for $\Phi_{>1}|_{\beta_j}$, and $\rho$ a \Resolution refutation or a placeholder, such that
	\begin{alphaenumerate}
		\item if $Q_1=\exists$, then $\alpha_j\cup\beta_j$ satisfies $\Phi_{\exists V_1}\wedge\bigwedge_{a=1}^{j-1}\Rex{\beta_a}$ for $j=1,\ldots,g$. $\mathcal{S}$ is a refutation slice if $\mathcal{S}_1,\ldots,\mathcal{S}_g$ are refutation slices and $\rho$ is a \Resolution refutation of $\Phi_{\exists V_1}\wedge\bigwedge_{a=1}^{g}\Rex{\beta_a}$; $\mathcal{S}$ is a verification slice if $\mathcal{S}_1,\ldots,\mathcal{S}_{g-1}$ are refutation slices and $\mathcal{S}_g$ is a verification slice (in this case $\rho$ is a placeholder);
		\item if $Q_1=\forall$, then $\beta_j$ is canonical for $\alpha_j$\footnote{We make this assumption because we explicitly do not consider failed assumptions in this article: a non-canonical $\beta_j$ on a universal block would delete clauses that are not satisfied by $\alpha_j$ from the formula $\Phi_{>1}|_{\beta_j}$ of the recursive call, which can be seen as a form of failed assumptions. The assumption corresponds to the maximal assignment $b_{\max}$ in~\cite{RabeT15}.} and $\alpha_j\cup\beta_j$ satisfies $\Phi_{\forall V_1}\wedge\bigwedge_{a=1}^{j-1}\Rall{\beta_a}$ for $j=1,\ldots,g$. $\mathcal{S}$ is a verification slice if $\mathcal{S}_1,\ldots,\mathcal{S}_g$ are verification slices and $\rho$ is a \Resolution refutation of $\Phi_{\forall V_1}\wedge\bigwedge_{a=1}^{g}\Rall{\beta_a}$; $\mathcal{S}$ is a refutation slice if $\mathcal{S}_1,\ldots,\mathcal{S}_{g-1}$ are verification slices and $\mathcal{S}_g$ is a refutation slice (in this case $\rho$ is a placeholder).
	\end{alphaenumerate}
	We set $|\mathcal{S}|:=g$ and say that $\mathcal{S}$ \emph{refers} to the slices $\mathcal{S}_1,\ldots,\mathcal{S}_g$. Furthermore, we set $\block(\mathcal{S}):=Q_1V_1$.
\end{enumerate}
For the propositional slices of the first two cases we also set $\block(\mathcal{S}):=\exists V_1$ if the prefix of $\Phi$ consists of the single block $\exists V_1$.
\end{definition}

Note that we do not include computations of the SAT solver (i.e.\ $\rho$) in the size of our slices. This is because first we want to concentrate on genuine QBF-hardness where SAT calls come for free~\cite{BHP20} and second  we do not specify the SAT solver that is used.

Next we define \emph{\caqe proofs} and distinguish between refutations (for false formulas) and verifications (for true formulas). Intuitively, a \caqe proof for a QCNF $\Phi$ is a sequence of slices, starting with the trivial cases of propositional formulas (one can consider these slices as axioms) and ending with a slice for $\Phi$ itself. From an algorithmic perspective, a proof is generated from back to front. The last slices of a proof correspond to the first quantifier blocks, while the first slices belong to the last quantifier block.

\begin{definition}[\caqe proofs]  \label{def:caqe}
A \emph{\caqe proof} $\kappa$ of a QCNF $\Phi$ is a sequence of slices $\kappa=(\mathcal{S}_1,\ldots,\mathcal{S}_e)$ such that $\mathcal{S}_e$ is a slice for $\Phi$ and each slice refers only to slices that occur earlier in the sequence. We set $|\kappa|:=\sum_{i=1}^{e}|\mathcal{S}_i|$.
The proof $\kappa$ is called a \emph{\caqe refutation} of $\Phi$ if $\mathcal{S}_e$ is a refutation slice, and a \emph{\caqe verification} of $\Phi$ if $\mathcal{S}_e$ is a verification slice.
\end{definition}

We can check in polynomial time whether a given sequence of slices is a \caqe refutation (or verification), since the conditions in \cref{Def:slices} are purely syntactic and whether a slice is a refutation or a verification slice only depends on the slices it refers to. Hence, \caqe is a proof system in the sense of~\cite{CR79} as soon as we have shown soundness (\cref{Cor:Soundness}) and completeness (\cref{Cor:Completeness}).

It is straightforward to check that \caqe proofs tightly model the basic algorithmic CAQE paradigm without failed assumptions.
\begin{theorem}  \label{thm:caqe-alg-ps}
	We call a \caqe proof \emph{tree-like} if every slice is referred to by at most one other slice. From each CAQE run without failed assumptions (that means the first \texttt{solve} procedure from~\cite{RabeT15}, where the SAT solver returns the maximal assignment $b_{\max}$ of the auxiliary variables on universal blocks) we can efficiently extract a tree-like \caqe proof. Conversely, each tree-like \caqe proof can be efficiently transformed into such a CAQE run.
\end{theorem}
\begin{proofsketch}
	Consider the recursion tree of a run of the \texttt{solve} procedure on a QCNF $\Phi$. Each call of \texttt{solve} gets a formula $\Psi$. This is $\Phi$ restricted by the assignments of the auxiliary variables that were chosen by the calls above it. In other words, $\Psi=\Phi_{>1}|_{\beta}$ for the current $\beta$ of the parent call (resp. $\Psi=\Phi$ for the top-level call). If $\Psi$ is propositional, the SAT solver returns either a \Resolution refutation $\pi$ or a satisfying assignment $\sigma$. This gives us the slice $(\Psi,\pi)$ or $(\Psi,\sigma)$. In the other case, the call iterates. In the $j^\text{th}$ iteration the SAT solver returns an assignment $\alpha_j\cup\beta_j$ of the clausal abstraction $\Psi_{Q_1V_1}$ together with the refinement clauses added so far, namely of $\Psi_{Q_1V_1}\wedge\bigwedge_{a<j}\Rex{\beta_a}$ (resp. $\Rall{\beta_a}$). After that, a recursive call on $\Psi_{>1}|_{\beta_j}$ follows. The iteration stops for one of two reasons. Either the recursive call returns the ``winning'' result for the player of the first block (true in the existential case, false in the universal case), or the SAT solver reports that the abstraction with all refinement clauses is unsatisfiable. In the latter case the SAT solver returns a \Resolution refutation $\rho$. If we collect $\Psi$, the $\alpha_j,\beta_j$, the slices of the recursive calls and $\rho$, we get a slice for $\Psi$. It follows directly from the description of the algorithm that this slice is a refutation slice if the call returns false and a verification slice if the call returns true. Now we list the slices of the recursion tree in post-order and obtain a \caqe proof of $\Phi$, which is tree-like because every call has exactly one parent. Its size is the number of SAT calls that returned an assignment plus the number of calls on propositional formulas that returned a refutation (these correspond to the slices of the first kind in \cref{Def:slices}, which have size $1$). Note that the SAT solver can return different satisfying assignments of the same abstraction, hence the run is not unique. In the universal case we assume that the SAT solver returns $b_{\max}$, which is the canonical assignment.

	For the other direction, let $\kappa$ be a tree-like \caqe proof of $\Phi$. We run \texttt{solve} with a SAT solver that answers each query with the assignment $\alpha_j\cup\beta_j$ (resp. the refutation $\rho$, $\pi$ or the assignment $\sigma$) that is stored in the corresponding slice of $\kappa$. By \cref{Def:slices}, these answers are correct answers of the SAT solver, and the recursive calls of the run follow the references between the slices of $\kappa$. If $\kappa$ is not tree-like, the same construction visits a slice once for each reference to it, so the run corresponds to the tree unfolding of $\kappa$. For a bounded number of quantifier blocks the unfolding is polynomial in $|\kappa|$, but in general it can be exponentially larger than $\kappa$.
\end{proofsketch}

 \caqe refutations only need to contain verification slices if the SAT solver did not succeed in immediately finding a $\beta$ on a universal block such that $\Phi_{>1}|_\beta$ is false. In most of our results we only need to consider a shortest \caqe refutation, for which we can assume that slices for universal blocks only contain one assignment (namely the correct assignment). This is because we get a smaller refutation if we delete the verification slices $\mathcal{S}_1,\ldots,\mathcal{S}_{g-1}$ (and all slices only referred to by them) from a refutation slice of a universal block. That means for such results we can assume that \caqe refutations do not contain verification slices.

Let us also point out a difference between the basic procedure and the implementation of CAQE. The implementation keeps one SAT solver per quantifier block for the whole run~\cite[Section~IV.A]{RabeT15}. Hence, the refinement clauses that were learned in one call are also available in all later calls for the same block, even if these calls work on different restricted formulas. In \cref{Def:slices}, however, the refutation $\rho$ of a slice may only use the refinement clauses of this slice, in accordance with the pseudocode of the first \texttt{solve} procedure of~\cite{RabeT15}. We leave the analysis of this reuse of refinements -- like that of failed assumptions -- to future work. In particular, our lower bounds refer to the basic procedure.

\begin{example}\label{Example:Refutation}
	Let $\Psi$ be the QCNF with the prefix $\exists d\forall e \exists x_1,x_2,f\forall u_1,u_2,w\exists y_1,y_2,z$ and the matrix
	\begin{align*}
		\small
		\begin{array}{lll}
			C_1=d\vee e \vee   x_1\vee u_1\vee y_1 & C_2=d\vee e \vee x_1\vee x_2\vee u_2\vee y_2 & C_3= d\vee e \vee \bar{y}_1\vee \bar{y}_2 \\
			C_4= d\vee e \vee \bar{x}_1\vee\bar{u}_1\vee y_1 &C_5=d\vee e \vee \bar{x}_1\vee\bar{u}_2\vee y_2 &  C_6=d\vee e \vee \bar{x}_2\vee\bar{u}_2\vee y_2 \\
			C_7=\bar{d}\vee \bar{e}\vee    f\vee w\vee z & C_8=\bar{d}\vee \bar{e}\vee   \bar{f}\vee \bar{w}\vee z &C_9=\bar{d}\vee \bar{e}\vee    \bar{z}
		\end{array}
	\end{align*}
	If we set $d\mapsto 0$ and $e\mapsto 0$ we end up with the formula from  \cref{Example:OptimalAssignment}, and if we set $d\mapsto 1$ and $e\mapsto 1$ we end up with the first equality formula $\mathtt{Eq}_1$ (cf.\ \cref{Def:Eq}).

	The sequence of slices $\kappa=(\mathcal{S}_1,\ldots,\mathcal{S}_{15})$ is a \caqe refutation of $\Psi$, with the slices defined as follows:
	\begin{align*}
		\mathcal{S}_1&=(\Psi|_{\bar{d}\wedge \bar{e}\wedge \bar{x}_1\wedge \bar{x}_2\wedge\bar{u}_1\wedge
			\bar{u}_2 },\pi_1)\\
		\mathcal{S}_2&=(\Psi|_{\bar{d}\wedge \bar{e}\wedge \bar{x}_1\wedge {x}_2\wedge\bar{u}_1\wedge
			{u}_2} ,\pi_2)\\
		\mathcal{S}_3&=(\Psi|_{\bar{d}\wedge \bar{e}\wedge {x}_1\wedge \bar{x}_2\wedge{u}_1\wedge
			{u}_2 },\pi_3)\\
		\mathcal{S}_4&=(\Psi|_{{d}\wedge {e}\wedge {f}\wedge
			{w} },\pi_4)\\
		\mathcal{S}_5&=(\Psi|_{{d}\wedge {e}\wedge\bar{f}\wedge
			\bar{w} },\pi_5)\\
		\mathcal{S}_6&=(\Psi|_{\bar{d}\wedge \bar{e}\wedge \bar{x}_1\wedge \bar{x}_2}, (\alpha_1=
		\bar{u}_1\wedge
		\bar{u}_2
		),(\beta_1),(\mathcal{S}_1),\rho_6  )\\
		\mathcal{S}_7&=(\Psi|_{\bar{d}\wedge \bar{e}\wedge \bar{x}_1\wedge {x}_2}, (\alpha_1=
		\bar{u}_1\wedge
		{u}_2
		),(\beta_1),(\mathcal{S}_2),\rho_7  )\\
		\mathcal{S}_8&=(\Psi|_{\bar{d}\wedge \bar{e}\wedge {x}_1\wedge \bar{x}_2}, (\alpha_1=
		{u}_1\wedge
		{u}_2
		),(\beta_1),(\mathcal{S}_3),\rho_8  )\\
		\mathcal{S}_{9}&=(\Psi|_{{d}\wedge {e}\wedge {f}}, (\alpha_1=
		{w}
		),(\beta_1),(\mathcal{S}_4),\rho_9  )\\
		\mathcal{S}_{10}&=(\Psi|_{{d}\wedge {e}\wedge\bar{f}}, (\alpha_1=
		\bar{w}
		),(\beta_1),(\mathcal{S}_5),\rho_{10}  )\\
		\mathcal{S}_{11}&=(\Psi|_{\bar{d}\wedge \bar{e}}, (\alpha_1=
		\bar{x}_1\wedge
		\bar{x}_2
		,\alpha_2=
		\bar{x}_1\wedge
		{x}_2
		,\alpha_3=
		x_1\wedge
		\bar{x}_2
		),(\beta_1,\beta_2,\beta_3),(\mathcal{S}_6,\mathcal{S}_7,\mathcal{S}_8),\rho_{11}  )\\
		\mathcal{S}_{12}&=(\Psi|_{d\wedge e}, (\alpha_1=
		[{f}]
		,\alpha_2=
		[\bar{f}]
		),(\beta_1,\beta_2),(\mathcal{S}_{9},\mathcal{S}_{10}) ,\rho_{12} )\\
		\mathcal{S}_{13}&=(\Psi|_{\bar{d}}, (\alpha_1=
		[\bar{e}]
		),(\beta_1),(\mathcal{S}_{11}),\rho_{13}  )\\
		\mathcal{S}_{14}&=(\Psi|_{d}, (\alpha_1=
		[e]
		),(\beta_1),(\mathcal{S}_{12}),\rho_{14}  )\\
		\mathcal{S}_{15}&=(\Psi, (\alpha_1=
		[\bar{d}]
		, \alpha_2=
		[d]
		),(\beta_1,\beta_2),(\mathcal{S}_{13},\mathcal{S}_{14}),\rho_{15}  )
	\end{align*}
	To keep the example readable, we omit variables that no longer occur in the matrix of the corresponding restricted formula. Quantifier blocks that become empty in this way are dropped as well. For example, in $\mathcal{S}_{12}$ we only consider assignments of $f$ because $x_1$ and $x_2$ no longer occur after the assignment $d\wedge e$. Strictly speaking, the omitted variables are assigned arbitrarily by the $\alpha$'s of their blocks -- for example, the $\alpha$'s of $\mathcal{S}_{11}$ also assign $f$, and the $\alpha$'s of $\mathcal{S}_6,\mathcal{S}_7,\mathcal{S}_8$ also assign $w$. This has no effect on the restricted formulas. All assignments $\beta$ are chosen canonically, such that they precisely indicate the effect of the corresponding $\alpha$. For example,  $\beta_1$ from the slice $\mathcal{S}_{15}$ is the assignment $b_1\wedge b_2\wedge b_3\wedge b_4\wedge b_5\wedge b_6\wedge \bar{b}_7\wedge \bar{b}_8\wedge \bar{b}_9$. The $\pi_i$ from the slices $\mathcal{S}_1,\ldots,\mathcal{S}_5$ are \Resolution refutations of the corresponding formulas and $\rho_{11},\rho_{12},\rho_{15}$ are \Resolution refutations according to   \cref{Def:slices}, whereas $\rho_6,\ldots,\rho_{10},\rho_{13},\rho_{14}$ are placeholders. Note that the slices of formulas starting with a universal quantifier, which are $\mathcal{S}_6,\mathcal{S}_7,\mathcal{S}_8,\mathcal{S}_{9},\mathcal{S}_{10},\mathcal{S}_{13},\mathcal{S}_{14}$, only consist of one $\alpha$ and one $\beta $ assignment. This is because we have always immediately picked the correct $\alpha$ that represents the universal winning strategy. In practice  we would potentially make a few incorrect guesses  before we pick a correct $\alpha$.

	We can depict \caqe proofs in a graph representation, where each node is a slice and we connect slices that refer to each other. Below we show a graph representation of $\kappa$. Each edge is labelled with the assignment $\alpha$ that leads to the recursive call. Slices with recursive calls are drawn in blue, and the propositional slices, which are handled by the SAT solver, are drawn in yellow. The rows correspond to the quantifier blocks of $\Psi$.

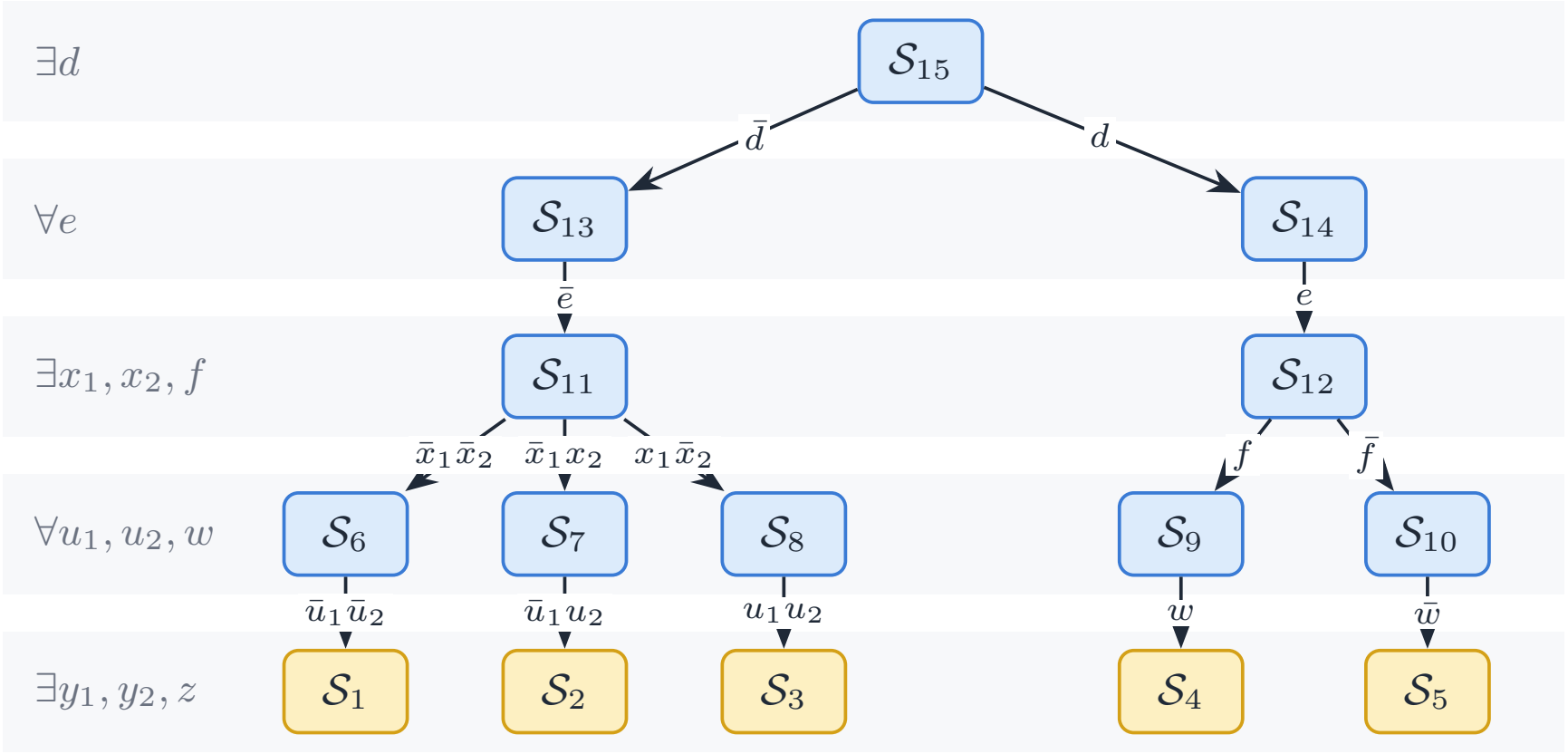
	\begin{center}
	\begin{tikzpicture}[fig, x=1cm, y=1.15cm,
			sl/.style={frec, minimum width=9mm, minimum height=6mm, inner sep=1.5pt},
			sp/.style={fsat, minimum width=9mm, minimum height=6mm, inner sep=1.5pt},
			al/.style={flabel, text=figInk}]
		\foreach \y/\lab in {0/{$\exists d$}, -1/{$\forall e$}, -2/{$\exists x_1,x_2,f$}, -3/{$\forall u_1,u_2,w$}, -4/{$\exists y_1,y_2,z$}} {
			\fill[figBand] (-6.1,\y-0.38) rectangle (5.3,\y+0.38);
			\node[anchor=west, font=\footnotesize, text=figGrey] at (-6.0,\y) {\lab};
		}
		\node[sl] (S15) at (0.6,0) {$\mathcal{S}_{15}$};
		\node[sl] (S13) at (-2,-1) {$\mathcal{S}_{13}$};
		\node[sl] (S14) at (3.4,-1) {$\mathcal{S}_{14}$};
		\node[sl] (S11) at (-2,-2) {$\mathcal{S}_{11}$};
		\node[sl] (S12) at (3.4,-2) {$\mathcal{S}_{12}$};
		\node[sl] (S6)  at (-3.6,-3) {$\mathcal{S}_{6}$};
		\node[sl] (S7)  at (-2,-3) {$\mathcal{S}_{7}$};
		\node[sl] (S8)  at (-0.4,-3) {$\mathcal{S}_{8}$};
		\node[sl] (S9)  at (2.5,-3) {$\mathcal{S}_{9}$};
		\node[sl] (S10) at (4.3,-3) {$\mathcal{S}_{10}$};
		\node[sp] (S1)  at (-3.6,-4) {$\mathcal{S}_{1}$};
		\node[sp] (S2)  at (-2,-4) {$\mathcal{S}_{2}$};
		\node[sp] (S3)  at (-0.4,-4) {$\mathcal{S}_{3}$};
		\node[sp] (S4)  at (2.5,-4) {$\mathcal{S}_{4}$};
		\node[sp] (S5)  at (4.3,-4) {$\mathcal{S}_{5}$};
		\draw[fedge] (S15) -- node[al, pos=0.45] {$\bar{d}$} (S13);
		\draw[fedge] (S15) -- node[al, pos=0.45] {$d$} (S14);
		\draw[fedge] (S13) -- node[al] {$\bar{e}$} (S11);
		\draw[fedge] (S14) -- node[al] {$e$} (S12);
		\draw[fedge] (S11) -- node[al, pos=0.5] {$\bar{x}_1\bar{x}_2$} (S6);
		\draw[fedge] (S11) -- node[al, pos=0.5] {$\bar{x}_1x_2$} (S7);
		\draw[fedge] (S11) -- node[al, pos=0.5] {$x_1\bar{x}_2$} (S8);
		\draw[fedge] (S12) -- node[al, pos=0.5] {$f$} (S9);
		\draw[fedge] (S12) -- node[al, pos=0.5] {$\bar{f}$} (S10);
		\draw[fedge] (S6) -- node[al] {$\bar{u}_1\bar{u}_2$} (S1);
		\draw[fedge] (S7) -- node[al] {$\bar{u}_1u_2$} (S2);
		\draw[fedge] (S8) -- node[al] {$u_1u_2$} (S3);
		\draw[fedge] (S9) -- node[al] {$w$} (S4);
		\draw[fedge] (S10) -- node[al] {$\bar{w}$} (S5);
	\end{tikzpicture}
	\end{center}

	One can observe that $\mathcal{S}_{11}$ only splits into three slices instead of four, although we have two variables remaining in the first block, namely $x_1$ and $x_2$. This is because only three of these four assignments are optimal assignments. As described in   \cref{Example:OptimalAssignment}, the assignment $x_1\wedge x_2$ does not need to be considered since it is worse than the previous assignment $x_1\wedge \bar{x}_2$, which is already handled in $\mathcal{S}_8$. This ensures that our proof, and therefore the graph representation, does not  necessarily  blow up -- although there are formulas where indeed all of the assignments need to be tested because all of them are optimal.

	Note that graph representations of \caqe proofs do not need to be trees. If we have two slices $\mathcal{S}$ and $\mathcal{S}'$ for the same formula, then we can omit $\mathcal{S}'$ and make all slices that refer to $\mathcal{S}'$ refer to $\mathcal{S}$ instead.
\end{example}

The following two lemmas relate the formulas $\Phi_{>1}|_{\beta_h}$ of the recursive calls to the restrictions $\Phi|_{\alpha_h}$ of the original formula. We will use them in the whole paper.

\begin{lemma}\label{Lemma:PhiBetaContainedInPhiAlpha}
	Let $\Phi$ be a QCNF that starts with a universal quantifier $\forall V_1$, and let $\alpha_h\in\langle V_1\rangle$ and $\beta_h\in\langle B\rangle$ be such that $\alpha_h\cup\beta_h$ satisfies $\Phi_{\forall V_1}$ (for example the assignments of a slice for $\Phi$, cf.\ \cref{Def:slices}). Then we have $\Phi_{>1}|_{ \beta_h}\subseteq \Phi|_{\alpha_h}$. Furthermore, if $\beta_h$ is canonical for $\alpha_h$ (in which case $\alpha_h\cup\beta_h$ satisfies $\Phi_{\forall V_1}$ by \cref{Lemma:alphaBetaSatisfies}), we even have $\Phi_{>1}|_{ \beta_h}= \Phi|_{\alpha_h}$. In particular, the latter holds for all assignments of a slice for $\Phi$ due to \cref{Def:slices}.
\end{lemma}
\begin{proof}
	Both formulas have the same prefix. Let $C_{q,>1}|_{\beta_h}$ be a clause of $\Phi_{>1}|_{\beta_h}$, that is, $\beta_h(b_q)=1$. Since $\alpha_h\cup\beta_h$ satisfies $C_{q,1}=\bigwedge_{\ell \in C_q,\, \var(\ell) \in V_1} (\bar{\ell} \vee \bar{b}_q)$, the assignment $\alpha_h$ falsifies all literals of $C_q$ over $V_1$. Hence, $\alpha_h$ does not satisfy $C_q$, and
	\begin{align*}
		C_{q,>1}|_{\beta_h} = \bigvee_{\substack{\ell \in C_q \\ \var(\ell) \not\in V_1}} \ell=C_q|_{\alpha_h}
	\end{align*}
	is a clause of $\Phi|_{\alpha_h}$. If $\beta_h$ is canonical for $\alpha_h$, then we have $\beta_h(b_q)=1$ for every clause $C_q$ that is not satisfied by $\alpha_h$. So conversely every clause $C_q|_{\alpha_h}$ of $\Phi|_{\alpha_h}$ is the clause $C_{q,>1}|_{\beta_h}$ of $\Phi_{>1}|_{\beta_h}$.
\end{proof}

\begin{lemma}\label{Lemma:PhiBetaContainedInPhiAlpha-exists}
	Let $\Phi$ be a QCNF that starts with an existential  quantifier $\exists V_1$, and let $\alpha_h\in\langle V_1\rangle$ and $\beta_h\in\langle B\rangle$ be such that $\alpha_h\cup\beta_h$ satisfies $\Phi_{\exists V_1}$ (for example the assignments of a slice for $\Phi$). Then we have $\Phi_{>1}|_{ \beta_h}\supseteq \Phi|_{\alpha_h}$.
	Furthermore, if $\beta_h$ is canonical for $\alpha_h$, we even have $\Phi_{>1}|_{ \beta_h}= \Phi|_{\alpha_h}$.
\end{lemma}
\begin{proof}
	Both formulas have the same prefix. Let $C_q|_{\alpha_h}$ be a clause of $\Phi|_{\alpha_h}$, that is, $\alpha_h$ does not satisfy $C_q$. Then $\alpha_h$ falsifies all literals of $C_q$ over $V_1$. Since $\alpha_h\cup\beta_h$ satisfies $C_{q,1}=\bigvee_{\ell \in C_q,\, \var(\ell) \in V_1}\ell\vee b_q$, we get $\beta_h(b_q)=1$. Hence,
	\begin{align*}
		C_q|_{\alpha_h}=\bigvee_{\substack{\ell \in C_q \\ \var(\ell) \not\in V_1}}\ell = C_{q,>1}|_{\beta_h}
	\end{align*}
	is a clause of $\Phi_{>1}|_{\beta_h}$. If $\beta_h$ is canonical for $\alpha_h$, then $\beta_h(b_q)=1$ only holds for clauses $C_q$ that are not satisfied by $\alpha_h$, so $\Phi_{>1}|_{\beta_h}$ contains no further clauses.
\end{proof}

Next we recall the $\farpr$ proof system introduced in~\cite{Tentrup17} to analyse CEGAR-type solving. We will show that $\farpr$, after adding an NP oracle,  simulates \caqe proofs, but is exponentially stronger (\Cref{Thm:SeparationCAQEfarpr}).

\begin{definition}[$\farpr$,~\cite{Tentrup17}]
	Let $\Phi=Q_1V_1\ldots Q_nV_n\cdot C_1\wedge \ldots \wedge C_k$ with $Q_i\in \{ \exists ,\forall  \}$ for each $i$ and $Q_j\neq Q_{j+1}$ for each $j$. A $\farpr$ proof of $\Phi$ is a sequence of sets $\mathcal{P}^a$, where $\mathcal{P}\subseteq \{ 1,\ldots,k \}$ and $a\in \{0,1,\ldots,n\}$, derived by the following three rules: the resolution rule  $\dfrac{\mathcal{P}_1^a\ldots \mathcal{P}_g^a\qquad \pi}{(\mathcal{P}_1\cup \ldots\cup \mathcal{P}_g)^{a-1}}$ where $Q_a=\exists$ and $\pi$ is a \Resolution refutation of $\bigwedge_{h=1}^g\bigvee_{q\in \mathcal{P}_h}C_q|_{V_a}$,  the reduction rule $\dfrac{\mathcal{P}^{a}}{\mathcal{P}^{a-1}}$ where $Q_a=\forall$ and there is no $u\in V_a$ with $u\in \bigvee_{q\in \mathcal{P}}C_q|_{V_a}$ and $\bar{u}\in \bigvee_{q\in \mathcal{P}}C_q|_{V_a}$,   and the axiom rule $\dfrac{}{\{q\}^n}$   with $q\in \{1,\ldots,k\}$.

	A $\farpr$ proof is called a \emph{$\farpr$ refutation} if we derived $\mathcal{P}^0$ for some $\mathcal{P}\subseteq \{1,\ldots,k\}$. The size of a $\farpr$ proof is the number of sets $\mathcal{P}^a$ and the sum of the sizes of all \Resolution refutations $\pi$ used in the proof. We call $\farpr^\proofsystem{NP}$ the version of $\farpr$ in which the $\pi$ are not counted.
\end{definition}

For the sake of convenience, we will often write $\mathcal{P}^{V_a}$ instead of $\mathcal{P}^{a}$ for $a>0$. That means the superscript names the block over which the next rule is applied. In the following proof we consider a QCNF with the prefix $\forall U_0\exists X_1\forall U_1\ldots\exists X_m\forall U_m\exists X_{m+1}$, where $U_0$ might be empty. Then the line $\mathcal{P}^0$ is $\mathcal{P}^{U_0}$ if $U_0$ is empty, and if not, we write $\mathcal{P}^{X_0}$ for it.

\begin{theorem}
	 \label{thm:farpr-caqe}
	Let $\kappa$ be a \caqe refutation of a QCNF $\Phi$ with $k$ clauses. Then we can efficiently construct a $\farpr^{\proofsystem{NP}}$ refutation $\nu$ of $\Phi$ with $|\nu|\leq 2|\kappa|+k$. Hence, $\farpr^\proofsystem{NP}$ p-simulates \caqe on false QCNFs.
\end{theorem}
\begin{proof}
	Let $\Phi$ have the prefix $\forall U_0\exists X_1\forall U_1\ldots\exists X_m\forall U_m\exists X_{m+1}$, where $U_0$ is allowed to be empty, and the clauses $C_1,\ldots,C_k$.
	Let $\kappa=(\mathcal{S}_1,\ldots,\mathcal{S}_e)$ and w.l.o.g. let $\kappa$ only consist of relevant slices (meaning that all slices have to be connected to $\mathcal{S}_e$ somehow) and let each slice of a universal block contain only one assignment, hence referring to one further slice only (cf.\ the discussion after \cref{thm:caqe-alg-ps}). In particular, all slices in $\kappa$ are refutation slices. The formulas of the axiom slices $\mathcal{S}=(\Psi,\pi)$ have the prefix $\exists X_{m+1}$, and we set $\block(\mathcal{S}):=\exists X_{m+1}$ for them. Now let $\mathcal{S}\in\kappa$ be a slice with $\block(\mathcal{S})=\exists X_i$ for some $i\in\{1,\ldots,m+1\}$. We will define a set $\mathcal{P}_{\mathcal{S}}\subseteq\{1,\ldots,k\}$ of clause indices by induction on $m+1-i$ and show that the line $\mathcal{P}_{\mathcal{S}}^{U_{i-1}}$ can be derived in $\farpr$. During the construction we maintain the following invariant:
	\begin{description}
		\item[($\ast$)] If $\mathcal{S}$ is a slice for the formula $\Psi$, then every $q\in\mathcal{P}_{\mathcal{S}}$ is the index of a clause that is present in $\Psi$.
	\end{description}
	Note that if $C_q$ is present in $\Psi$, then the clause of $\Psi$ with index $q$ is simply $C_q$ without the literals of the variables that are already assigned. In particular, its literals from $X_i$ are precisely the literals of $C_q|_{X_i}$.

	The invariant allows us to continue with a reduction step. If $\mathcal{S}$ is not the last slice of $\kappa$, then it is referred to by a slice $(\Psi',(\alpha'),(\beta'),(\mathcal{S}),\rho')$ of the block $\forall U_{i-1}$. As $\beta'$ is canonical for $\alpha'$, we have $\Psi=\Psi'_{>1}|_{\beta'}=\Psi'|_{\alpha'}$ by \cref{Lemma:PhiBetaContainedInPhiAlpha}. By ($\ast$), for every $q\in\mathcal{P}_{\mathcal{S}}$ the clause of $\Psi'$ with index $q$ is not satisfied by $\alpha'$, which means that $\alpha'$ falsifies all literals of $C_q$ from $U_{i-1}$. Hence, the clauses $C_q$ with $q\in \mathcal{P}_{\mathcal{S}}$ do not contain complementary literals from $U_{i-1}$, and the reduction step $\dfrac{\mathcal{P}_{\mathcal{S}}^{U_{i-1}}}{\mathcal{P}_{\mathcal{S}}^{X_{i-1}}}$ is valid.

	For $i=m+1$, the slice $\mathcal{S}$ is an axiom slice of the form $\mathcal{S}=(\Psi,\pi)$, where $\pi$ is a \Resolution  refutation of the propositional formula $\Psi$. Let $\mathcal{P}_{\mathcal{S}}:=\{ q_1,\ldots,q_a \}$ be the set of indices of all clauses that are present in $\Psi$. As we have seen above, $\Psi=C_{q_1}|_{X_{m+1}}\wedge\ldots\wedge C_{q_a}|_{X_{m+1}}$. Therefore,
	\begin{align*}
		\dfrac{ \{q_1\}^{X_{m+1}}\ldots \{q_a\}^{X_{m+1}} \qquad\pi }{\{ q_1,\ldots,q_a \}^{U_m}}
	\end{align*}
	is a valid resolution step in $\farpr$, where the lines $\{q_h\}^{X_{m+1}}$ are axioms.

	For $i\leq m$, let
	$\mathcal{S}=\left( \Psi, (\alpha_1,\ldots,\alpha_g),(\beta_1,\ldots,\beta_g),(\mathcal{S}_{j_1},\ldots,\mathcal{S}_{j_g}), \rho\right)$
	with
	$\block(\mathcal{S})=\exists X_{i}$. This slice refers to the slices $\mathcal{S}_{j_1},\ldots,\mathcal{S}_{j_g}$ with $\block(\mathcal{S}_{j_1})=\ldots=\block(\mathcal{S}_{j_g})=\forall U_{i}$. Each $\mathcal{S}_{j_h}$ refers to exactly one slice $\mathcal{S}'_{j_h}$ with $\block(\mathcal{S}'_{j_h})=\exists X_{i+1}$. Inductively, we have already constructed the sets $\mathcal{P}_{\mathcal{S}'_{j_1}},\ldots,\mathcal{P}_{\mathcal{S}'_{j_g}}$ and derived the lines $\mathcal{P}_{\mathcal{S}'_{j_h}}^{U_i}$, and with the reduction steps from above also the lines $\mathcal{P}_{\mathcal{S}'_{j_h}}^{X_i}$. We set $\mathcal{P}_{\mathcal{S}}:=\bigcup_{h=1}^g\mathcal{P}_{\mathcal{S}'_{j_h}}$. The invariant ($\ast$) holds for $\mathcal{P}_{\mathcal{S}}$ because each clause that is present in the formula of $\mathcal{S}'_{j_h}$ is also present in $\Psi$.

	We claim that $\bigwedge_{h=1}^g\bigvee_{q\in \mathcal{P}_{\mathcal{S}'_{j_h}}}C_q|_{X_i}$ is unsatisfiable. Assume for the sake of contradiction that this formula is satisfied by an assignment $\alpha\in \langle X_i\rangle$.  Let $\beta\in \langle B \rangle$ be the assignment that is canonical for $\alpha$ with respect to $\Psi$. Due to \cref{Lemma:alphaBetaSatisfies}, $\alpha\cup\beta$ satisfies $\Psi_{\exists X_i}$.
	Now, for each $h$ there exists some $q\in \mathcal{P}_{\mathcal{S}'_{j_h}}$ such that $\alpha$ satisfies $C_q|_{X_i}$, and hence the clause $C_q$ of $\Psi$. Therefore $\beta(b_q)=0$. On the other hand, $C_q$ is present in the formula of $\mathcal{S}'_{j_h}$, and this formula is a subformula of $\Psi_{>1}|_{\beta_h}$, so $\beta_h(b_q)=1$. Using \cref{Lemma:RefinementReformulation}, we conclude that $\beta$ satisfies $\Rex{\beta_h}$ for all $h=1,\ldots,g$. But this means that $\alpha\cup \beta$ satisfies $\Psi_{\exists X_i}\wedge \bigwedge_{h=1}^{g}\Rex{\beta_h}$, which contradicts the existence of the \Resolution refutation $\rho$ in the refutation slice $\mathcal{S}$.

	Hence, $\bigwedge_{h=1}^g\bigvee_{q\in \mathcal{P}_{\mathcal{S}'_{j_h}}}C_q|_{X_i}$ is unsatisfiable and it has a \Resolution refutation $\pi$.\footnote{We do not claim that $\pi$ can be obtained from $\rho$ efficiently. The refutation $\rho$ uses the auxiliary $B$-variables, and eliminating them might increase the size (cf.\ the discussion after this proof). As we use an NP oracle, we do not need to construct $\pi$.}  In particular,
	\begin{align*}
		\dfrac{ \mathcal{P}^{X_i}_{\mathcal{S}'_{j_1}}\ldots \mathcal{P}^{X_i}_{\mathcal{S}'_{j_g}}\qquad \pi }{\left( \bigcup_{h=1}^g\mathcal{P}_{\mathcal{S}'_{j_h}}\right) ^{U_{i-1}}}
	\end{align*}
	is a valid resolution step in $\farpr$.

	Finally, let $\mathcal{S}$ be the slice of the block $\exists X_1$ in $\kappa$. If $U_0$ is empty, then $\mathcal{S}$ is the last slice of $\kappa$, and the line $\mathcal{P}_{\mathcal{S}}^{U_0}=\mathcal{P}_{\mathcal{S}}^{0}$ completes the $\farpr$ refutation of $\Phi$. If not, then $\mathcal{S}$ is referred to by the last slice of $\kappa$, and the reduction step from above gives us $\mathcal{P}_{\mathcal{S}}^{X_0}=\mathcal{P}_{\mathcal{S}}^{0}$.

	It remains to consider the size. The refutation $\nu$ uses at most $k$ axioms and at most two further lines for each slice of an existential block in $\kappa$. Since each slice has size at least $1$, we get $|\nu|\leq 2|\kappa|+k$. Obviously, the construction can be computed in polynomial time -- note that the refutations $\pi$ are not part of a $\farpr^\proofsystem{NP}$ refutation.
\end{proof}

We would like to comment that it is unclear whether $\farpr$ simulates the non-oracle version of \caqe, or even the general algorithmic CAQE paradigm. While the \Resolution refutations $\pi$ in $\farpr$ proofs somehow correspond to the refutations $\rho$ generated by a SAT solver during a \caqe run, there might not be a polynomial-time transformation between those two. The reason is that the latter uses the auxiliary $B$-variables from \caqe, while the former does not. Resolving over those $B$-variables first would effectively eliminate them, but this might result in an exponential blow-up in the proof size. On the other hand, redefining $\farpr$ by allowing $B$-variables in the refutations~$\pi$ could cause other problems: In~\cite{Tentrup17} it was proven that $\farpr$ is equivalent to \levordqres. For the first direction, one can argue that all $\pi$ simply have to be stuck together in order to obtain a \levordqres refutation. Allowing $B$-variables in $\pi$ could render this simulation impossible.

We avoid these potential issues by adding an NP oracle and disregarding \Resolution refutations both for $\farpr$ and for \caqe. Whether the simulation of \caqe by $\farpr$ also holds without NP oracles remains open. Note that such a simulation would not directly imply \cref{thm:farpr-caqe}, since a \caqe refutation of small oracle size may contain large \Resolution refutations $\rho$.

\section{Strategy extraction and soundness of CAQE}\label{Sec:LowerBoundsStrategySize}
In this section we will show how to perform strategy extraction on \caqe proofs. Strategy extraction is a common technique not only for obtaining lower bounds, but also for proving  the soundness of a proof system. More precisely, from a given \caqe refutation we  extract a universal winning strategy, whereas from a \caqe verification we extract an existential winning strategy. This results in  strategy size as a lower bound for the size of \caqe proofs.

From now on, if not otherwise specified, we will always assume that $\Phi$ is a QCNF with  prefix $\forall U_0\exists X_1\forall U_1\exists X_2\forall U_2 \ldots \exists X_m \forall U_m \exists X_{m+1}$, where $U_0$ is allowed to be empty. For the sake of readability, we will denote $X:=X_1\cup \ldots\cup X_{m+1}$ and $U:=U_0\cup \ldots \cup U_m$. Furthermore, for a slice $\mathcal{S}$ of a \caqe proof of $\Phi$ with $\block(\mathcal{S})=QV$ we simply write $\block(\mathcal{S})=V$. In particular, $\block(\mathcal{S})=X_{m+1}$ for the propositional slices $\mathcal{S}=(\Psi,\pi)$ and $\mathcal{S}=(\Psi,\sigma)$, whose formula $\Psi$ has the prefix $\exists X_{m+1}$ (cf.\ \cref{Def:slices}).

Both extraction procedures rest on the following lemma. It says that the assignments of a refutation slice of an existential block cover all assignments of this block, and dually for verification slices of universal blocks.

\begin{lemma}\label{Lemma:Coverage}
	Let $\mathcal{S}$ be a slice of the form $(\Psi,(\alpha_1,\ldots,\alpha_g),(\beta_1,\ldots,\beta_g),(\mathcal{S}_1,\ldots,\mathcal{S}_g),\rho)$ with $\block(\mathcal{S})=QV$. Furthermore, let $\gamma\in\langle V\rangle$ and let $\beta_\gamma$ be canonical for $\gamma$ with respect to $\Psi$.
	\begin{enumerate}
		\item If $Q=\exists$ and $\mathcal{S}$ is a refutation slice, then there is some $h$ with $\beta_h^{-1}(1)\subseteq\beta_\gamma^{-1}(1)$. For every such $h$ we have $\Psi_{>1}|_{\beta_h}\subseteq\Psi|_\gamma$, and every clause of $\Psi$ that is satisfied by $\gamma$ is also satisfied by $\alpha_h$.
		\item If $Q=\forall$ and $\mathcal{S}$ is a verification slice, then there is some $h$ with $\beta_h^{-1}(0)\subseteq\beta_\gamma^{-1}(0)$. For every such $h$ we have $\Psi_{>1}|_{\beta_h}\supseteq\Psi|_\gamma$, and every clause of $\Psi$ that is satisfied by $\alpha_h$ is also satisfied by $\gamma$.
	\end{enumerate}
	In both cases, if $\alpha_h$ and $\gamma$ satisfy the same clauses of $\Psi$, then $\beta_h=\beta_\gamma$ and therefore $\Psi_{>1}|_{\beta_h}=\Psi|_\gamma$.
\end{lemma}
\begin{proof}
	Let $Q=\exists$. If there were no such $h$, then $\beta_\gamma$ would satisfy $\bigwedge_{a=1}^g\Rex{\beta_a}$ by \cref{Lemma:RefinementReformulation}. As $\gamma\cup\beta_\gamma$ satisfies $\Psi_{\exists V}$ by \cref{Lemma:alphaBetaSatisfies}, this contradicts the existence of the refutation $\rho$ of $\Psi_{\exists V}\wedge\bigwedge_{a=1}^g\Rex{\beta_a}$. Now let $h$ be as in the statement. Then every clause of $\Psi_{>1}|_{\beta_h}$ is a clause of $\Psi_{>1}|_{\beta_\gamma}$, and we have $\Psi_{>1}|_{\beta_\gamma}=\Psi|_\gamma$ by \cref{Lemma:PhiBetaContainedInPhiAlpha-exists}. If $\gamma$ satisfies $C_r$, then $\beta_\gamma(b_r)=0$ and therefore $\beta_h(b_r)=0$. Since $\alpha_h\cup\beta_h$ satisfies $C_{r,1}$, the assignment $\alpha_h$ satisfies $C_r$. If $\alpha_h$ and $\gamma$ satisfy the same clauses, then conversely $\beta_h(b_r)=0$ implies that $\alpha_h$ and hence $\gamma$ satisfies $C_r$, which means $\beta_\gamma(b_r)=0$. So we get $\beta_h=\beta_\gamma$.

	The case $Q=\forall$ is dual. Here the existence of $h$ follows in the same way with the clauses $\Rall{\beta_a}$, and the assignments $\beta_h$ are canonical by \cref{Def:slices}. Hence, $\beta_h^{-1}(0)\subseteq\beta_\gamma^{-1}(0)$ says that every clause that is satisfied by $\alpha_h$ is also satisfied by $\gamma$. Moreover, every clause of $\Psi_{>1}|_{\beta_\gamma}$ is a clause of $\Psi_{>1}|_{\beta_h}$, and $\Psi_{>1}|_{\beta_\gamma}=\Psi|_\gamma$ by \cref{Lemma:PhiBetaContainedInPhiAlpha}. If $\alpha_h$ and $\gamma$ satisfy the same clauses, then $\beta_h=\beta_\gamma$, because both assignments are canonical.
\end{proof}

\begin{proposition}\label{Prop:ExtractUniWinStratFromCAQE}
	From a \caqe refutation $\kappa$  of a QCNF $\Phi$
	 we can efficiently extract a universal winning strategy $s$ such that
	$|\Img(s_{U_j})|\leq
	|\{\mathcal{S}\in \kappa :\: \block(\mathcal{S})=U_j\}|
	\leq |\kappa|$
	for all $j=0,\ldots,m$ with $U_j\neq\emptyset$ (if $U_0$ is empty, we ignore the trivial function $s_{U_0}$). By summing over all universal blocks, we even obtain
	$|s|\leq |\kappa|$.
\end{proposition}
\begin{proof}
	For every refutation slice $\mathcal{S}\in\kappa$ we define a universal winning strategy $t^{\mathcal{S}}$ for the formula $\Psi$ of $\mathcal{S}$, by induction along the references of $\kappa$. Recall that by \cref{Def:slices} the last slice referred to by a refutation slice of a universal block is a refutation slice, and all slices referred to by a refutation slice of an existential block are refutation slices.
	\begin{itemize}
		\item If $\mathcal{S}=(\Psi,\pi)$, then $\Psi$ is an unsatisfiable propositional formula, and $t^{\mathcal{S}}$ is the empty strategy.
		\item Let $\mathcal{S}=(\Psi,(\alpha_1,\ldots,\alpha_g),(\beta_1,\ldots,\beta_g),(\mathcal{S}_1,\ldots,\mathcal{S}_g),\rho)$ with $\block(\mathcal{S})=\forall V$. Then $\mathcal{S}_g$ is a refutation slice for $\Psi_{>1}|_{\beta_g}$, and we have $\Psi_{>1}|_{\beta_g}=\Psi|_{\alpha_g}$ by \cref{Lemma:PhiBetaContainedInPhiAlpha}. The strategy $t^{\mathcal{S}}$ plays $\alpha_g$ and then follows $t^{\mathcal{S}_g}$.
		\item Let $\mathcal{S}=(\Psi,(\alpha_1,\ldots,\alpha_g),(\beta_1,\ldots,\beta_g),(\mathcal{S}_1,\ldots,\mathcal{S}_g),\rho)$ with $\block(\mathcal{S})=\exists V$. For an assignment $\gamma\in\langle V\rangle$ of the existential player let $\beta_\gamma$ be canonical for $\gamma$ with respect to $\Psi$, and let $h$ be minimal with $\beta_h^{-1}(1)\subseteq\beta_\gamma^{-1}(1)$. Such an $h$ exists by \cref{Lemma:Coverage}, and we have $\Psi_{>1}|_{\beta_h}\subseteq\Psi|_\gamma$. After $\gamma$, the strategy $t^{\mathcal{S}}$ follows $t^{\mathcal{S}_h}$. This strategy is winning for $\Psi_{>1}|_{\beta_h}$ and therefore also for $\Psi|_\gamma$, since an assignment that falsifies a clause of $\Psi_{>1}|_{\beta_h}$ falsifies the matrix of $\Psi|_\gamma$.
	\end{itemize}
	By induction, $t^{\mathcal{S}}$ is a universal winning strategy for the formula of $\mathcal{S}$. In particular, for the last slice $\mathcal{S}_e$ of $\kappa$ we obtain a universal winning strategy $s:=t^{\mathcal{S}_e}$ for $\Phi$, and $\Phi$ is indeed false.

	For each $j=0,\ldots,m$, every value of $s_{U_j}$ is the last universal assignment $\alpha_g$ of a slice $\mathcal{S}\in\kappa$ with $\block(\mathcal{S})=U_j$. This gives
	$|\Img(s_{U_j})|\leq |\{\mathcal{S}\in \kappa :\: \block(\mathcal{S})=U_j\}|$. Finally, the construction is efficient: given $\kappa$ and $\gamma\in\langle X\rangle$, we only have to follow one path of slices through $\kappa$ in order to compute $s(\gamma)$.
\end{proof}

For true QCNFs and \caqe verifications, an analogous result holds.

\begin{proposition}\label{Prop:ExtractExiWinStratFromCAQE}
	From a \caqe verification $\kappa$ of a QCNF $\Phi$ we can efficiently extract an existential  winning strategy $s$ such that
	$|\Img(s_{X_j})|\leq
	|\{\mathcal{S}\in \kappa:\: \block(\mathcal{S})=X_j\}|
	\leq |\kappa|$
	for $j=1,\ldots,m+1$. By summing over all existential blocks, we even obtain
		$	|s|\leq |\kappa|$.
\end{proposition}
\begin{proof}
	The proof is dual to the proof of \cref{Prop:ExtractUniWinStratFromCAQE}. For every verification slice $\mathcal{S}\in\kappa$ we define an existential winning strategy $t^{\mathcal{S}}$ for the formula $\Psi$ of $\mathcal{S}$. By \cref{Def:slices}, the last slice referred to by a verification slice of an existential block is a verification slice, and all slices referred to by a verification slice of a universal block are verification slices.
	\begin{itemize}
		\item If $\mathcal{S}=(\Psi,\sigma)$, then $t^{\mathcal{S}}$ plays $\sigma$, which we extend arbitrarily to an assignment of all variables of $X_{m+1}$. This assignment satisfies $\Psi$.
		\item Let $\mathcal{S}=(\Psi,(\alpha_1,\ldots,\alpha_g),(\beta_1,\ldots,\beta_g),(\mathcal{S}_1,\ldots,\mathcal{S}_g),\rho)$ with $\block(\mathcal{S})=\exists V$. Then $\mathcal{S}_g$ is a verification slice for $\Psi_{>1}|_{\beta_g}$, and we have $\Psi_{>1}|_{\beta_g}\supseteq\Psi|_{\alpha_g}$ by \cref{Lemma:PhiBetaContainedInPhiAlpha-exists}. The strategy $t^{\mathcal{S}}$ plays $\alpha_g$ and then follows $t^{\mathcal{S}_g}$, which satisfies all clauses of $\Psi_{>1}|_{\beta_g}$ and therefore all clauses of $\Psi|_{\alpha_g}$.
		\item Let $\mathcal{S}=(\Psi,(\alpha_1,\ldots,\alpha_g),(\beta_1,\ldots,\beta_g),(\mathcal{S}_1,\ldots,\mathcal{S}_g),\rho)$ with $\block(\mathcal{S})=\forall V$. For an assignment $\gamma\in\langle V\rangle$ of the universal player let $\beta_\gamma$ be canonical for $\gamma$ with respect to $\Psi$, and let $h$ be minimal with $\beta_h^{-1}(0)\subseteq\beta_\gamma^{-1}(0)$. Such an $h$ exists by \cref{Lemma:Coverage}, and we have $\Psi_{>1}|_{\beta_h}\supseteq\Psi|_\gamma$. After $\gamma$, the strategy $t^{\mathcal{S}}$ follows $t^{\mathcal{S}_h}$, which satisfies all clauses of $\Psi_{>1}|_{\beta_h}$ and therefore all clauses of $\Psi|_\gamma$.
	\end{itemize}
	By induction, $t^{\mathcal{S}}$ is an existential winning strategy for the formula of $\mathcal{S}$. In particular, $s:=t^{\mathcal{S}_e}$ is an existential winning strategy for $\Phi$, and $\Phi$ is indeed true.

	For each $j=1,\ldots,m+1$, every value of $s_{X_j}$ is read off a slice $\mathcal{S}\in\kappa$ with $\block(\mathcal{S})=X_j$: it is the last existential assignment $\alpha_g$ of $\mathcal{S}$ for $j\leq m$, and the assignment $\sigma$ for $j=m+1$. Therefore,
	$|\Img(s_{X_j})|\leq |\{\mathcal{S}\in \kappa :\: \block(\mathcal{S})=X_j\}|$. As before, the construction is efficient.
\end{proof}

As a corollary we obtain the soundness of the proof system. Completeness will follow from the upper bounds in the next section (\cref{Cor:Completeness}).

\begin{corollary}\label{Cor:Soundness}
	\caqe is sound on false and on true QCNFs. In other words, if a QCNF $\Phi$ has a \caqe refutation, then $\Phi$ is false, and if $\Phi$ has a \caqe verification, then $\Phi$ is true.
\end{corollary}

\section{Lower and upper bounds for proof size via CAQE width}\label{Sec:CAQEwidth}

Next we want to prove both lower and upper bounds for the size of \caqe proofs, for which we introduce a measure called \emph{CAQE width}. The main idea is as follows: We already know that CAQE  only tests assignments if they are not worse than previously tested ones. This means that we can minimise the number of assignments  to be tested by only picking  optimal ones. Hence, as an upper bound we can count the number of these optimal assignments. On the other hand, since we need to test at least all optimal assignments, this also yields a lower bound.

The intricate part is to generalise the notion of optimal assignments to formulas with more than one quantifier alternation, as we can only check optimality on the first block. To do this, we  define \emph{optimal moves} on the first variable block, and \emph{optimal chains} as a sequence of optimal moves for several quantifier alternations.

\begin{definition}[optimal moves]\label{Def:OptimalMoves}
	Let $\Phi$ be a QCNF with matrix $C_1\wedge \ldots \wedge C_k$ such that its prefix starts with the quantifier block $QV$.
	In case $Q=\exists$, an assignment $\alpha\in \langle V\rangle$ is called an \emph{optimal move} if and only if the following holds:
		For each $\alpha'\in \langle V\rangle$ and each clause $C_r$ with $\alpha \nvDash C_r$ and $\alpha'\vDash C_r$ there exists a clause $C_q$ with $\alpha \vDash C_q$ and $\alpha'\nvDash C_q$. In other words, there is no assignment $\alpha'\in \langle V \rangle$ such that $\alpha'$ satisfies all clauses satisfied by $\alpha$, plus at least one more clause.

		Analogously, if $Q=\forall$ an assignment $\alpha\in \langle V\rangle$ is called an \emph{optimal move} if and only if the following holds:
	For each $\alpha'\in \langle V\rangle$ and each clause  $C_r$ with $\alpha \vDash C_r$ and $\alpha'\nvDash C_r$ there is a clause $C_q$ with $\alpha \nvDash C_q$ and $\alpha'\vDash C_q$. In other words, there is no assignment $\alpha'\in \langle V \rangle$ such that $\alpha$ satisfies all clauses that are satisfied by $\alpha'$, plus at least one more clause.
\end{definition}

For a (partial) assignment $\tau$ of the variables of $\Phi$ we denote by $\sat(\tau):=\{r\in\{1,\ldots,k\}:\:\tau\vDash C_r\}$ the set of indices of the clauses of $\Phi$ that are satisfied by $\tau$. We will use this notation throughout the rest of the paper, always with respect to the original clauses $C_1,\ldots,C_k$ of the QCNF $\Phi$ under consideration. In the existential case, an assignment $\alpha\in\langle V\rangle$ is then an optimal move if and only if $\sat(\alpha)$ is a $\subseteq$-maximal element of $\{\sat(\alpha'):\:\alpha'\in\langle V\rangle\}$, and in the universal case this holds if and only if $\sat(\alpha)$ is a $\subseteq$-minimal element of this set. Every assignment $\alpha'\in\langle V\rangle$ is dominated by an optimal move, i.e.\ in the existential case there is an optimal move $\alpha$ with $\sat(\alpha')\subseteq\sat(\alpha)$, and dually in the universal case.

Optimal moves can now be generalised to optimal chains. An important aspect of optimal chains is that they always rely on a given winning strategy $s$. This is because we need to consider quantifier alternations and we need to know how the variables are assigned that are quantified between the blocks on which we want to detect optimal moves. Here we use the $\game$ notation from \cref{Def:game}.

\begin{definition}[optimal chains]\label{Def:OptimalChains}
	Let $\Phi$ be a false QCNF
	and $s$ be  a universal winning strategy for $\Phi$. An assignment $(\eta_1,\ldots,\eta_j)\in \langle X_1\cup \ldots\cup X_j\rangle$ for $j\leq m$ is called an \emph{optimal chain up to block $j$ with respect to $s$} if and only if each $\eta_i\in \langle X_i\rangle$ is an optimal move for $\Phi|_{\game(s,\eta_1\cup\ldots\cup \eta_{i-1},U_{i-1})}$ for each $i=1,\ldots,j$.\footnote{Note that it is not necessary to consider optimal moves on the block $X_{m+1}$ because \caqe would simply use a SAT solver at this point. In general, we are only interested in optimal moves if at least one quantifier block with opposite quantification follows.}

	Analogously, in the case where $\Phi$ is true and $s$ is an existential winning strategy, an assignment $(\eta_0,\ldots,\eta_j)\in \langle U_0\cup \ldots\cup U_j\rangle$ for $j\leq m$ is called an \emph{optimal  chain up to block $j$ with respect to $s$} if and only if each $\eta_i\in \langle U_i\rangle$ is an optimal move for $\Phi|_{\game(s,\eta_0\cup\ldots\cup \eta_{i-1},X_{i})}$ for each $i=0,\ldots,j$.\footnote{For $i=0$ we simply set $X_i=\emptyset$, i.e.\ $\eta_0$ has to be an optimal move for $\Phi$ itself.}

	We denote the set of all optimal  chains up to block $j$ with respect to $s$ as $K_j(\Phi,s)$. Furthermore, we set $K(\Phi,s):=K_m(\Phi,s)$. For $j\leq 0$ (false case) and $j\leq -1$ (true case) we set $K_j(\Phi,s):=\{\emptyset\}$, which means that the only optimal chain up to a non-existing block is the empty assignment.
\end{definition}

We recall the definition of maximal and minimal elements with respect to the subset relation.
	Let $\mathfrak{A}$ be a family of sets. Then we define $\max_\subseteq (\mathfrak{A}):=\{ M\in \mathfrak{A}:\: \not\exists N\in \mathfrak{A}:\: M \subsetneq N  \}$ and $\min_\subseteq (\mathfrak{A}):=\{ M\in \mathfrak{A}:\: \not\exists N\in \mathfrak{A}:\: N \subsetneq M \}$.

\begin{definition}[CAQE width]\label{Def:cw}
	Let $\Phi$ be a QCNF.
 If $\Phi$ is false,  we define
	\begin{center}
		$	\cw^-(\Phi):= \min\limits_{s}\max\limits_{1\leq j\leq m}\max\limits_{\tilde{\gamma}\in K_{j-1}(\Phi,s)}\left| \max_\subseteq \{  \{ r:\: \text{$\game(s,\gamma,X_j)\vDash C_r$}   \} :\: \tilde{\gamma}\subseteq \gamma\in \langle X\rangle    \}      \right|$
	\end{center}
	and if $\Phi$ is true we define
		\begin{center}
		$\cw^+(\Phi):=  \min\limits_{s}\max\limits_{0\leq j\leq m}\max\limits_{\tilde{\gamma}\in K_{j-1}(\Phi,s)}\left| \min_\subseteq \{  \{ r:\: \game(s,\gamma,U_j)\vDash C_r   \} :\: \tilde{\gamma}\subseteq \gamma\in \langle U\rangle  \}      \right|$
			\end{center}
	as the \emph{CAQE width} of $\Phi$, where $s$ ranges over all universal (resp. existential) winning strategies for $\Phi$.\footnote{For $j\leq 1$ we have $K_{j-1}(\Phi,s)=\{\emptyset\}$, so the inner maximum then ranges over all assignments $\gamma$.} If $m=0$, then the prefix of a false QCNF $\Phi$ contains no existential block that is followed by a universal block, and in this case we set $\cw^-(\Phi):=1$.
\end{definition}

	Computing the CAQE width can be interpreted as a game between two players: the strategist and the optimiser. In the case where $\Phi$ is false, the strategist controls the universal variables and the optimiser controls the existential variables. The task of the strategist is to pick a universal winning strategy for $\Phi$ at the beginning and assign  the variables of $\Phi$ along that strategy throughout the whole game. On the other hand, before assigning some block $X_j$ the optimiser counts the number  of optimal moves on that block under the side condition that different moves that have the same effect\footnote{We will denote this as the \emph{fruition} later.} on the formula only count once. Let $p_j$ be that number. Then the optimiser chooses an optimal move to play it on the formula. The goal of the optimiser is to maximise $\max_j p_j$ by assigning $X$ suitably, while the goal of the strategist is to minimise $\max_j p_j$ by choosing the best strategy $s$ at the beginning. If both players play optimally, we end up with $\max_j p_j=\cw^-(\Phi)$.

	The case where $\Phi$ is true is similar. There the strategist controls the existential variables and picks an existential winning strategy, whereas the optimiser assigns the universal variables. At the end we obtain $\max_j p_j=\cw^+(\Phi)$.

	In the proofs we will mostly use the CAQE width in the following local form. Let $\Psi$ be a QCNF that we obtain from $\Phi$ by a restriction, with clauses indexed by a set $J\subseteq\{1,\ldots,k\}$ and a prefix that starts with the block $QV$. For $\alpha\in\langle V\rangle$ we denote by $\sat_\Psi(\alpha)\subseteq J$ the set of indices of the clauses of $\Psi$ that are satisfied by $\alpha$, and we define
	\begin{align*}
		w(\Psi):=\begin{cases}
			\left|\max_\subseteq\{\sat_\Psi(\alpha):\:\alpha\in\langle V\rangle\}\right| & \text{if } Q=\exists,\\
			\left|\min_\subseteq\{\sat_\Psi(\alpha):\:\alpha\in\langle V\rangle\}\right| & \text{if } Q=\forall,
		\end{cases}
	\end{align*}
	which is the number of optimal moves of $\Psi$ up to their effect on the clauses. Now let $\Phi$ be false, let $s$ be a universal winning strategy, $\tilde{\gamma}\in K_{j-1}(\Phi,s)$ and $\Psi:=\Phi|_{\game(s,\tilde{\gamma},U_{j-1})}$. For every $\gamma\in\langle X\rangle$ with $\tilde{\gamma}\subseteq\gamma$ we have $\game(s,\gamma,X_j)=\game(s,\tilde{\gamma},U_{j-1})\cup\gamma_j$. Hence, the sets $\{r:\:\game(s,\gamma,X_j)\vDash C_r\}$ with $\tilde{\gamma}\subseteq\gamma$ are exactly the sets $I\cup\sat_\Psi(\alpha)$ for $\alpha\in\langle X_j\rangle$, where $I$ is the fixed set of clauses satisfied by $\game(s,\tilde{\gamma},U_{j-1})$. As a consequence, the number of $\subseteq$-maximal sets in \cref{Def:cw} is $w(\Psi)$, and we get
	\begin{align}\label{eq:cwlocal}
		\cw^-(\Phi)=\min_{s}\,\max_{1\leq j\leq m}\,\max_{\tilde{\gamma}\in K_{j-1}(\Phi,s)}w\bigl(\Phi|_{\game(s,\tilde{\gamma},U_{j-1})}\bigr).
	\end{align}
	In the same way we obtain $\cw^+(\Phi)=\min_{s}\max_{0\leq j\leq m}\max_{\tilde{\gamma}\in K_{j-1}(\Phi,s)}w\bigl(\Phi|_{\game(s,\tilde{\gamma},X_{j})}\bigr)$ for a true QCNF $\Phi$.

\begin{example}\label{Example:cw}
	Let $\Phi$ be the QCNF from \cref{Example:OptimalAssignment}. The first block of $\Phi$ is existential and $m=1$, so the CAQE width of $\Phi$ does not depend on the strategy $s$, and $\cw^-(\Phi)$ is simply the number of $\subseteq$-maximal sets among the sets $\sat(\alpha)$ for $\alpha\in\langle x_1,x_2\rangle$. We already computed these sets in \cref{Example:OptimalAssignment}: they are $\sat(\alpha_1)=\{4,5,6\}$, $\sat(\alpha_2)=\{2,4,5\}$ and $\sat(\alpha_3)=\{1,2,6\}$, whereas $\sat(\alpha_4)=\{1,2\}\subsetneq\sat(\alpha_3)$. This gives $\cw^-(\Phi)=3$.
\end{example}

There are formulas for which we need to count all assignments for  CAQE width. Examples are the equality formulas, in which each assignment of the first block is optimal by  symmetry.

\begin{definition}[equality formulas~\cite{BBH19,BPBSAT22journal}]\label{Def:Eq}
		The false QCNF $\mathtt{Eq}_n$ consists of the prefix $	\exists x_1,\ldots,x_n$ $\forall u_1,\ldots,u_n\exists t_1,\ldots,t_n$
		and the matrix
			$x_i\vee u_i\vee t_i$,\ \  $\bar{x}_i\vee \bar{u}_i\vee t_i$,\ \   $ \bar{t}_1\vee\ldots\vee\bar{t}_n$ for $i=1,\ldots,n$.
	\end{definition}
	We use the form of the equality formulas from~\cite{BPBSAT22journal,BPB22IJCAIjournal}. The original definition in~\cite{BBH19} is the same up to renaming $t_i\mapsto\bar{t}_i$.

\begin{proposition}\label{Prop:EqCW}
	We have $\cw^-(\mathtt{Eq}_n)=2^n$.
\end{proposition}
\begin{proof}
	Let $\gamma_1,\gamma_2\in \langle \{x_1,\ldots,x_n\}\rangle$ be two assignments with $\gamma_1\neq \gamma_2$. Then there exists an $i$ with $\gamma_1(x_i)\neq \gamma_2(x_i)$. Hence, $x_i\vee u_i\vee t_i$ is satisfied by $\gamma_1$ but not by $\gamma_2$, and $\bar{x}_i\vee \bar{u}_i\vee t_i$ is satisfied by $\gamma_2$ but not by $\gamma_1$, or vice versa. That means all assignments of the first block are optimal. The first block is existential and $m=1$, hence the CAQE width does not depend on the strategy and we obtain $\cw^-(\mathtt{Eq}_n)=|\langle \{x_1,\ldots,x_n\}\rangle |=2^n$.
\end{proof}

Next we define  \emph{fruitions}. These are projections of assignments that tell us which clauses of a formula $\Phi$ are satisfied by these assignments. Like optimal chains, they are defined across several quantifier alternations and hence need a fixed winning strategy $s$ as input. Intuitively, while assignments are purely syntactic, their fruitions encode the effect of their assignments on the formula. We will need fruitions for the decision lists in Sections~\ref{Sec:UDL} and~\ref{Sec:CAQEandQCDCL}.

\begin{definition}[fruitions]\label{Def:Fruitions}
	Let $\Phi$ be a  QCNF.

	If $\Phi$ is false and $s$ is a universal winning strategy, then for each assignment $\sigma=(\sigma_1,\ldots,\sigma_j)\in \langle X_1\cup \ldots\cup X_j\rangle$ with $j\leq m$ we call
		$\abs(\sigma):=(I_1,\ldots,I_j)$
	the \emph{fruition of $\sigma$}
	with
	$	I_i:=\{ r:\: \game(s,\sigma,X_i)\vDash C_r \}$,
for $i=1,\ldots,j$.

Analogously, if $\Phi$ is true and $s$ is an existential winning strategy, then for each assignment $\sigma=(\sigma_0,\ldots,\sigma_j)\in \langle U_0\cup \ldots\cup U_j\rangle$ with $j\leq m$ we call
$	\abs(\sigma):=(I_0,\ldots,I_j)$
the \emph{fruition of $\sigma$}
with
$I_i:=\{ r:\: \game(s,\sigma,U_{i})\vDash C_r \}$,
for $i=0,\ldots,j$.

We denote $F_j(\Phi,s):=\{ \abs(\sigma):\: \sigma\in K_{j}(\Phi,s) \}$  and set $F(\Phi,s):=F_m(\Phi,s)$. We will sometimes refer to the elements of $F_j(\Phi,s)$ as \emph{optimal fruitions (up to block $j$ with respect to $s$)}. Note that $I_1\subseteq I_2\subseteq\ldots$ for every fruition, since assignments are only extended along the game.
\end{definition}


We can now  prove an upper bound for \caqe proof size using  CAQE width. The main idea is to construct \caqe proofs by always choosing optimal moves for testing, and their number is  bounded by the CAQE width. As a side result, we also prove the completeness of \caqe by showing how refutations or verifications can always be constructed.

The following lemma shows that the optimal moves of a formula, one for each effect, always form a slice.

\begin{lemma}\label{Lemma:SliceFromOptimalMoves}
	Let $\Psi$ be a QCNF that is not propositional and whose prefix starts with the block $\exists V$. Let $\alpha_1,\ldots,\alpha_g\in\langle V\rangle$ be such that the sets $\sat_\Psi(\alpha_1),\ldots,\sat_\Psi(\alpha_g)$ are pairwise different and are exactly the $\subseteq$-maximal elements of $\{\sat_\Psi(\alpha):\:\alpha\in\langle V\rangle\}$, and let $\beta_i$ be canonical for $\alpha_i$. Then
	\begin{bracketenumerate}
		\item $\alpha_i\cup\beta_i$ satisfies $\Psi_{\exists V}\wedge\bigwedge_{a\neq i}\Rex{\beta_a}$ for $i=1,\ldots,g$,
		\item $\Psi_{\exists V}\wedge\bigwedge_{a=1}^{g}\Rex{\beta_a}$ is unsatisfiable, and
		\item $\Psi_{>1}|_{\beta_i}=\Psi|_{\alpha_i}$ for $i=1,\ldots,g$.
	\end{bracketenumerate}
	Hence, if $\mathcal{S}_i$ is a refutation slice for $\Psi|_{\alpha_i}$ for $i=1,\ldots,g$ and $\rho$ is a \Resolution refutation of the formula in (2), then $(\Psi,(\alpha_1,\ldots,\alpha_g),(\beta_1,\ldots,\beta_g),(\mathcal{S}_1,\ldots,\mathcal{S}_g),\rho)$ is a refutation slice for $\Psi$ of size $g=w(\Psi)$.

	The dual statement holds if the prefix of $\Psi$ starts with $\forall V$. Here we replace $\subseteq$-maximal by $\subseteq$-minimal, $\Psi_{\exists V}$ by $\Psi_{\forall V}$, $\Rex{\beta_a}$ by $\Rall{\beta_a}$, and refutation slices by verification slices.
\end{lemma}
\begin{proof}
	By \cref{Lemma:alphaBetaSatisfies}, $\alpha_i\cup\beta_i$ satisfies $\Psi_{\exists V}$. For $a\neq i$ the sets $\sat_\Psi(\alpha_i)$ and $\sat_\Psi(\alpha_a)$ are different and $\subseteq$-maximal, so there is some $r\in\sat_\Psi(\alpha_i)\setminus\sat_\Psi(\alpha_a)$. Then $\beta_i(b_r)=0$ and $\beta_a(b_r)=1$, hence $\beta_i$ satisfies the literal $\bar{b}_r$ of $\Rex{\beta_a}$. This shows (1).

	For (2), let $\tilde{\alpha}\cup\tilde{\beta}$ be an arbitrary assignment that satisfies $\Psi_{\exists V}$. We choose $a$ with $\sat_\Psi(\tilde{\alpha})\subseteq\sat_\Psi(\alpha_a)$. For every $r$ with $\beta_a(b_r)=1$, the clause $C_r$ is not satisfied by $\alpha_a$ and therefore not by $\tilde{\alpha}$. As $\tilde{\alpha}\cup\tilde{\beta}$ satisfies $C_{r,1}$, we get $\tilde{\beta}(b_r)=1$. Hence, $\tilde{\beta}$ falsifies $\Rex{\beta_a}$.

	Statement (3) holds by \cref{Lemma:PhiBetaContainedInPhiAlpha-exists}, and (1)--(3) are exactly the conditions of \cref{Def:slices} for a refutation slice.

	In the dual case we argue in the same way. For (1) we take some $r\in\sat_\Psi(\alpha_a)\setminus\sat_\Psi(\alpha_i)$, so that $\beta_i$ satisfies the literal $b_r$ of $\Rall{\beta_a}$. For (2) we choose $a$ with $\sat_\Psi(\alpha_a)\subseteq\sat_\Psi(\tilde{\alpha})$. For every $r$ with $\beta_a(b_r)=0$ the assignment $\tilde{\alpha}$ satisfies some literal $\ell$ of $C_r$, and the clause $\bar{\ell}\vee\bar{b}_r$ of $C_{r,1}$ gives us $\tilde{\beta}(b_r)=0$. Hence, $\tilde{\beta}$ falsifies $\Rall{\beta_a}$. For (3) we use \cref{Lemma:PhiBetaContainedInPhiAlpha}. Note that the assignments $\beta_i$ are canonical, as \cref{Def:slices} requires for universal blocks.
\end{proof}

\begin{theorem}\label{Prop:CAQEwidthUpperBoundFalse}
For each false QCNF $\Phi$ there exists a \caqe refutation of size at most $(2m+2)\cdot\cw^-(\Phi)^{m}$. In particular, $\Phi$ has a \caqe refutation of size $\mathcal{O}( \cw^-(\Phi)^{m})$ if $\cw^-(\Phi)\geq 2$.
\end{theorem}
\begin{proof}
	Let $s$ be a universal winning strategy that attains the minimum in \cref{Def:cw}, and let $c:=\cw^-(\Phi)$. By \eqref{eq:cwlocal}, we have $w(\Phi|_{\game(s,\tilde{\gamma},U_{j-1})})\leq c$ for all $1\leq j\leq m$ and all $\tilde{\gamma}\in K_{j-1}(\Phi,s)$.

	We define refutation slices $\mathcal{S}^{X}(\tilde{\gamma})$ for the formulas $\Phi|_{\game(s,\tilde{\gamma},U_{j-1})}$ with $\tilde{\gamma}\in K_{j-1}(\Phi,s)$ and $1\leq j\leq m+1$, and refutation slices $\mathcal{S}^{U}(\tilde{\gamma})$ for the formulas $\Phi|_{\game(s,\tilde{\gamma},X_{j})}$ with $\tilde{\gamma}\in K_{j}(\Phi,s)$ and $0\leq j\leq m$. All these formulas are false, because $s$ is a universal winning strategy. The definition is by recursion, starting from the last quantifier block.
	\begin{itemize}
		\item Let $\tilde{\gamma}\in K_m(\Phi,s)$. Then $\Phi|_{\game(s,\tilde{\gamma},U_m)}$ is an unsatisfiable propositional formula, and we set $\mathcal{S}^{X}(\tilde{\gamma}):=(\Phi|_{\game(s,\tilde{\gamma},U_m)},\pi)$ for a \Resolution refutation $\pi$ of this formula.
		\item Let $\tilde{\gamma}\in K_j(\Phi,s)$ with $0\leq j\leq m$ and $\Psi:=\Phi|_{\game(s,\tilde{\gamma},X_j)}$ (for $j=0$ only if $U_0$ is not empty). Let $\alpha:=s_{U_j}(\tilde{\gamma})$ and let $\beta$ be canonical for $\alpha$. Then $\alpha\cup\beta$ satisfies $\Psi_{\forall U_j}$ by \cref{Lemma:alphaBetaSatisfies}, and we have $\Psi_{>1}|_\beta=\Psi|_\alpha=\Phi|_{\game(s,\tilde{\gamma},U_j)}$ by \cref{Lemma:PhiBetaContainedInPhiAlpha}. Hence, $\mathcal{S}^{U}(\tilde{\gamma}):=(\Psi,(\alpha),(\beta),(\mathcal{S}^{X}(\tilde{\gamma})),\rho)$ with a placeholder $\rho$ is a refutation slice for $\Psi$ of size $1$.
		\item Let $\tilde{\gamma}\in K_{j-1}(\Phi,s)$ with $1\leq j\leq m$ and $\Psi:=\Phi|_{\game(s,\tilde{\gamma},U_{j-1})}$. We choose $\alpha_1,\ldots,\alpha_g\in\langle X_j\rangle$ as in \cref{Lemma:SliceFromOptimalMoves}, that means one optimal move of $\Psi$ for each $\subseteq$-maximal set $\sat_\Psi(\alpha)$. Then $g=w(\Psi)\leq c$, and for each $i$ we have $\tilde{\gamma}\cup\alpha_i\in K_j(\Phi,s)$ and $\Psi|_{\alpha_i}=\Phi|_{\game(s,\tilde{\gamma}\cup\alpha_i,X_j)}$. By \cref{Lemma:SliceFromOptimalMoves},
		\begin{align*}
			\mathcal{S}^{X}(\tilde{\gamma}):=\left( \Psi, (\alpha_1,\ldots,\alpha_g),(\beta_1,\ldots,\beta_g),(\mathcal{S}^{U}(\tilde{\gamma}\cup\alpha_1),\ldots,\mathcal{S}^{U}(\tilde{\gamma}\cup\alpha_g)),\rho\right)
		\end{align*}
		with the canonical assignments $\beta_i$ and a suitable \Resolution refutation $\rho$ is a refutation slice for $\Psi$ of size at most $c$.
	\end{itemize}
	Let $\kappa$ consist of the slice for $\Phi$ -- this is $\mathcal{S}^{U}(\emptyset)$, or $\mathcal{S}^{X}(\emptyset)$ if $U_0$ is empty -- and of all slices that can be reached from it via references, in an order such that each slice occurs after the slices it refers to. Then $\kappa$ is a \caqe refutation of $\Phi$.

	Since each slice $\mathcal{S}^{X}(\tilde{\gamma})$ refers to at most $c$ slices and each slice $\mathcal{S}^{U}(\tilde{\gamma})$ refers to only one slice, $\kappa$ contains for each $j=1,\ldots,m$ at most $c^{j-1}$ slices of the block $\exists X_j$, each of size at most $c$, and for each $j=0,\ldots,m$ at most $c^j$ slices of the block $\forall U_j$, each of size $1$. In addition, there are at most $c^m$ propositional slices of size $1$. We obtain $|\kappa|\leq 1+\sum_{j=1}^{m}(c^{j}+c^j)+c^m\leq (2m+2)\cdot c^m$.\footnote{If a slice for the same formula is constructed several times, we can also identify these slices, which only decreases the size.} If $c\geq 2$, then $\sum_{j=0}^m c^j\leq 2c^m$ and hence $|\kappa|\leq 5c^m\in\mathcal{O}(c^m)$.
\end{proof}

An analogous result holds for true QCNFs.

\begin{proposition}\label{Prop:CAQEwidthUpperBoundTrue}
	For each true QCNF $\Phi$ there exists a \caqe verification $\kappa$ of size at most $(2m+2)\cdot\cw^+(\Phi)^{m+1}$. In particular, $|\kappa|\in \mathcal{O}( \cw^+(\Phi)^{m+1})$ if $\cw^+(\Phi)\geq 2$.\footnote{If the first block $U_0$ is empty, we even get $|\kappa|\leq (2m+2)\cdot \cw^+(\Phi)^{m}$.}
\end{proposition}
\begin{proof}
	The proof is dual to the proof of \cref{Prop:CAQEwidthUpperBoundFalse}. Let $s$ be an existential winning strategy that attains the minimum in \cref{Def:cw}, and let $c:=\cw^+(\Phi)$. Then we have $w(\Phi|_{\game(s,\tilde{\gamma},X_{j})})\leq c$ for all $0\leq j\leq m$ and all $\tilde{\gamma}\in K_{j-1}(\Phi,s)$.

	For $\tilde{\gamma}\in K_{j-1}(\Phi,s)$ we define verification slices $\mathcal{S}^{X}(\tilde{\gamma})$ for the formulas $\Phi|_{\game(s,\tilde{\gamma},U_{j-1})}$ with $1\leq j\leq m+1$, and verification slices $\mathcal{S}^{U}(\tilde{\gamma})$ for the formulas $\Phi|_{\game(s,\tilde{\gamma},X_{j})}$ with $0\leq j\leq m$. All these formulas are true, because $s$ is an existential winning strategy. Again, the definition is by recursion, starting from the last quantifier block.
	\begin{itemize}
		\item Let $\tilde{\gamma}\in K_m(\Phi,s)$. Then $\Phi|_{\game(s,\tilde{\gamma},U_m)}$ is a satisfiable propositional formula, and we set $\mathcal{S}^{X}(\tilde{\gamma}):=(\Phi|_{\game(s,\tilde{\gamma},U_m)},\sigma)$ for a satisfying assignment $\sigma$ of this formula.
		\item Let $\tilde{\gamma}\in K_{j-1}(\Phi,s)$ with $1\leq j\leq m$ and $\Psi:=\Phi|_{\game(s,\tilde{\gamma},U_{j-1})}$. Let $\alpha:=s_{X_j}(\tilde{\gamma})$ and let $\beta$ be canonical for $\alpha$. Then $\alpha\cup\beta$ satisfies $\Psi_{\exists X_j}$ by \cref{Lemma:alphaBetaSatisfies}, and we have $\Psi_{>1}|_\beta=\Psi|_\alpha=\Phi|_{\game(s,\tilde{\gamma},X_j)}$ by \cref{Lemma:PhiBetaContainedInPhiAlpha-exists}. Hence, $\mathcal{S}^{X}(\tilde{\gamma}):=(\Psi,(\alpha),(\beta),(\mathcal{S}^{U}(\tilde{\gamma})),\rho)$ with a placeholder $\rho$ is a verification slice for $\Psi$ of size $1$.
		\item Let $\tilde{\gamma}\in K_{j-1}(\Phi,s)$ with $0\leq j\leq m$ and $\Psi:=\Phi|_{\game(s,\tilde{\gamma},X_{j})}$ (for $j=0$ only if $U_0$ is not empty). We choose $\alpha_1,\ldots,\alpha_g\in\langle U_j\rangle$ as in the dual case of \cref{Lemma:SliceFromOptimalMoves}, that means one optimal move of $\Psi$ for each $\subseteq$-minimal set $\sat_\Psi(\alpha)$. Then $g=w(\Psi)\leq c$, and for each $i$ we have $\tilde{\gamma}\cup\alpha_i\in K_j(\Phi,s)$ and $\Psi|_{\alpha_i}=\Phi|_{\game(s,\tilde{\gamma}\cup\alpha_i,U_j)}$. By \cref{Lemma:SliceFromOptimalMoves},
		\begin{align*}
			\mathcal{S}^{U}(\tilde{\gamma}):=\left( \Psi, (\alpha_1,\ldots,\alpha_g),(\beta_1,\ldots,\beta_g),(\mathcal{S}^{X}(\tilde{\gamma}\cup\alpha_1),\ldots,\mathcal{S}^{X}(\tilde{\gamma}\cup\alpha_g)),\rho\right)
		\end{align*}
		with the canonical assignments $\beta_i$ and a suitable \Resolution refutation $\rho$ is a verification slice for $\Psi$ of size at most $c$.
	\end{itemize}
	Let $\kappa$ consist of the slice for $\Phi$ -- this is $\mathcal{S}^{U}(\emptyset)$, or $\mathcal{S}^{X}(\emptyset)$ if $U_0$ is empty -- and of all slices that can be reached from it via references, in an order such that each slice occurs after the slices it refers to. Then $\kappa$ is a \caqe verification of $\Phi$.

	Since each slice $\mathcal{S}^{U}(\tilde{\gamma})$ refers to at most $c$ slices and each slice $\mathcal{S}^{X}(\tilde{\gamma})$ refers to only one slice, $\kappa$ contains at most $c^j$ slices of the block $\forall U_j$ for $j=0,\ldots,m$, each of size at most $c$, at most $c^j$ slices of the block $\exists X_j$ for $j=1,\ldots,m$, each of size $1$, and at most $c^{m+1}$ propositional slices of size $1$. Hence, $|\kappa|\leq\sum_{j=0}^{m}c^{j+1}+\sum_{j=1}^{m}c^j+c^{m+1}\leq (2m+2)\cdot c^{m+1}$, and for $c\geq2$ we get $|\kappa|\leq 2c^{m+1}+c^{m+1}+c^{m+1}=4c^{m+1}$. If $U_0$ is empty, the same count gives $|\kappa|\leq (2m+2)\cdot c^{m}$.
\end{proof}

From \cref{Prop:CAQEwidthUpperBoundFalse,Prop:CAQEwidthUpperBoundTrue} we conclude:

\begin{corollary}\label{Cor:Completeness}
	\caqe is complete on false and on true QCNFs: every false QCNF has a \caqe refutation and every true QCNF has a \caqe verification. Together with \cref{Cor:Soundness}, \caqe is a proof system for false QCNFs and for true QCNFs.
\end{corollary}

Next we show that  CAQE width can be used for lower bounds as well. The idea is simple: As the CAQE width in a sense counts the number of optimal assignments throughout the formula, and since CAQE only terminates once no more assignments can be tested that are not worse than the previously tested ones, we need to test at least the optimal assignments before returning the truth value.

\begin{proposition}\label{Prop:LowerBoundCAQEwidth}
	For each   \caqe refutation $\kappa$ of a false QCNF $\Phi$ we have $\cw^-(\Phi)\leq |\kappa|$.
	Analogously, for each true QCNF $\Phi$ and \caqe verification $\kappa$ of $\Phi$ we have $\cw^+(\Phi)\leq |\kappa|$.
\end{proposition}
\begin{proof}
	We first consider the case that $\Phi$ is false. For $m=0$ we have $\cw^-(\Phi)=1\leq|\kappa|$, so let $m\geq 1$. Let $s_\kappa$ be the universal winning strategy that we extract from $\kappa$ in the proof of \cref{Prop:ExtractUniWinStratFromCAQE}. By \eqref{eq:cwlocal}, there are some $j\in\{1,\ldots,m\}$ and an optimal chain $\delta=(\eta_1,\ldots,\eta_{j-1})\in K_{j-1}(\Phi,s_\kappa)$ such that
	\begin{align*}
		\cw^-(\Phi)\leq w(\Psi)\qquad\text{for}\qquad \Psi:=\Phi|_{\game(s_\kappa,\delta,U_{j-1})}.
	\end{align*}
	If the existential player plays $\delta$, then the extraction of $s_\kappa$ follows a path of slices through $\kappa$ that leads to a slice $\mathcal{S}$ of the block $\exists X_j$. We show that $\mathcal{S}$ is a slice for $\Psi$. More generally, the slice of the block $\forall U_i$ on this path is a slice for $\Phi|_{\game(s_\kappa,\delta,X_i)}$, and the slice of the block $\exists X_i$ is a slice for $\Phi|_{\game(s_\kappa,\delta,U_{i-1})}$. This holds at the beginning, since the last slice of $\kappa$ is a slice for $\Phi$. In a slice of a universal block with the formula $\Psi'$, the extraction plays the last assignment $\alpha_g$ and continues with the slice for $\Psi'_{>1}|_{\beta_g}$, which is $\Psi'|_{\alpha_g}$ by \cref{Lemma:PhiBetaContainedInPhiAlpha}. In a slice of an existential block $X_i$ with the formula $\Psi'=\Phi|_{\game(s_\kappa,\delta,U_{i-1})}$, the extraction selects some $h$ with $\beta_h^{-1}(1)\subseteq\beta_\gamma^{-1}(1)$, where $\beta_\gamma$ is canonical for $\eta_i$. By \cref{Lemma:Coverage}, $\alpha_h$ satisfies all clauses of $\Psi'$ that are satisfied by $\eta_i$. As $\eta_i$ is an optimal move for $\Psi'$, the two assignments satisfy the same clauses, and \cref{Lemma:Coverage} gives us $\Psi'_{>1}|_{\beta_h}=\Psi'|_{\eta_i}$ for the formula of the next slice.

	Now let $\alpha_1,\ldots,\alpha_g$ be the assignments of $\mathcal{S}$, and let $M$ be a $\subseteq$-maximal element of $\{\sat_\Psi(\alpha):\:\alpha\in\langle X_j\rangle\}$, say $M=\sat_\Psi(\gamma)$. By \cref{Lemma:Coverage}, there is some $h$ with $M\subseteq\sat_\Psi(\alpha_h)$, and we get $M=\sat_\Psi(\alpha_h)$ by the maximality of $M$. Therefore, different $\subseteq$-maximal sets belong to different assignments of $\mathcal{S}$, and we conclude
	\begin{align*}
		\cw^-(\Phi)\leq w(\Psi)\leq |\mathcal{S}|\leq |\kappa|.
	\end{align*}

	The case for true formulas $\Phi$ works analogously, with the roles of the existential and the universal blocks swapped. We use the existential winning strategy $s_\kappa$ from \cref{Prop:ExtractExiWinStratFromCAQE} and obtain some $j\in\{0,\ldots,m\}$ and an optimal chain $\delta\in K_{j-1}(\Phi,s_\kappa)$ with $\cw^+(\Phi)\leq w(\Phi')$ for $\Phi':=\Phi|_{\game(s_\kappa,\delta,X_{j})}$ (if $j=0$ and $U_0$ is empty, the bound is trivial). If the universal player plays $\delta$, the extraction arrives at a verification slice $\mathcal{S}$ of the block $\forall U_{j}$. There is only one difference. By \cref{Lemma:PhiBetaContainedInPhiAlpha-exists} and \cref{Lemma:Coverage}, the formulas of the slices on this path contain the corresponding formulas $\Phi|_{\game(s_\kappa,\delta,\cdot)}$, but they may contain further clauses. Hence, $\mathcal{S}$ is a slice for a formula $\Psi\supseteq\Phi'$. This does not affect the bound. Let $M$ be a $\subseteq$-minimal element of $\{\sat_{\Phi'}(\alpha):\:\alpha\in\langle U_j\rangle\}$, say $M=\sat_{\Phi'}(\gamma)$. By \cref{Lemma:Coverage}, there is an assignment $\alpha_h$ of $\mathcal{S}$ such that every clause of $\Psi$ that is satisfied by $\alpha_h$ is also satisfied by $\gamma$. In particular, we have $\sat_{\Phi'}(\alpha_h)\subseteq M$ and hence $\sat_{\Phi'}(\alpha_h)= M$ by the minimality of $M$. Again, different sets $M$ belong to different assignments of $\mathcal{S}$, which gives $\cw^+(\Phi)\leq w(\Phi')\leq|\mathcal{S}|\leq|\kappa|$.
\end{proof}

We now define new QBFs $\mathtt{Copy}_n$ consisting of $n$ disjoint copies of $\mathtt{Eq}_1$:

\begin{definition}\label{Def:Copy}
	The false QCNF $\mathtt{Copy}_n$ has prefix $\exists X\forall U \exists T$ with $X=\{x_1,\ldots,x_n\}$, $U=\{u_1,\ldots,u_n\}$ and $T=\{t_1,\ldots,t_n\}$ and  clauses $x_i\vee u_i\vee t_i$, $\bar{x}_i\vee \bar{u}_i\vee t_i$ and $(\bar{t}_i)$ for $i=1,\ldots,n$.
\end{definition}

These QBFs witness an exponential  separation between \caqe and $\farpr$:

\begin{theorem}\label{Thm:SeparationCAQEfarpr}
	$\mathtt{Copy}_n$ needs \caqe refutations of size at least $2^n$, but has constant-size $\farpr$ (and $\farpr^\proofsystem{NP}$) refutations.
\end{theorem}
\begin{proof}
	For the \caqe lower bound, we can simply compute the CAQE width of $\mathtt{Copy}_n$. We obtain $\cw^-(\mathtt{Copy}_n)=2^n$ by the same argument as in   \cref{Prop:EqCW}. Two different assignments of $X$ differ in some $x_i$, so one of them satisfies $x_i\vee u_i\vee t_i$ but not $\bar{x}_i\vee\bar{u}_i\vee t_i$, and vice versa for the other one. Hence, each \caqe refutation of $\mathtt{Copy}_n$ has at least size $2^n$ by   \cref{Prop:LowerBoundCAQEwidth}. For the $\farpr$ refutation, let $C_1=x_1\vee u_1\vee t_1$, $C_2=\bar{x}_1\vee \bar{u}_1 \vee t_1$ and $C_3=(\bar{t}_1)$. Then we can do the  resolution steps
	\begin{align*}
		\dfrac{\{1 \}^T \qquad\{3\}^T\qquad \pi}{\{1,3\}^U} \qquad\text{and}\qquad	\dfrac{\{2\}^T \qquad\{3\}^T\qquad \pi}{\{2,3\}^U},
	\end{align*}
	where $\pi$ is the \Resolution refutation of $(t_1)\wedge (\bar{t}_1)$; note that $\bigvee_{q\in\{1\}}C_q|_T=(t_1)$ and $\bigvee_{q\in\{3\}}C_q|_T=(\bar{t}_1)$. The clauses $C_1,C_3$ (resp. $C_2,C_3$) do not contain complementary literals from $U$, and therefore we can reduce both sets
	\begin{align*}
		\dfrac{\{1,3\}^U}{\{1,3\}^X}\qquad  \text{and}\qquad \dfrac{\{2,3\}^U}{\{2,3\}^X}
	\end{align*}
	and finish with one last resolution step
	\begin{align*}
		\dfrac{\{1,3\}^X\qquad\{2,3\}^X\qquad\pi'}{\{1,2,3\}^0},
	\end{align*}
	where $\pi'$ is the \Resolution refutation of $(x_1)\wedge (\bar{x}_1)=\bigvee_{q\in\{1,3\}}C_q|_X\wedge\bigvee_{q\in\{2,3\}}C_q|_X$. This $\farpr$ refutation consists of $8$ lines and uses \Resolution refutations of constant size, independently of $n$.
\end{proof}

Consequently, \caqe is simulated by  $\farpr^{\proofsystem{NP}}$, but exponentially weaker. Hence the system $\farpr$ is too strong to capture the basic version of CAQE without failed assumptions. Whether $\farpr$ correctly captures the strength of full CAQE with failed assumptions remains open.

\section{CAQE vs \texorpdfstring{$\qresnp$}{Q-Res with NP oracle}: CAQE width and unified decision lists}\label{Sec:UDL}

In this section we want to show that \caqe is simulated by $\qresnp$. While versions of \qres (without the NP oracle) underlie proof systems for \qcdcl~\cite{BohmB24-JAIR}, the connection between $\qresnp$ and \caqe appears nontrivial. This is why we will make use of \emph{decision lists}, which represent strategies and even  characterise $\qresnp$ on bounded quantifier alternation~\cite{BeyersdorffBMP23}.

\begin{definition}[decision lists,~\cite{DBLP:journals/ml/Rivest87,BeyersdorffBMP23}]
A \emph{multi-output decision list} $\mathcal{M}$ from $X$ to $U$, where $X$ and $U$ are disjoint sets of variables, is a sequence of pairs
$	(D_1, \mu_1),\ldots,(D_z,\mu_z)$
where
\begin{itemize}
	\item $D_i$ are cubes with $\var(D_i)\subseteq X$ and $\bigvee_{i=1}^z D_i\equiv \top$,
	\item  $\mu_i\in \langle U\rangle$
\end{itemize}
for $i=1,\ldots,z$. The size of $\mathcal{M}$ is defined as $|\mathcal{M}|:=z$.

$\mathcal{M}$ computes the function from $\langle X\rangle$ to $\langle U\rangle$ that maps each $\gamma\in \langle X \rangle$ to $\mu_{i_\gamma}$, where
	$i_\gamma:=\min\{ i:\: \gamma \vDash D_i \}$.
 If $\mathcal{M}$ computes a universal winning strategy for a QCNF $\Phi$, then $\mathcal{M}$ is called a \emph{unified decision list (UDL)} for $\Phi$. That means the function $\gamma\mapsto \mathcal{M}(\gamma)|_{U_j}$ only depends on $\gamma|_{X_1\cup\ldots\cup X_j}$ for each $j$, and the resulting strategy is winning. In that case, we call $\mathcal{M}$ a \udl refutation of $\Phi$.  We will use the name of the UDL $\mathcal{M}$ to refer to the computed function.
\end{definition}

\begin{theorem}[{\cite[Theorem~4.2 and Corollary~5.6]{BeyersdorffBMP23}}]\label{Prop:UDLcharacterizationQResNP}
	 $\qresnp\equiv_p \udl$ on bounded quantifier alternation.
\end{theorem}
Instead of proving that $\qresnp$ simulates \caqe directly, we show that UDLs are bounded by the CAQE width.
\begin{theorem}\label{Prop:cwUDL}
	Let $\Phi$ be a false QCNF. Then there exists a UDL $\mathcal{M}$ for $\Phi$ with $|\mathcal{M}|\leq \cw^-(\Phi)^m$.
\end{theorem}
\begin{proof}
	Let $s$ be a universal winning strategy for $\Phi$ that attains the minimum in \cref{Def:cw}. For $j\leq m$ and an assignment $\sigma$ of $X_1\cup\ldots\cup X_j$ (or of a superset of these variables) we abbreviate $\sat_j(\sigma):=\{r:\:\game(s,\sigma,X_j)\vDash C_r\}$, which is the set of clauses satisfied in the game up to block $X_j$.

	We fix an arbitrary linear order on each $\langle X_j\rangle$, and lexicographically minimal moves and chains will always refer to these orders.

	Let $\sigma=(\sigma_1,\ldots,\sigma_{m+1})\in \langle X \rangle$ be arbitrary. We define an optimal chain $\eta(\sigma):=(\eta_1,\ldots,\eta_m)\in K(\Phi,s)$ inductively: Let $\eta_1\in \langle X_1 \rangle$ be the lexicographically minimal optimal move for $\Phi|_{\game(s,\emptyset,U_0)}$ such that $\sat_1(\sigma_1)\subseteq \sat_1(\eta_1)$. This is possible as either $\sigma_1$ is already an optimal move, or there exists an optimal move $\eta_1$ that satisfies all clauses satisfied by $\sigma_1$ (cf.\ the discussion after \cref{Def:OptimalMoves}). Assume that we already  constructed $\eta_1,\ldots,\eta_j$. Then let $\eta_{j+1}\in \langle X_{j+1}\rangle$ be the lexicographically minimal optimal move for $\Phi|_{\game(s,\eta_1\cup\ldots\cup\eta_j,U_j)}$ such that $\sat_{j+1}(\eta_1\cup\ldots\cup \eta_j\cup  \sigma_{j+1})\subseteq \sat_{j+1}(\eta_1\cup\ldots\cup \eta_{j+1})$. Such an $\eta_{j+1}$ exists by the same argument. Note that $\eta_1,\ldots,\eta_j$ only depend on $\sigma_1,\ldots,\sigma_j$ by construction. We have now constructed $\eta(\sigma)=(\eta_1,\ldots,\eta_m)\in K(\Phi,s)$.

	We define a strategy $\tilde{s}=(\tilde{s}_{U_0},\ldots,\tilde{s}_{U_m})$ by $\tilde{s}_{U_j}(\sigma_1\cup\ldots\cup\sigma_j):=s_{U_j}(\eta_1\cup\ldots\cup \eta_j)$, where $(\eta_1,\ldots,\eta_j)$ is the prefix of $\eta(\sigma)$. As we noted above, $\eta_1,\ldots,\eta_j$ only depend on $\sigma_1,\ldots,\sigma_j$, so this is indeed a strategy. It remains to show that it is in fact a winning strategy, meaning that for all $\sigma\in \langle X\rangle$ the assignment $\sigma\cup \tilde{s}(\sigma)$ has to falsify some clause $C_q$.

	Let $\sigma=(\sigma_1,\ldots,\sigma_{m+1})\in \langle X\rangle$ be arbitrary, let $\eta(\sigma)=(\eta_1,\ldots,\eta_m)$ and consider the assignment $\gamma:=(\eta_1,\ldots,\eta_m,\sigma_{m+1})\in\langle X\rangle$. The value $s(\gamma)$ only depends on $\eta_1,\ldots,\eta_m$, so we have $s(\gamma)=\tilde{s}(\sigma)$.
	We  know that $s$ is a winning strategy, hence some $C_q$ is falsified by $\gamma\cup s(\gamma)$. Assume, for the sake of contradiction, that $C_q$ is satisfied by $\sigma\cup \tilde{s}(\sigma)=\sigma \cup s(\gamma)$.
	Since $\gamma\cup s(\gamma)$ and $\sigma\cup s(\gamma)$ agree on $U$ and on $X_{m+1}$, there is some $j\leq m$ such that $\sigma_j\vDash C_q$. But then we have $q\in \sat_j(\eta_1\cup\ldots\cup \eta_{j-1}\cup \sigma_j)\subseteq \sat_j(\eta_1\cup\ldots\cup \eta_{j})$, and therefore  $\game(s,\eta_1\cup\ldots\cup \eta_{j} ,X_j)\subseteq\gamma\cup s(\gamma)$ would also satisfy $C_q$. This is a contradiction to the fact that $C_q$  is falsified by $\gamma\cup s(\gamma)$.
	Hence $\tilde{s}$ is indeed a winning strategy for $\Phi$. Intuitively, $\tilde{s}$ projects all input assignments to better  optimal, lexicographically minimal assignments and applies $s$ to these optimal assignments.

	Let $H:=\{\eta(\sigma):\:\sigma\in\langle X\rangle\}\subseteq K(\Phi,s)$, and let $\gamma_1<\gamma_2<\ldots<\gamma_z$ be the elements of $H$ in lexicographic order. We first show $z\leq\cw^-(\Phi)^m$. For $j\leq m$ let $H_j$ be the set of the prefixes $(\eta_1,\ldots,\eta_j)$ of the chains in $H$. We have $|H_0|=1$, so it suffices to show that every $\theta=(\eta_1,\ldots,\eta_{j-1})\in H_{j-1}$ has at most $\cw^-(\Phi)$ extensions $(\theta,\eta_j)\in H_j$. By the definition of $\eta(\sigma)$, each such $\eta_j$ is the lexicographically minimal optimal move for $\Phi|_{\game(s,\theta,U_{j-1})}$ with $\sat_j(\theta\cup\sigma_j)\subseteq\sat_j(\theta\cup\eta_j)$, for some $\sigma_j\in\langle X_j\rangle$. Let $(\theta,\eta_j)$ and $(\theta,\eta'_j)$ be two such extensions with $\sat_j(\theta\cup\eta_j)=\sat_j(\theta\cup\eta'_j)$. Then $\eta'_j$ is one of the optimal moves among which $\eta_j$ was chosen, so $\eta_j\leq\eta'_j$, and in the same way $\eta'_j\leq\eta_j$. Hence, different extensions have different sets $\sat_j(\theta\cup\eta_j)$. Since the $\eta_j$ are optimal moves, these sets are $\subseteq$-maximal elements of $\{\sat_j(\gamma):\:\theta\subseteq\gamma\in\langle X\rangle\}$. As we have $\theta\in K_{j-1}(\Phi,s)$ and $s$ attains the minimum in \cref{Def:cw}, there are at most $\cw^-(\Phi)$ of them.

	For $i=1,\ldots,z$ let $\xi_i:=\abs(\gamma_i)$.
	We now define $\mathcal{M}$ as
	$(D_1,\mu_1),\ldots,(D_z,\mu_z)$
	with
	\begin{align*}
		D_i:=\bigwedge_{j=1}^m\bigwedge_{r\not\in I_j}\neg \bigl(C_r|_{X_1\cup\ldots\cup X_j}\bigr)	\qquad\text{and}\qquad \mu_i:={s}(\gamma_i)
	\end{align*}
	for $i=1,\ldots,z$, where $(I_1,\ldots,I_m)=\xi_i$. In other words, for each $j$ we go through all indices $r$ that do not appear in the $j^\text{th}$ set $I_j$ of $\xi_i$. Here $\neg(C|_{V})$ denotes the cube consisting of the negations of all literals of $C|_V$, which is the empty cube if $C|_V$ is empty. So $D_i$ is a cube over $X_1\cup\ldots\cup X_m$, and an assignment $\sigma\in\langle X\rangle$ satisfies $D_i$ if and only if $\sigma_1\cup\ldots\cup\sigma_j$ satisfies no clause $C_r$ with $r\notin I_j$, for all $j=1,\ldots,m$.
	It remains to show that $\mathcal{M}$ is a UDL, which we will do by proving that $\mathcal{M}$ computes $\tilde{s}$.

	\begin{claim}\label{Claim:UDL}
		For each $\sigma\in\langle X\rangle$ and $i\in\{1,\ldots,z\}$ the following are equivalent:
		\begin{romanenumerate}
			\item $i$ is minimal such that $\sigma\vDash D_i$;
			\item $\gamma_i=\eta(\sigma)$.
		\end{romanenumerate}
	\end{claim}
	\begin{claimproof}[Proof of the claim]
		Let $\sigma\in\langle X\rangle$ be fixed. We first show the following two facts for an arbitrary $i$ with $\xi_i=(I_1,\ldots,I_m)$ and $\gamma_i=(\eta'_1,\ldots,\eta'_m)$.

		\emph{Fact A: If $\sigma\vDash D_i$, then $\sat_j(\eta'_1\cup\ldots\cup \eta'_{j-1}\cup  \sigma_{j})\subseteq \sat_j(\eta'_1\cup\ldots\cup \eta'_{j})=I_j$ for all $j=1,\ldots,m$.}
		Let $q\in \sat_j(\eta'_1\cup\ldots\cup \eta'_{j-1}\cup  \sigma_{j})$. If $C_q$ is satisfied by the universal part of $\game(s,\eta'_1\cup\ldots\cup \eta'_{j-1}\cup  \sigma_{j}, X_{j})$ or by $\eta'_1\cup\ldots\cup\eta'_{j-1}$, then $C_q$ is also satisfied by $\game(s,\eta'_1\cup\ldots\cup \eta'_{j}, X_{j})$ -- the two assignments agree outside of $X_j$. In the other case, $\sigma_j$ satisfies $C_q$, hence $\sigma_1\cup\ldots\cup\sigma_j$ satisfies $C_q$ as well. Since $\sigma\vDash D_i$, we have $q\in I_j$. In both cases, we get $q\in I_j= \sat_j(\eta'_1\cup\ldots\cup \eta'_{j})$.

		\emph{Fact B: If $\gamma_i=\eta(\sigma)$, then $\sigma\vDash D_i$.}
		Let $j\in\{1,\ldots,m\}$ and $q\not \in I_j$ be arbitrary. We  want to show that $\sigma_1\cup\ldots\cup \sigma_{j}$ does not satisfy $C_q$. Assume, for the sake of contradiction, that it does. Then there exists a minimal $a\leq j$ such that $\sigma_a$ satisfies $C_q$. By the definition of $\eta(\sigma)=\gamma_i$, this implies $q\in\sat_a(\eta'_1\cup\ldots\cup\eta'_{a-1}\cup\sigma_a)\subseteq\sat_a(\eta'_1\cup\ldots\cup\eta'_a)=I_a$. But $I_a\subseteq I_j$, which contradicts the choice of $q$.
		We conclude that $\sigma$ falsifies $C_q|_{X_1\cup\ldots\cup X_j}$ for all $q\notin I_j$ and all $j$, which means that $\sigma\vDash D_i$.

		Now let $i^\ast$ be the index with $\gamma_{i^\ast}=\eta(\sigma)$, which exists because $\eta(\sigma)\in H$.
		By Fact B, we have $\sigma\vDash D_{i^\ast}$. Let $i\leq i^\ast$ be arbitrary with $\sigma\vDash D_i$ and $\gamma_i=(\eta'_1,\ldots,\eta'_m)$. From $\gamma_i\in K(\Phi,s)$ we know that each $\eta'_j$ is an optimal move for $\Phi|_{\game(s,\eta'_1\cup\ldots\cup\eta'_{j-1},U_{j-1})}$, and Fact A tells us that $\sat_j(\eta'_1\cup\ldots\cup \eta'_{j-1}\cup  \sigma_{j})\subseteq \sat_j(\eta'_1\cup\ldots\cup \eta'_{j})$ for each $j$. Let us compare $\gamma_i$ with $\eta(\sigma)$ block by block. If they do not differ in any block, then $\gamma_i=\eta(\sigma)$ and $i=i^\ast$. In the other case, let $j$ be the first block where they differ. Then $\eta'_1=\eta_1,\ldots,\eta'_{j-1}=\eta_{j-1}$, and by the definition of $\eta_j$ as the lexicographically minimal optimal move with the above property we obtain $\eta_j<\eta'_j$. But then $\eta(\sigma)<\gamma_i$ and $i^\ast<i$, which is a contradiction to $i\leq i^\ast$. Therefore, $i^\ast$ is the minimal index with $\sigma\vDash D_{i^\ast}$, which proves the equivalence of (i) and (ii).
	\end{claimproof}

	By the claim, for each $\sigma\in\langle X\rangle$ there is a minimal $i$ with $\sigma\vDash D_i$, hence $\bigvee_{i=1}^zD_i\equiv \top$. The claim also gives us $\mathcal{M}(\sigma)=\mu_{i_\sigma}=s(\gamma_{i_\sigma})=s(\eta(\sigma))=\tilde{s}(\sigma)$, so $\mathcal{M}$ computes exactly $\tilde{s}$. We already know that $\tilde{s}$ is a universal winning strategy, therefore $\mathcal{M}$ is a UDL with $|\mathcal{M}|=z\leq \cw^-(\Phi)^m$.
\end{proof}

Because of the lower bound of \caqe proofs via CAQE width (\cref{Prop:LowerBoundCAQEwidth}) and the UDL characterisation of $\qresnp$ (\cref{Prop:UDLcharacterizationQResNP}), as well as the fact that the QBFs $\mathtt{Copy}_n$ are obviously easy for $\qresnp$ but hard for \caqe (cf.\ \cref{Thm:SeparationCAQEfarpr}) we conclude:

\begin{corollary}\label{Cor:QresNPsimulatesCAQE}
	$\qresnp$ simulates \caqe on bounded quantifier alternation, but \caqe is exponentially weaker than $\qresnp$ as witnessed by $\mathtt{Copy}_n$.
\end{corollary}
\begin{proof}
	Let $\kappa$ be a \caqe refutation of a false QCNF $\Phi$ with $m$ universal blocks (after a possibly empty block $U_0$). Using \cref{Prop:LowerBoundCAQEwidth,Prop:cwUDL}, we know that $\Phi$ has a UDL of size at most $\cw^-(\Phi)^m\leq|\kappa|^m$, which is polynomial in $|\kappa|$ for bounded $m$. Now \cref{Prop:UDLcharacterizationQResNP} gives us that $\Phi$ has a $\qresnp$ refutation of size polynomial in $|\kappa|$.

	We will now show the separation. By \cref{Thm:SeparationCAQEfarpr}, $\mathtt{Copy}_n$ requires \caqe refutations of size $2^n$. On the other hand, $\mathcal{M}=([x_1],u_1\wedge\bar{u}_2\wedge\ldots\wedge\bar{u}_n),([\bar{x}_1],\bar{u}_1\wedge\bar{u}_2\wedge\ldots\wedge\bar{u}_n)$ is a UDL of size $2$ for $\mathtt{Copy}_n$. It computes the strategy $u_1\mapsto x_1$ and $u_i\mapsto 0$ for $i\geq 2$, under which the clauses $x_1\vee u_1\vee t_1$, $\bar{x}_1\vee\bar{u}_1\vee t_1$ and $(\bar{t}_1)$ cannot be satisfied simultaneously. Hence, $\mathtt{Copy}_n$ has $\qresnp$ refutations of size polynomial in $n$. In fact, we can directly obtain a \qres refutation of constant size from the $\farpr$ refutation in \cref{Thm:SeparationCAQEfarpr}.
\end{proof}

We remark that the simulation in \cref{Cor:QresNPsimulatesCAQE} is a simulation in terms of proof size. The UDL in \cref{Prop:cwUDL} is constructed from an optimal strategy $s$ and from the lexicographically minimal optimal chains, and we do not claim that it can be computed from $\kappa$ in polynomial time.

\section{CAQE vs QCDCL}\label{Sec:CAQEandQCDCL}

Finally, we compare \caqe and \qcdcl. We will show that there is a separation between \caqe and \qcdcl in one direction. For the other direction, at least on a specific but relevant class of QCNFs, namely $\Sigma_3^b$ formulas, \qcdcl can simulate \caqe proofs (after a simple preprocessing step). We utilise the \qcdcl proof system framework   used in~\cite{BPBSAT22journal,BohmB24-JAIR,BBM24journal} that proof-theoretically tightly captures the algorithmic QCDCL paradigm. Intuitively, a \qcdcl proof is a sequence of tuples of the form $(\mathcal{T}_i,C_i,\pi_i)$. The $\mathcal{T}_i$ are called \emph{trails}, which are (ordered) assignments containing all decided and propagated literals in the order in which they were assigned. Each $C_i$  is the clause learned from the trail $\mathcal{T}_i$ and  $\pi_i$ is the derivation of the learned clause $C_i$ from the axioms and previously learned clauses (e.g. via \qres). Below we recall the precise definitions from~\cite{BBM24journal}.

We have already shown that the equality formulas have exponential  CAQE width, hence they need exponential-size \caqe proofs by \cref{Prop:LowerBoundCAQEwidth}. We recall that $\mathtt{Eq}_n$ have
\qcdcl refutations of quadratic size if cube learning is used~\cite[Proposition~5.2]{BPB22IJCAIjournal} -- the construction given there uses $\mathcal{O}(n)$ trails of length $\mathcal{O}(n)$. These refutations learn cubes from total assignments; without cube learning, $\mathtt{Eq}_n$ requires exponential-size \qcdcl refutations~\cite{BB23-LMCS}. The \qcdcl proof system that we recall below (\cref{Def:QCDCLproofSystems}) includes cube learning. Hence we obtain:

\begin{theorem}  \label{thm:eq-sep}
	$\mathtt{Eq}_n$ require  \caqe refutations of size at least $2^n$ and have  \qcdcl refutations (with cube learning) of size $\mathcal{O}(n^2)$. Hence, \caqe does not simulate \qcdcl.
\end{theorem}
\begin{proof}
	Due to \cref{Prop:EqCW} we have $\cw^-(\mathtt{Eq}_n)=2^n$. Hence, each \caqe refutation of $\mathtt{Eq}_n$ has size at least $2^n$ by \cref{Prop:LowerBoundCAQEwidth}. The \qcdcl refutations of size $\mathcal{O}(n^2)$ follow from the construction in the proof of~\cite[Proposition~5.2]{BPB22IJCAIjournal}, as explained above.
\end{proof}

Before we turn to the other direction, let us briefly recall how \qcdcl works~\cite{ZM02,sathandbookqbf}. Like \cdcl, the algorithm builds up a partial assignment -- the \emph{trail} -- by alternately deciding variables and performing unit propagation. There are two differences to the propositional case. First, decisions have to respect the prefix: only variables from the leftmost quantifier block that is not yet completely assigned may be decided. Second, unit propagation is combined with universal reduction. A clause is considered unit if all its literals except one existential literal are falsified, where universal literals quantified right of this existential literal are ignored. In particular, universal literals are never propagated from clauses. If the trail falsifies a clause, we have run into a \emph{conflict}. As in \cdcl, the algorithm then analyses the conflict: it resolves the falsified clause with the antecedent clauses of propagated literals in reverse trail order, learns the resulting clause (after universal reduction) and backtracks to an earlier point of the trail. In this way, the learned clauses form a derivation in long-distance Q-resolution, and the QCNF is refuted as soon as the empty clause is learned. Symmetrically, \qcdcl can learn cubes from trails that satisfy the matrix, which is needed to verify true QCNFs. Cube learning is only needed for the separation in \cref{thm:eq-sep}; all refutations that we construct ourselves in this section learn clauses only.

One major difference between \caqe and \qcdcl is that \qcdcl does not directly call a SAT solver. Instead, \qcdcl itself is a natural extension of \cdcl and therefore contains all \cdcl functionality. This also means that it is not as trivial to discard propositional hardness from \qcdcl as it is with \caqe where computations by the SAT oracle are not counted towards \caqe proof size. Hence we need to define explicitly which steps of \qcdcl could be performed by \cdcl as well. Informally, a \cdcl step in a \qcdcl proof $\iota$ is a trail of $\iota$ (together with its learned clause) that can occur in a \cdcl proof as well. Such a \cdcl  step occurs if the formula under the current assignment (i.e.\ the trail) contains a subformula that is propositional, unsatisfiable and only contains variables from the (currently) first existential quantifier block. In such a situation, \qcdcl can run into a conflict by propagating and deciding variables from this subformula only without any reductions. The subsequent clause learning will therefore only perform operations that are also doable in \cdcl if we do not overstep the point at which the unsatisfiable subformula appeared. In order to formalise this, we first need to recall the \qcdcl proof system from~\cite{BBM24journal}.

\begin{definition}[quantification level]
	Let $\Phi:=Q_1V_1Q_2V_2\ldots Q_dV_d\cdot \varphi$ be a QBF with the quantifier blocks $V_1,\ldots,V_d$ (pairwise disjoint sets of variables) and quantifiers $Q_i\in \{ \exists,\forall \}$ such that $Q_j\neq Q_{j+1}$ for all $j$.  For some variable $v$ from $\Phi$ we set $\lv(v):=i$ if $v\in V_i$. Furthermore, we set $\lv(\bar{v}):=\lv(v)$. For literals $\ell_1,\ell_2$ we write $\ell_1<_\Phi \ell_2$ (or simply $\ell_1<\ell_2$ if the formula is obvious) if $\lv(\ell_1)<\lv(\ell_2)$.
\end{definition}

\begin{definition}[QDNF, AQBF]
	A \emph{QDNF} is a QBF $\mathcal{Q}\cdot \psi$ such that $\psi$ is  a disjunction of cubes (\emph{DNF}). An \emph{augmented QBF} (\emph{AQBF}) is a QBF of the form $\Phi=\mathcal{Q}\cdot (\varphi\vee \psi)$, such that $\varphi$ is a conjunction of clauses (CNF) and $\psi$ is a disjunction of cubes (DNF). We denote $\mathfrak{C}(\Phi):=\varphi$ and $\mathfrak{D}(\Phi):=\psi$.
\end{definition}

\begin{definition}[(Long-distance) Q-resolution and Q-consensus]

	Let $C_1$ and $C_2$ be two clauses (cubes). Let $\ell$ be a literal.

	The \emph{resolvent} of $C_1\vee \ell$ and $C_2\vee \bar{\ell}$ over $\ell$ is defined as
	$$
	(C_1\vee\ell)\resop{\ell}(C_2\vee\bar{\ell}):=C_1\vee C_2
	$$
	({resp. } $(C_1\wedge\ell)\resop{\ell}(C_2\wedge\bar{\ell}):=C_1\wedge C_2 )$.

	Let $C:=\ell_1\vee\ldots\vee \ell_n$ be a clause from a QCNF or AQBF $\Phi$ such that $\ell_i\leq_\Phi\ell_j$ for all $1\leq i<j\leq n$. Let $p\in\{1,\ldots,n+1\}$ be minimal such that all of $\ell_p,\ldots,\ell_n$ are universal (so $p=n+1$ if $\ell_n$ is existential). Then we can perform a \emph{universal reduction} step and obtain
	\begin{align*}
		\red_\Phi^\forall(C):=\ell_1\vee\ldots\vee\ell_{p-1}.
	\end{align*}

	Analogously, we  perform \emph{existential reduction}  on cubes, which we denote as $\red_\Phi^\exists(C)$.

	If it is clear whether $C$ is a clause or a cube, we can just write $\red_\Phi(C)$ or even $\red(C)$, if the QBF $\Phi$ is also obvious. 	We will write $\red(\Phi)=\red_\Phi(\Phi)$, if we reduce all clauses
	of the AQBF $\Phi$ according to its prefix.

	Long-distance resolution goes back to the solver of Zhang and Malik~\cite{ZM02}; Balabanov and Jiang~\cite{Balabanov12} formalised it as the proof system \ldqres (\ldqcon), which extends \qres (\qcon) proofs by replacing the resolution rule by
	\begin{itemize}
		\item \emph{Resolution (long-distance):} $C_i=C_j\resop{x}_\Phi C_k$ with $x$ existential (universal)
		and  $j,k<i$. The resolvent $C_i$ is allowed to contain tautologies such as $u\vee \bar{u}$ (resp. contradictions $u\wedge \bar{u}$).
		If there is such a universal (existential) variable $u$ that occurs in $C_j$ and $C_k$ with opposite polarities, then we  require $x<_\Phi u$.
	\end{itemize}
\end{definition}

\begin{definition}[trails,~\cite{BBM24journal}]
		A \emph{trail} $\mathcal{T}$ for a QCNF  or AQBF
	$\Phi$ is a (finite) sequence of pairwise distinct  literals from $\Phi$, including the empty literals $\bot$ and $\top$.  Each two literals in $\mathcal{T}$ have to correspond to pairwise distinct variables from $\Phi$.
	In general, a trail has the form
	\begin{align} \label{eq:trail}
		\begin{gathered}
			\mathcal{T}=(p_{(0,1)},\ldots, p_{(0,g_0)};\mathbf{d_{1}},p_{(1,1)},\ldots,p_{(1,g_{1})};\ldots; \mathbf{d_r},p_{(r,1)},\ldots ,p_{(r,g_r)}  )\text{,}
		\end{gathered}
	\end{align}
	where the $d_i$ are \emph{decision literals} and $p_{(i,j)}$ are \emph{propagated literals}. Decision literals are written in \textbf{boldface}. We use a semicolon before each decision to mark the end of a decision level. If one of the empty literals $\bot$ or $\top$ is contained in $\mathcal{T}$, then it has to be the last literal $p_{(r,g_r)}$. In this case, we say that $\mathcal{T}$ has \emph{run into a conflict}.

	Trails can be interpreted as non-tautological sets of literals, and therefore as (partial) assignments. We define the \emph{length} $|\mathcal{T}|$ of a trail as the number of its literals, where we do not count the empty literals $\bot$ and $\top$.
	We  write $x<_\mathcal{T}y$ if $x,y\in \mathcal{T}$ and $x$ is left of $y$ in $\mathcal{T}$. Furthermore, we write $x\leq_\mathcal{T}y$ if $x<_\mathcal{T}y$ or $x=y$.

	As trails are produced gradually from left to right in an algorithm, we define $\mathcal{T}[i,j]$ for $i\in \{0,\ldots,r\}$ and $j\in \{0,\ldots,g_i\}$ as the subtrail that contains all literals from $\mathcal{T}$ up to (and excluding) $p_{(i,j)}$ (resp. $d_i$, if $j=0$) in the same order. Intuitively, $\mathcal{T}[i,j]$ is the state of the trail before we assigned the literal at the point $[i,j]$ (which is $p_{(i,j)}$ or $d_i$).

	For each propagated literal $p_{(i,j)}$ in a trail $\mathcal{T}$ the formula must contain a clause or a cube that caused this propagation by becoming a unit clause/cube under the current trail assignment. We call such a  clause (cube)
	the \emph{antecedent clause (cube)} of $p_{(i,j)}$ and denote it by $\ante_{\mathcal{T}}(p_{(i,j)})$.

	We call a trail $\mathcal{T}$ \emph{natural} for a formula $\Phi$, if for each $i\in\{1,\ldots,r \}$ the formula $\red(\Phi|_{\mathcal{T}[i,0]})$ contains no unit or empty constraints. Furthermore,  the formula $\red(\Phi|_{\mathcal{T}[i,j]})$  must not contain empty constraints for each $i\in\{0,\ldots,r\}$, $j\in \{1,\ldots,g_i\}$, except $[i,j]=[r,g_r]$. Intuitively, this means that decisions are only made if there are no more propagations possible on the same decision level, and conflicts have to be taken care of immediately. Note that the first condition says nothing about the empty trail $\mathcal{T}[0,0]$: the formula $\Phi$ itself may contain unit clauses (for example learned unit clauses), and these are propagated on decision level $0$.
\end{definition}

\begin{definition}[learnable constraints,~\cite{BBM24journal}]\label{Def:LearnableConstraints}

	Let $\mathcal{T}$ be a
	trail for $\Phi$
	of the form \eqref{eq:trail}
	with $p_{(r,g_r)}\in \{\bot,\top\}$.
	Starting with $\ante_{\mathcal{T}}(\bot)$ (resp. $\ante_\mathcal{T}(\top)$) we reversely resolve with the antecedent clauses (cubes)  that were used to propagate the existential (universal) variables, until we stop at some point. Literals that were propagated via cubes (clauses) will be interpreted as decisions. If a resolution step cannot be performed at some point due to a missing pivot, we simply skip that antecedent. The clause (cube) we so derive, after universal (existential) reduction, is a \emph{learnable constraint}.
	We denote the sequence  of learnable constraints  by $\mathfrak{L}(\mathcal{T})$.

	We can also learn cubes from trails that did not run into conflict. If $\mathcal{T}$ is a total assignment of the variables from $\Phi$, then we define the set of learnable constraints as the set of cubes

	$\mathfrak{L}(\mathcal{T}):=	\{ \red_\Phi^\exists(D):\: D\subseteq \mathcal{T}\text{ and }D\text{ satisfies }\mathfrak{C}(\Phi)  \}$.

\end{definition}

\begin{definition}[QCDCL proof system,~\cite{BBM24journal}]\label{Def:QCDCLproofSystems}

	A $\qcdcl$ proof $\iota$ from a QCNF $\Phi=\mathcal{Q}\cdot \phi$ of a clause or cube $C$ is a sequence of triples

	$$
	\iota:=[(\mathcal{T}_i,C_i,\pi_i)]_{i=1}^n,
	$$

	where $C_n=C$, each $\mathcal{T}_i$ is a trail for the current formula $\Phi_i$ (defined below) such that
	\begin{itemize}
		\item for each decision $d_i$ we have that  $\lv_{\Phi|_{\mathcal{T}[i,0]}}(d_i)=1$ (i.e.,  decisions are level-ordered: only variables from the leftmost block that is not yet completely assigned may be decided),
		\item reductions during unit propagation have to be performed all the time --  formally,  for each propagated literal $p_{(i,j)}\in \mathcal{T}$ we need  $\red(\ante_{\mathcal{T}}(p_{(i,j)})|_{\mathcal{T}[i,j]})=(p_{(i,j)})$ (or $[\bar{p}_{(i,j)}]$),
		\item unit clauses (cubes) can only propagate existential (universal) literals -- formally, this means that if $p_{(i,j)}$ is existential (universal), then $\ante_{\mathcal{T}}(p_{(i,j)})$ has to be a clause (cube),
	\end{itemize}
	 each $C_i\in \mathfrak{L}(\mathcal{T}_i)$ is one of the constraints we can learn from the trail, and $\pi_i$ is the \ldqres or \ldqcon  derivation of $C_i$ we get by performing the steps in   \cref{Def:LearnableConstraints}.
	We define $\pf(\iota)$ as the extracted proof of $C$ that we get by sticking together suitable $\pi_i$.

	We set $\Phi_1:=\Phi$ and obtain $\Phi_{j+1}$ by adding the clause $C_j$ to $\mathfrak{C}(\Phi_j)$ (resp. adding the cube $C_j$ to $\mathfrak{D}(\Phi_j)$).
	We require that $\mathcal{T}_1$ is a natural trail and for each $2\leq i\leq n$ there is a point $[a_i,b_i]$ such that $\mathcal{T}_i[a_i,b_i]=\mathcal{T}_{i-1}[a_i,b_i]$ and $\mathcal{T}_i\backslash\mathcal{T}_i[a_i,b_i]$ has to be a natural   trail for $\Phi_i|_{\mathcal{T}_i[a_i,b_i]}$. This process is called \emph{backtracking}.
	If $\mathcal{T}_{i-1}[a_i,b_i]=\emptyset$, then this is also called a \emph{restart}.

	For $C=(\bot)$, $\iota$ is a $\qcdcl$ \emph{refutation} of $\Phi$. For $C=\left[\top\right]$, $\iota$ is a $\qcdcl$ \emph{verification} of $\Phi$.
	The \emph{trail size} of $\iota$ is defined as $|\iota|:=\sum_{i=1}^{n}|\mathcal{T}_i|$. Note that all \qcdcl refutations constructed in this paper only learn clauses. In particular, they do not learn cubes from total assignments.

\end{definition}

\begin{definition}[\cdcl steps]

	\label{Def:CDCLsteps}
	Let $\iota=[(\mathcal{T}_i,C_i,\pi_i)]_{i=1}^n$ be a \qcdcl refutation of a false QCNF $\Phi$. If there exist an existential quantifier block $\exists Y$ of $\Phi$ and a point $[a,b]$ in some $\mathcal{T}_j$ such that
	\begin{enumerate}
		\item all variables left of $Y$ are already assigned in $\mathcal{T}_j$ prior to $[a,b]$,
		\item  $\bigwedge_{\substack{C\in \red(\Phi_j|_{\mathcal{T}_j[a,b]}),{\var(C)\subseteq Y}}}C$
		is an unsatisfiable propositional formula (where $\red$ denotes the universal reduction of all clauses, which is performed anyway during unit propagation),
		\item after $[a,b]$ only variables from $Y$ get assigned,
		\item the point $[a,b]$ was not overstepped,  neither by the learning of $C_j$, nor by backtracking from $\mathcal{T}_j$ to $\mathcal{T}_{j+1}$ -- that means the derivation $\pi_j$ of $C_j$ only uses antecedent clauses of literals that were assigned after $[a,b]$, and $\mathcal{T}_{j+1}$ starts with $\mathcal{T}_j[a,b]$ (if $j<n$),
	\end{enumerate}
	then $(\mathcal{T}_j,C_j,\pi_j)$ is called a \emph{\cdcl step} in $\iota$. For $P:=\{ j:\: (\mathcal{T}_j,C_j,\pi_j) \text{ is a \cdcl step} \}$, we define $|\iota|_{\cdcl}:=\sum_{j\in P}|\mathcal{T}_j|$. Hence  $|\iota|-|\iota|_{\cdcl}$ measures ``genuine'' QCDCL size, discarding pure \cdcl runs.
\end{definition}

To ensure that such \cdcl steps are cleanly detectable, we require that  trails follow a certain structure. A way to achieve this is by imposing the \emph{XT-property} on the QCNF~\cite{BBSAT2021journal}. Informally, this XT-property causes the trails to be fully level-ordered (from $X$- to $U$- to $T$-variables), which is important so that propagations do not get mixed up too much. Originally, the XT-property was designed to obtain \qcdcl lower bounds (cf.~\cite{BBSAT2021journal}) as level-ordered trails turn out to be disadvantageous for QCDCL  runtime. Ironically, this XT-property will now help us to obtain upper bounds for \qcdcl.

\begin{definition}[XT-property,~\cite{BBSAT2021journal}]\label{Def:XTproperty}
	A $\Sigma_3^b$ QCNF with the prefix $\exists X\forall U\exists T$ fulfils the \emph{XT-property} if no two clauses $C_1,C_2$ with $\var(C_1),\var(C_2)\subseteq T$ (``T-clauses'') are resolvable and if  it contains no unit clauses $(\ell)$ with $\ell\in T$ (``unit T-clauses'') and no clauses $C$ with $\var(C)\cap X\neq \emptyset$, $\var(C)\cap T\neq \emptyset$ and $\var(C)\cap U= \emptyset$ (``XT-clauses'').
\end{definition}

Note that the XT-property still holds after learning a clause (cf.\ \cite{BBSAT2021journal}). This means that a formula with the XT-property will continue to fulfil it for the entire \qcdcl run. We can now show:

\begin{theorem}\label{Prop:QCDCLupperBoundViaCW}
	Let $\Phi$ be a false $\Sigma_3^b$ QCNF that fulfils the XT-property. Then there exists a \qcdcl refutation $\iota$ with $|\iota|-|\iota|_{\cdcl}\leq \cw^-(\Phi) \cdot |\var(\Phi)| $.
\end{theorem}
\begin{proof}
	Let $\Phi=\exists X\forall U\exists T\cdot C_1\wedge\ldots\wedge C_k$. We can assume that no clause of $\Phi$ consists of universal literals only. Otherwise $\red(\Phi)$ contains the empty clause, and the trail $(\bot)$ alone is a \qcdcl refutation $\iota$ with $|\iota|=0$. We call a clause \emph{$T$-free} if it contains no literal over $T$. For a $T$-free clause $C$ we have $\red(C)=C|_X$, because all universal literals of $C$ are quantified right of its existential literals.

	Let us first collect what we need from the XT-property. Since the XT-property is preserved by clause learning, it holds for all formulas $\Phi_j$ that occur during the run we construct. In particular, every clause of $\Phi_j$ that contains a literal over $T$ either contains a literal over $U$ as well, or it is a $T$-clause with at least two literals. Hence, as long as no variable from $T$ is assigned, such a clause can neither become unit nor empty. That means while only variables from $X$ are assigned, unit propagation only assigns literals over $X$, and the antecedent clauses and conflict clauses are $T$-free. \hfill $(\mathrm{P})$

	Now let
	\begin{align*}
		\mathcal{M}= (D_1,\mu_1),\ldots,(D_z,\mu_z)
	\end{align*}
	be the UDL that we have constructed in the proof of \cref{Prop:cwUDL} from a universal winning strategy $s$. Let us recall this construction for the case of a $\Sigma_3^b$ formula, where $m=1$ and $U_0=\emptyset$. Here we have $\game(s,\sigma,X)=\sigma$ for every $\sigma\in\langle X\rangle$, so the optimal chains with respect to $s$ are exactly the optimal moves $\sigma\in\langle X\rangle$ of the block $\exists X$, and the fruition of $\sigma$ is $(\sat(\sigma))$. The elements $\gamma_1<\ldots<\gamma_z$ of the set $H$ from that proof are therefore optimal moves in $\langle X\rangle$. They have pairwise different sets $I_i:=\sat(\gamma_i)$, as shown there, and we have $\xi_i=(I_i)$. Furthermore,
	\begin{align*}
		D_i=\bigwedge_{r\notin I_i}\neg\bigl(C_r|_X\bigr)\qquad\text{and}\qquad\mu_i=s(\gamma_i)
	\end{align*}
	for $i=1,\ldots,z$. By \cref{Prop:cwUDL} we have $z\leq \cw^-(\Phi)^m=\cw^-(\Phi)$. Because $\mathcal{M}$ is a UDL, we know that $\bigvee_{i=1}^zD_i\equiv\top$, hence $\bigwedge_{i=1}^z \neg D_i$ is unsatisfiable. Note that $\neg D_i$ is a clause over $X$, and that $C_r|_X\subseteq\neg D_i$ for every $r\notin I_i$.

	The plan is as follows. We construct the refutation $\iota$ in $z$ stages. In stage $i$ we make sure that the current formula contains a $T$-free clause $E_i$ with $\red(E_i)\subseteq \neg D_i$, and we use at most one trail that is not a \cdcl step for this. Once all these clauses are present, the formula $\bigwedge_{i=1}^z\red(E_i)$ is an unsatisfiable CNF over $X$, and we can learn the empty clause with \cdcl steps. During the construction we maintain the following invariant for every learned clause $L$:
	\begin{description}
		\item[($\ast$)] There is a nonempty set $R(L)\subseteq\{1,\ldots,k\}$ such that $L|_X=\bigcup_{r\in R(L)}C_r|_X$.
	\end{description}
	In other words, $L$ contains all $X$-literals of the original clauses it stems from, and no other $X$-literals. For an original clause we set $R(C_r):=\{r\}$. The invariant is preserved as long as the pivots of the resolution steps are variables from $T$: $X$-literals are never removed by such resolution steps, and they are never removed by universal reduction.

	The crucial consequence of ($\ast$) is the following. Let $L$ be an original or learned clause that fulfils ($\ast$), and assume that $\gamma_i$ falsifies all literals of $L|_X$. Then $\gamma_i$ falsifies $C_r|_X$ for every $r\in R(L)$. As $\gamma_i$ only assigns variables from $X$, it does not satisfy $C_r$, and we get $r\notin\sat(\gamma_i)=I_i$. Therefore, $D_i$ contains $\neg(C_r|_X)$, which means $C_r|_X\subseteq\neg D_i$. We conclude
	\begin{align}\label{eq:keyobservation}
		L|_X\subseteq\neg D_i\qquad\text{whenever $L$ fulfils ($\ast$) and $\gamma_i$ falsifies all literals of $L|_X$.}
	\end{align}

	We now describe stage $i$. Let $\Phi_j$ be the current formula, that is, $\Phi$ together with the clauses learned so far, all of which fulfil ($\ast$). If the empty clause has already been learned, we are done. Otherwise we build a natural trail as follows. We perform all possible unit propagations. Whenever no propagation is possible and $X$ is not yet completely assigned, we decide the next unassigned variable of $X$ with the value that $\gamma_i$ gives to it. By $(\mathrm{P})$, only literals over $X$ are propagated in this phase, and only $T$-free clauses are involved. We say that the trail is \emph{consistent} with $\gamma_i$ if all its literals belong to $\gamma_i$. We distinguish two cases.

	In the first case, a conflict occurs or a literal $\ell$ with $\bar{\ell}\in\gamma_i$ is propagated at some point while the trail is still consistent with $\gamma_i$. Let $C$ be the conflict clause (resp. the antecedent clause of $\ell$). By $(\mathrm{P})$, $C$ is $T$-free and $\red(C)=C|_X$. All literals of $C|_X$ are falsified by $\gamma_i$: the literals other than $\ell$ are falsified by the trail, which is consistent with $\gamma_i$, and we have $\bar{\ell}\in\gamma_i$. Since $C$ fulfils ($\ast$), \eqref{eq:keyobservation} gives us $\red(C)=C|_X\subseteq\neg D_i$. Hence, the current formula already contains a clause with the required property, and we set $E_i:=C$. In this case we do not add any trail to $\iota$ -- the trail we have considered only served to find the clause $E_i$.

	In the second case, no such event occurs. Then the trail assigns all variables of $X$ according to $\gamma_i$, no other variables are assigned, and no conflict has occurred. We continue the trail by deciding the variables of $U$ according to $\mu_i$, one after the other, each time followed by all possible propagations, until a conflict occurs or all of $U$ is assigned. Universal literals are never propagated from clauses, so the universal part of the trail is contained in $\mu_i$. Assume first that all of $U$ gets assigned without a conflict, and let $\mathcal{P}$ be the trail at this point. The strategy $s$ is winning, hence $\Phi|_{\gamma_i\cup\mu_i}$ is an unsatisfiable CNF over $T$. The clauses of $\red(\Phi_j|_{\mathcal{P}})$ over $T$ therefore form an unsatisfiable propositional CNF $\Psi$ as well: they include all clauses of $\Phi|_{\gamma_i\cup\mu_i}$ restricted by the literals over $T$ that were propagated in $\mathcal{P}$. That means the point $[a,b]$ at the end of $\mathcal{P}$ fulfils the first two conditions of \cref{Def:CDCLsteps} with $Y:=T$. We now run \cdcl on $\Psi$: we decide variables from $T$ and propagate until a conflict occurs; we learn a clause by resolving the conflict clause with the antecedent clauses of the literals over $T$ that were propagated after $[a,b]$, in reverse order of the trail, until the clause is asserting with respect to the decision levels after $[a,b]$; and we backtrack to the end of $\mathcal{P}$ or to a later decision level. Hence $[a,b]$ is never overstepped and all these trails are \cdcl steps. Note that the learned clauses contain literals over $X$ and $U$ as well, namely the $X$- and $U$-literals of the clauses involved, all of which are falsified by $\gamma_i\cup\mu_i$. In particular, no complementary universal literals occur in these resolution steps, and the invariant ($\ast$) is preserved because all pivots are variables from $T$. Since $\Psi$ is unsatisfiable, we arrive after finitely many \cdcl steps at a trail $\mathcal{T}^\ast$ in which a conflict occurs before any variable from $T$ is decided, that is, on a universal decision level. If a conflict already occurs while we decide the variables of $U$, we let $\mathcal{T}^\ast$ be this trail, and there are no \cdcl steps in this stage at all.

	In $\mathcal{T}^\ast$, all literals over $T$ are propagated. Let $C$ be the conflict clause of $\mathcal{T}^\ast$. We now resolve $C$ with the antecedent clauses of all propagated literals over $T$ in reverse order of the trail. Each antecedent clause has all its $X$-literals falsified by $\gamma_i$ and all its $U$-literals falsified by $\mu_i$ -- its universal literals cannot be reduced, since the clause contains a literal over $T$. The conflict clause $C$ also has all its $X$-literals falsified by $\gamma_i$, and each of its $U$-literals is falsified by $\mu_i$ or unassigned (in the latter case it was removed by reduction). Hence no complementary universal literals occur in these resolution steps, and we obtain a clause over $X\cup U$. Universal reduction removes all its literals over $U$, and we learn the resulting clause $E_i$ over $X$, whose literals are all falsified by $\gamma_i$. This is a learnable constraint in the sense of \cref{Def:LearnableConstraints}: we stop the reverse resolution before we reach the antecedents of the literals over $X$. The clause $E_i$ fulfils ($\ast$), because all pivots were variables from $T$. By \eqref{eq:keyobservation}, we get $E_i\subseteq\neg D_i$. After learning $E_i$, we restart, that is, the next trail begins with the empty trail. The trail $\mathcal{T}^\ast$ is the only trail of this stage that is possibly not a \cdcl step.

	After stage $z$, the current formula contains $T$-free clauses $E_1,\ldots,E_z$ with $\red(E_i)\subseteq\neg D_i$ for $i=1,\ldots,z$. As $\bigwedge_{i=1}^z\neg D_i$ is unsatisfiable, the CNF $\bigwedge_{i=1}^z\red(E_i)$ over $X$ is unsatisfiable, and so is the CNF that consists of all clauses of $\red(\Phi_j)$ over $X$. We refute it by \cdcl steps $\mathcal{V}_1,\ldots,\mathcal{V}_c$ for some $c\in\N$: we decide and propagate variables from $X$ only -- by $(\mathrm{P})$ no other variables get assigned -- and learn clauses by resolving over variables from $X$. Complementary universal literals may occur in these resolution steps, for instance when we resolve $x\vee u$ with $\bar{x}\vee\bar{u}$, but this is allowed by the long-distance resolution rule, as all universal variables are quantified right of $X$, and the universal literals are removed by reduction afterwards. These trails are \cdcl steps with $Y:=X$ and the point $[a,b]=[0,0]$ at the beginning of the trail: the first condition of \cref{Def:CDCLsteps} is void, the second one holds since the clauses of $\red(\Phi_j)$ over $X$ are unsatisfiable, and the beginning of the trail is never overstepped.

	All in all, $\iota$ consists of the trails of the stages $1,\ldots,z$ followed by $\mathcal{V}_1,\ldots,\mathcal{V}_c$, and the only trails that are not \cdcl steps are the trails $\mathcal{T}^\ast$ of the stages that fall into the second case. There are at most $z\leq\cw^-(\Phi)$ of them. Each trail assigns every variable at most once, hence it has length at most $|\var(\Phi)|$ (recall that $\bot$ is not counted). We conclude that $|\iota|-|\iota|_{\cdcl}\leq\cw^-(\Phi)\cdot|\var(\Phi)|$.
\end{proof}

\begin{remark}\label{Rem:XTPropertyPreprocessing}
	If $\Phi=\exists X\forall U \exists T\cdot C_1\wedge\ldots\wedge C_k$ is a false $\Sigma_3^b$ QCNF that does not fulfil the XT-property, then we can transform it into a false $\Sigma_3^b$ QCNF with the XT-property as follows: Let $w$ be a new variable. We set $C'_r:=w\vee C_r$ for all $r=1,\ldots,k$ and define $\Phi':=\exists X\forall U\forall w\exists T\cdot C'_1\wedge\ldots\wedge C'_k$. The QCNF $\Phi'$ is false, because we get $\Phi$ back if we set $w$ to $0$, and it fulfils the XT-property, because every clause contains the universal literal $w$. Since we do not alter the first block, we even get $\cw^-(\Phi)=\cw^-(\Phi')$. By allowing this preprocessing step, we can extend the upper bound from   \cref{Prop:QCDCLupperBoundViaCW} to arbitrary false $\Sigma_3^b$ QCNFs.
\end{remark}

\begin{corollary}\label{Cor:QCDCLsimulatesCAQE}
	\qcdcl simulates \caqe on false $\Sigma_3^b$ QCNFs that fulfil the XT-property, under the condition that  \cdcl steps in \qcdcl are not counted. More precisely, every \caqe refutation $\kappa$ of such a QCNF $\Phi$ can be turned into a \qcdcl refutation $\iota$ of $\Phi$ with $|\iota|-|\iota|_{\cdcl}\leq |\kappa|\cdot|\var(\Phi)|$.
	By adding the preprocessing step as described in    \cref{Rem:XTPropertyPreprocessing}, \qcdcl even simulates \caqe on all  false $\Sigma_3^b$ QCNFs. In this case, $\iota$ is a \qcdcl refutation of the preprocessed QCNF $\Phi'$ with $|\iota|-|\iota|_{\cdcl}\leq |\kappa|\cdot(|\var(\Phi)|+1)$.
\end{corollary}
\begin{proof}
	By \cref{Prop:LowerBoundCAQEwidth}, we have $\cw^-(\Phi)\leq|\kappa|$, and the first bound follows from \cref{Prop:QCDCLupperBoundViaCW}. For the second statement, \cref{Rem:XTPropertyPreprocessing} gives us $\cw^-(\Phi')=\cw^-(\Phi)\leq|\kappa|$ and $|\var(\Phi')|=|\var(\Phi)|+1$, so we can apply \cref{Prop:QCDCLupperBoundViaCW} to $\Phi'$.
\end{proof}

As in \cref{Cor:QresNPsimulatesCAQE}, this is a simulation in terms of proof size. The refutation $\iota$ is constructed from the UDL of \cref{Prop:cwUDL}, which we obtain from an optimal strategy, and we do not claim that $\iota$ can be computed from $\kappa$ in polynomial time.

We conclude this section with an example that illustrates the construction in the proof of \cref{Prop:QCDCLupperBoundViaCW}.

\begin{example}\label{Example:QCDCL}
Let $\Phi$ be the QCNF from   \cref{Example:OptimalAssignment}   with prefix $\exists x_1,x_2\forall u_1,u_2\exists y_1,y_2$ and the matrix

\[\begin{array}{lll}
	C_1=x_1\vee u_1\vee y_1 & C_2=x_1\vee x_2\vee u_2\vee y_2 & C_3=\bar{y}_1\vee \bar{y}_2 \\
	C_4=\bar{x}_1\vee\bar{u}_1\vee y_1 &C_5=\bar{x}_1\vee\bar{u}_2\vee y_2 &  C_6=\bar{x}_2\vee\bar{u}_2\vee y_2.
\end{array}\]
Note that $\Phi$ fulfils the XT-property: there are no XT-clauses, and the only T-clause is $C_3$, which is not a unit clause. Let $s$  be the winning strategy for $\Phi$ that sets $s(\gamma):=(u_1\mapsto (x_1)|_\gamma,u_2\mapsto (x_1\vee x_2)|_\gamma)$ for $\gamma\in \langle x_1,x_2\rangle$. We now want to construct a UDL $\mathcal{M}$ of size at most $\cw^-(\Phi)=3$ and a \qcdcl refutation $\iota$ for $\Phi$ with $|\iota|-|\iota|_\cdcl\leq \cw^-(\Phi)\cdot |\var(\Phi)|=3\cdot 6=18$.

We start with computing the set of all optimal  chains $K(\Phi,s)$. Since $\Phi$ is a $\Sigma_3^b$ formula, $K(\Phi,s)$ does not depend on $s$ and only consists of optimal moves on the first quantifier block $\exists x_1,x_2$. In   \cref{Example:OptimalAssignment} we have already computed all optimal moves for $\Phi$, hence we obtain $K(\Phi,s)=\{ \gamma_1,\gamma_2,\gamma_3 \}$ with $\gamma_1=\bar{x}_1\wedge \bar{x}_2$, $\gamma_2=\bar{x}_1\wedge x_2$ and $\gamma_3=x_1\wedge \bar{x}_2$. Next we compute their fruitions, which are $\abs(\gamma_1)=(\{ 4,5,6 \})=:\xi_1$, because $\gamma_1$ satisfies precisely $C_4$, $C_5$ and $C_6$, and $\abs(\gamma_2)=(\{ 2,4,5 \})=:\xi_2$ and $\abs(\gamma_3)=(\{1,2,6\})=:\xi_3$. Because $\Phi$ is a $\Sigma_3^b$ formula, each fruition only consists of one set of indices. Note that the fruition of the  assignment $x_1\wedge x_2$ is $(\{ 1,2 \})$, which is non-optimal because $\{1,2\}$ is a proper subset of $\{1,2,6\}$. We conclude $F(\Phi,s)=\{\xi_1,\xi_2,\xi_3\}$. In this example, we have as many optimal fruitions as optimal moves. In general, we might have $|K(\Phi,s)|>|F(\Phi,s)|$ because different optimal chains may have the same fruitions.

Now we can construct the cubes for our UDL by negating the clauses whose indices do not appear in the corresponding fruition, hence
\begin{align*}
	D_1&=\bigwedge_{r\not\in \{ 4,5,6\}}\neg C_r|_X=(\neg C_1\wedge \neg C_2\wedge \neg C_3)|_X=\bar{x}_1\wedge \bar{x}_2\\
	D_2&=\bigwedge_{r\not\in \{2,4,5\}}\neg C_r|_X=(\neg C_1\wedge \neg C_3\wedge \neg C_6)|_X=\bar{x}_1\wedge x_2\\
	D_3&=\bigwedge_{r\not\in \{ 1,2,6\}}\neg C_r|_X=(\neg C_3\wedge \neg C_4\wedge \neg C_5)|_X=[x_1].
\end{align*}
We  define the assignments $\mu_1,\mu_2,\mu_3\in \langle u_1,u_2\rangle$ according to the strategy $s$, hence $\mu_1=\bar{u}_1\wedge \bar{u}_2$, $\mu_2=\bar{u}_1\wedge u_2$ and $\mu_3=u_1\wedge u_2$. We do not need to alter $s$ since it is the only winning strategy for $\Phi$ anyway. It is easy to see that
\begin{align*}
	\mathcal{M}=(D_1,\mu_1),(D_2,\mu_2),(D_3,\mu_3)=(\bar{x}_1\wedge \bar{x}_2,\bar{u}_1\wedge \bar{u}_2),(\bar{x}_1\wedge x_2,\bar{u}_1\wedge u_2),(x_1,u_1\wedge u_2)
\end{align*}
is a UDL that computes precisely the strategy $s$.

At the end we want to construct a \qcdcl refutation of $\Phi$ consisting of at most $3$ trails that are not \cdcl steps, and for this we follow the proof of \cref{Prop:QCDCLupperBoundViaCW}.\footnote{In the notation of the trails, we suppress an empty decision level $0$.} In stage $1$ we assign $X$ according to $\gamma_1$ and then $U$ according to $\mu_1$. By the construction in the proof, the first trail $\mathcal{T}_{1,1}$ looks as follows -- decisions are blue, propagated literals are white with their antecedent clauses below, and the conflict is red:

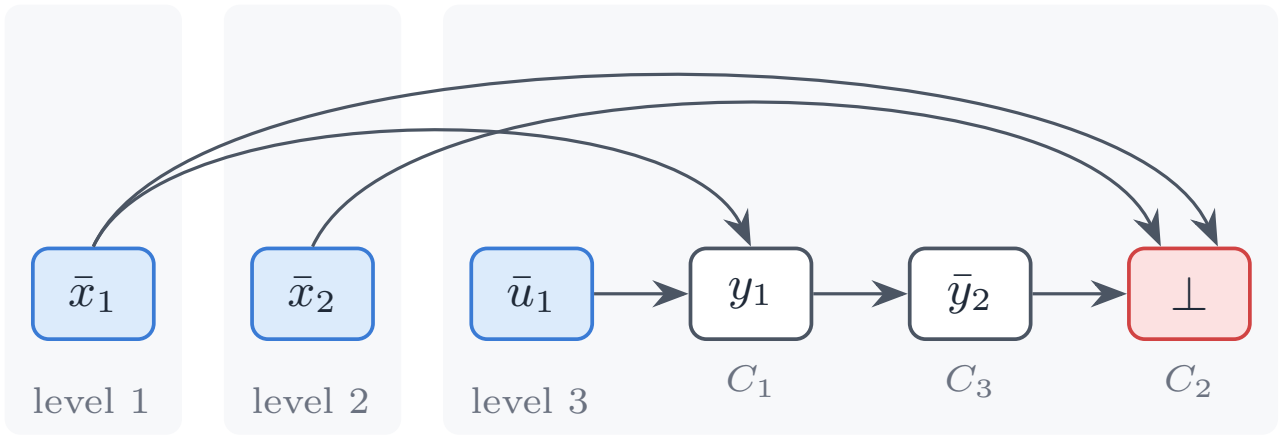
\begin{center}
\begin{tikzpicture}[fig, x=1.45cm,
		td/.style={frec, minimum width=8mm, minimum height=6mm, inner sep=1.5pt},
		tp/.style={fwhite, minimum width=8mm, minimum height=6mm, inner sep=1.5pt},
		tc/.style={fconf, minimum width=8mm, minimum height=6mm, inner sep=1.5pt},
		imp/.style={fedge, draw=figNeutralDraw, line width=0.6pt}]
	\node[td] (x1) at (0,0) {$\bar{x}_1$};
	\node[td] (x2) at (1,0) {$\bar{x}_2$};
	\node[td] (u1) at (2,0) {$\bar{u}_1$};
	\node[tp] (y1) at (3,0) {$y_1$};
	\node[tp] (y2) at (4,0) {$\bar{y}_2$};
	\node[tc] (bot) at (5,0) {$\bot$};
	\begin{scope}[on background layer]
		\fill[figBand, rounded corners=3pt] ($(x1.south west)+(-0.12,-0.62)$) rectangle ($(x1.north east)+(0.12,1.6)$);
		\node[font=\scriptsize, text=figGrey, anchor=south west] at ($(x1.south west)+(-0.08,-0.6)$) {level 1};
		\fill[figBand, rounded corners=3pt] ($(x2.south west)+(-0.12,-0.62)$) rectangle ($(x2.north east)+(0.12,1.6)$);
		\node[font=\scriptsize, text=figGrey, anchor=south west] at ($(x2.south west)+(-0.08,-0.6)$) {level 2};
		\fill[figBand, rounded corners=3pt] ($(u1.south west)+(-0.12,-0.62)$) rectangle ($(bot.north east)+(0.12,1.6)$);
		\node[font=\scriptsize, text=figGrey, anchor=south west] at ($(u1.south west)+(-0.08,-0.6)$) {level 3};
	\end{scope}
	\draw[imp] (u1) -- (y1);
	\draw[imp] (x1.north) to[out=65,in=115,distance=1.11cm] ($(y1.north)+(0.00,0)$);
	\draw[imp] (y1) -- (y2);
	\draw[imp] (y2) -- (bot);
	\draw[imp] (x2.north) to[out=65,in=115,distance=1.38cm] ($(bot.north)+(-0.13,0)$);
	\draw[imp] (x1.north) to[out=65,in=115,distance=1.65cm] ($(bot.north)+(0.13,0)$);
	\node[font=\scriptsize, text=figGrey, below=1pt of y1] {$C_1$};
	\node[font=\scriptsize, text=figGrey, below=1pt of y2] {$C_3$};
	\node[font=\scriptsize, text=figGrey, below=1pt of bot] {$C_2$};
\end{tikzpicture}
\end{center}
\begin{align*}
	\mathcal{T}_{1,1}=(\mathbf{\bar{x}_1};\mathbf{\bar{x}_2};\mathbf{\bar{u}_1},y_1,\bar{y}_2,\bot)
\end{align*}
Here the conflict occurs even before $u_2$ is decided, since the clause $C_2$ becomes the unit clause $(u_2)$, which is reduced to the empty clause. We do not need to perform any \cdcl steps and can immediately learn the clause $E_1=x_1\vee x_2=\neg D_1$ by resolving $C_2$ with the antecedents $C_3$ and $C_1$ of $\bar{y}_2$ and $y_1$ and reducing the universal literals $u_1,u_2$. In general, we may need some \cdcl steps before we can learn a subclause of $\neg D_1$, but these \cdcl steps would be ignored for the proof size anyway. We continue with stage $2$ and the trail $\mathcal{T}_{2,1}$:

\begin{center}
\begin{tikzpicture}[fig, x=1.45cm,
		td/.style={frec, minimum width=8mm, minimum height=6mm, inner sep=1.5pt},
		tp/.style={fwhite, minimum width=8mm, minimum height=6mm, inner sep=1.5pt},
		tc/.style={fconf, minimum width=8mm, minimum height=6mm, inner sep=1.5pt},
		imp/.style={fedge, draw=figNeutralDraw, line width=0.6pt}]
	\node[td] (x1) at (0,0) {$\bar{x}_1$};
	\node[tp] (x2) at (1,0) {$x_2$};
	\node[td] (u1) at (2,0) {$\bar{u}_1$};
	\node[tp] (y1) at (3,0) {$y_1$};
	\node[tp] (y2) at (4,0) {$\bar{y}_2$};
	\node[tc] (bot) at (5,0) {$\bot$};
	\begin{scope}[on background layer]
		\fill[figBand, rounded corners=3pt] ($(x1.south west)+(-0.12,-0.62)$) rectangle ($(x2.north east)+(0.12,1.6)$);
		\node[font=\scriptsize, text=figGrey, anchor=south west] at ($(x1.south west)+(-0.08,-0.6)$) {level 1};
		\fill[figBand, rounded corners=3pt] ($(u1.south west)+(-0.12,-0.62)$) rectangle ($(bot.north east)+(0.12,1.6)$);
		\node[font=\scriptsize, text=figGrey, anchor=south west] at ($(u1.south west)+(-0.08,-0.6)$) {level 2};
	\end{scope}
	\draw[imp] (x1) -- (x2);
	\draw[imp] (u1) -- (y1);
	\draw[imp] (x1.north) to[out=65,in=115,distance=1.11cm] ($(y1.north)+(0.00,0)$);
	\draw[imp] (y1) -- (y2);
	\draw[imp] (y2) -- (bot);
	\draw[imp] (x2.north) to[out=65,in=115,distance=1.38cm] ($(bot.north)+(0.00,0)$);
	\node[font=\scriptsize, text=figGrey, below=1pt of x2] {$E_1$};
	\node[font=\scriptsize, text=figGrey, below=1pt of y1] {$C_1$};
	\node[font=\scriptsize, text=figGrey, below=1pt of y2] {$C_3$};
	\node[font=\scriptsize, text=figGrey, below=1pt of bot] {$C_6$};
\end{tikzpicture}
\end{center}
\begin{align*}
	\mathcal{T}_{2,1}=(\mathbf{\bar{x}_1},{{x}_2};\mathbf{\bar{u}_1},y_1,\bar{y}_2,\bot)
\end{align*}
where $x_2$ is propagated by the learned clause $E_1$, which is consistent with $\gamma_2$. This time the conflict occurs with $C_6$. Resolving $C_6$ with the antecedents $C_3$ and $C_1$ of $\bar{y}_2$ and $y_1$ and applying reduction, we get $E_2=x_1\vee\bar{x}_2=\neg D_2$, again without any \cdcl steps. Finally, stage $3$ uses the trail $\mathcal{T}_{3,1}$:

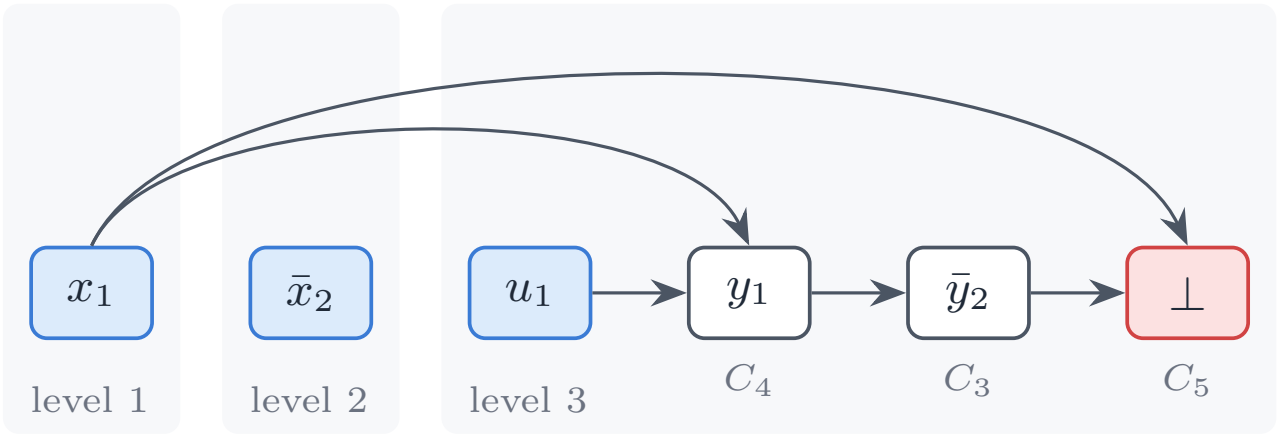
\begin{center}
\begin{tikzpicture}[fig, x=1.45cm,
		td/.style={frec, minimum width=8mm, minimum height=6mm, inner sep=1.5pt},
		tp/.style={fwhite, minimum width=8mm, minimum height=6mm, inner sep=1.5pt},
		tc/.style={fconf, minimum width=8mm, minimum height=6mm, inner sep=1.5pt},
		imp/.style={fedge, draw=figNeutralDraw, line width=0.6pt}]
	\node[td] (x1) at (0,0) {$x_1$};
	\node[td] (x2) at (1,0) {$\bar{x}_2$};
	\node[td] (u1) at (2,0) {$u_1$};
	\node[tp] (y1) at (3,0) {$y_1$};
	\node[tp] (y2) at (4,0) {$\bar{y}_2$};
	\node[tc] (bot) at (5,0) {$\bot$};
	\begin{scope}[on background layer]
		\fill[figBand, rounded corners=3pt] ($(x1.south west)+(-0.12,-0.62)$) rectangle ($(x1.north east)+(0.12,1.6)$);
		\node[font=\scriptsize, text=figGrey, anchor=south west] at ($(x1.south west)+(-0.08,-0.6)$) {level 1};
		\fill[figBand, rounded corners=3pt] ($(x2.south west)+(-0.12,-0.62)$) rectangle ($(x2.north east)+(0.12,1.6)$);
		\node[font=\scriptsize, text=figGrey, anchor=south west] at ($(x2.south west)+(-0.08,-0.6)$) {level 2};
		\fill[figBand, rounded corners=3pt] ($(u1.south west)+(-0.12,-0.62)$) rectangle ($(bot.north east)+(0.12,1.6)$);
		\node[font=\scriptsize, text=figGrey, anchor=south west] at ($(u1.south west)+(-0.08,-0.6)$) {level 3};
	\end{scope}
	\draw[imp] (u1) -- (y1);
	\draw[imp] (x1.north) to[out=65,in=115,distance=1.11cm] ($(y1.north)+(0.00,0)$);
	\draw[imp] (y1) -- (y2);
	\draw[imp] (y2) -- (bot);
	\draw[imp] (x1.north) to[out=65,in=115,distance=1.65cm] ($(bot.north)+(0.00,0)$);
	\node[font=\scriptsize, text=figGrey, below=1pt of y1] {$C_4$};
	\node[font=\scriptsize, text=figGrey, below=1pt of y2] {$C_3$};
	\node[font=\scriptsize, text=figGrey, below=1pt of bot] {$C_5$};
\end{tikzpicture}
\end{center}
\begin{align*}
	\mathcal{T}_{3,1}=(\mathbf{x_1};\mathbf{\bar{x}_2};\mathbf{{u}_1},y_1,\bar{y}_2,\bot)
\end{align*}
where the conflict with $C_5$ occurs before $u_2$ is decided. We resolve $C_5$ with the antecedents $C_3$ and $C_4$ of $\bar{y}_2$ and $y_1$, apply reduction and get $E_3=(\bar{x}_1)=\neg D_3$. In general, it is not guaranteed that we learn the full clauses $\neg D_i$. Instead, we can only guarantee to learn a subclause of the current $\neg D_i$ (or to find such a clause among the clauses already present). Now the clauses $E_1,E_2,E_3$ over $X=\{x_1,x_2\}$ form an unsatisfiable CNF, because $\bigwedge_{i=1}^3\neg D_i$ is unsatisfiable. In the final trail $\mathcal{V}_1=(\bar{x}_1,x_2,\bot)$, the literals $\bar{x}_1$ and $x_2$ are propagated by $E_3$ and $E_1$, and the conflict occurs with $E_2$. This trail is a \cdcl step, and we learn the empty clause $(\bot)$ by resolving $E_2$ with $E_1$ and $E_3$. In general, we might need several \cdcl steps at this point.

For the \qcdcl refutation $\iota=(\mathcal{T}_{1,1},\mathcal{T}_{2,1},\mathcal{T}_{3,1},\mathcal{V}_1)$ we have constructed, we conclude $|\iota|-|\iota|_\cdcl=|\mathcal{T}_{1,1}|+|\mathcal{T}_{2,1}|+|\mathcal{T}_{3,1}|=15\leq 3\cdot |\var(\Phi)|$.
\end{example}

\section{Conclusion}
\label{sec:concl}

We have proven the possibly surprising result that, in a theoretical setting in which SAT calls and purely propositional CDCL steps are not counted, the basic version of CAQE (without failed assumptions) is weaker than QCDCL on $\Sigma_3^b$ formulas (after a simple preprocessing step) and weaker than Q-resolution on all formulas of bounded quantifier alternation (cf.\    \cref{Fig:SimulationOrderCAQE}), while in practice CAQE outperforms almost all competing QBF solvers. This indicates that optimisations such as failed assumptions might play an important role for the performance of the solver. In future work, our lower and upper bound techniques could be used to prove separations between different variants of CAQE formally, and hence justify the use of modifications on a proof complexity level. 
 Furthermore,  the fact that we discarded propositional hardness in both approaches may be a very relevant factor in comparing these two solvers. It can be assumed that  direct usage of a SAT solver is a huge advantage for CAQE, while QCDCL as an extension of CDCL utilises SAT solving in a rather intricate and hidden way. Given the relatively simple setup of CAQE (Figure~\ref{Fig:CAQE}) this also means that CAQE can at least in principle use the latest state-of-the-art SAT solver without any changes to its implementation, while QCDCL does not have access to the latest SAT technology without major implementation effort.

We have also shown that, discounting propositional hardness, the proof size of \caqe is essentially determined by the number of optimal moves, counted up to their effect on the clauses.  Further research into this direction may have the potential to not only improve the performance of CAQE itself, but also lead to decision heuristics for QCDCL.

\bibliography{compl,FurtherInsidesECCC,furtherRefs,someJournalRefs,caqefa}

\end{document}